# Generalized failure law for landslides, rockbursts, glacier breakoffs, and volcanic eruptions

Qinghua Lei[1]*, Didier Sornette[2]

[1] Department of Earth Sciences, Uppsala University; Uppsala, 752 36, Sweden.

[2] Institute of Risk Analysis, Prediction and Management, Academy for Advanced Interdisciplinary Studies, Southern University of Science and Technology; Shenzhen, 518055, China.

*Corresponding author. Email: qinghua.lei@geo.uu.se

**Abstract**

Catastrophic failures have momentous impact in many scientific and technological fields but remain challenging to understand and predict. One key difficulty lies in the burstiness of rupture phenomena, which typically involve a series of progressively shorter quiescent phases punctuated by sudden bursts, rather than a smooth continuous progression. This seemingly erratic pattern challenges the conventional power law assumption of continuous scale invariance. Here, we propose a generalized material failure law based on the log-periodic power law, which better captures the discrete scale invariance inherent in intermittent rupture dynamics. Our method's superiority is demonstrated through testing on 109 historical geohazard events, including landslides, rockbursts, glacier breakoffs, and volcanic eruptions. The results indicate that our method is general and robust, offering significant potential to forecast catastrophic failures.

**Introduction**

Catastrophic failure occurs in a wide spectrum of geological materials such as rock, soil, and ice, driving various extreme geohazards like landslides, rockbursts, glacier breakoffs, and volcanic eruptions that jeopardize life and property (*1–4*). It is thus crucial to develop the best possible predictive understanding of these rupture phenomena, which is a fundamental goal of many scientific and technological disciplines, such as geology, seismology, glaciology, volcanology, mechanics, and engineering. Various empirical and physical approaches have been proposed to describe geomaterial failure, with the power law singularity (PLS) model and the derived inverse rate technique (*5–7*) being widely adopted for time-to-failure analysis of geohazard events (*8–15*). Over the past decades, great efforts have also been devoted to develop and deploy high-precision monitoring technologies to observe various geohazards (*3*, *13*, *16*, *17*). However, only a limited number of catastrophic events have been successfully predicted so far. One major uncertainty stems from the sporadic nature of rupture in heterogeneous materials, typically marked by a sequence of progressively shorter quiescent phases interspersed with sudden intense bursts (*18–21*), rather than a smooth continuous progression of deformation and damage. This seemingly erratic pattern complicates failure predictions as it challenges the continuous scale invariance assumed by the simple power law (*5–8*).

To address this issue, we propose a generalized failure law based on the log-periodic power law singularity (LPPLS) model. Amounting mathematically to a generalization of the power law exponent from real to complex numbers, this new failure law captures the partial break of continuous scale invariance to discrete scale invariance (*22*) that is inherent to the intermittent dynamics of damage and rupture processes in heterogeneous materials. We demonstrate the superiority of our LPPLS model over the conventional PLS model using real datasets (geodetic observations, geophysical records, and geochemical measurements) of 109 historical catastrophic events, including landslides, rockbursts, glacier breakoffs, and volcanic eruptions.

**Theory**

The response of a material prior to a catastrophic failure is generically governed by the following nonlinear dynamic equation (*5, 6*):

$$\ddot{\Omega} = \eta \dot{\Omega}^{\alpha}, \text{ with } \alpha > 1, \tag{1}$$

where $\Omega$ is an observable quantity (e.g., displacement, strain, energy release, gas emission), $\eta$ is a constant, and $\alpha$ is the exponent defining the degree of nonlinearity. The condition of $\alpha > 1$ ensures the presence of positive feedbacks (*23*), where the instantaneous growth rate of $\dot{\Omega}$ defined as $d(\ln \dot{\Omega})/dt$ increases super-linearly as a function of $\dot{\Omega}$. This corresponds to a super-exponential dynamic ending with a finite-time singularity at which a catastrophic failure occurs. This is seen by integrating equation (1), which yields:

$$\dot{\Omega} = \kappa (t_c - t)^{-\xi}, \text{ with } \xi > 0, \tag{2}$$

where $\kappa = (\xi/\eta)^{\xi}$, $\xi = 1/(\alpha - 1)$, $t$ is time, and $t_c$ is the time of failure determined from the initial condition $\dot{\Omega}(t = t_0) = \dot{\Omega}_0$. Here, $\xi > 0$ (for $\alpha > 1$) ensures that $\dot{\Omega}$ exhibits a singular behavior at $t = t_c$. A further integration of equation (2) leads to the PLS formulation (*5–8*) (see Materials and Methods):

$$\Omega(t) = A + B(t_c - t)^m, \text{ with } m < 1, \tag{3}$$

where $A$ and $B = -\kappa/m$ are constants, and $m = 1 - \xi$ is a critical exponent. Equation (3) is the general solution of $\Omega$ for $\alpha > 1$, $\xi > 0$, and thus $m < 1$ including $m = 0$ (see Materials and Methods), corresponding to an acceleration up to $t_c$. For $\alpha > 2$, $0 < \xi < 1$ and thus $0 < m < 1$, $\dot{\Omega}$ diverges at $t_c$ but $\Omega$ converges to the finite value $A$; for $1 < \alpha < 2$, $\xi > 1$ and thus $m < 0$, both $\dot{\Omega}$ and $\Omega$ diverge at $t_c$.

We now consider a generalized scenario where the critical exponent is extended from real to complex numbers. Then, the first-order Fourier expansion of the general solution of $\Omega$ leads to the following LPPLS formula (*19, 22*) (see Materials and Methods):

$$\Omega(t) = A + \{B + C \cos[\omega \ln(t_c - t) - \phi]\}(t_c - t)^m, \text{ with } m < 1, \tag{4}$$

expressing a log-periodic correction to the power law scaling, where $C$ is a constant, $\omega$ is the angular log-periodic frequency, and $\phi$ is a phase shift. Incorporating a discrete hierarchy of time scales (see Materials and Methods), the LPPLS model captures the non-monotonous dynamics of the system with a geometric increase in burst frequency on the approach to $t_c$. A mechanism for this log-periodicity relies on the cascade of Mullins-Sekerka instabilities of competing growing cracks (*24*) with a relative amplitude typically on the order of $10^{-1}$ (*22*), driven by the localized and threshold nature of the mechanics of rupture in heterogeneous materials (*20, 21*). The LPPLS model is applicable to both constant and varying loading conditions, as demonstrated in the previous experimental and numerical studies (*18, 20, 25–27*).

**Applications**

We extensively test the LPPLS versus PLS models in the diverse contexts of landslides, rockbursts, glacier breakoffs, and volcanic eruptions. We implement a stable and robust model calibration scheme, briefly described as follows (see Materials and Methods for detailed procedures). First, the Lagrange regularization approach (*28*) is employed to endogenously detect the onset of the failure crisis, based on which the optimal time window is defined; here, the end of this time window is fixed at the last available data point prior to the failure. Then, the optimal parameter values for the LPPLS or PLS model are determined by minimizing the sum of the squares of the residuals, which quantifies the difference between the model and data (*29*). The LPPLS model has seven parameters $t_c$, $m$, $\omega$, $\phi$, $A$, $B$, and $C$ (with the four linear parameters $A$, $B$, $C\cos\phi$, and $C\sin\phi$ slaved to the three nonlinear ones $t_c$, $m$, and $\omega$; see Materials and Methods), while the PLS model has four parameters $t_c$, $m$, $A$, and $B$ (with the two linear parameters $A$ and $B$ slaved to the two nonlinear ones $t_c$ and $m$; see Materials and Methods).

We quantitatively examine the performance of the two models based on a comprehensive list of evaluation metrics (see Materials and Methods). We first analyze the frequency distribution and empirical cumulative distribution function (eCDF) of residuals and further compute the normalized root mean square error (NRMSE). These metrics enable us to quantify the mismatch between the models and data. We also compute the normalized Akaike information criterion (NAIC) and normalized Bayesian information criterion (NBIC), which reward the goodness-of-fit while introducing a penalty term for the number of parameters. This enables an assessment of the relative quality of the two competing models having different numbers of parameters, with the one having the lower NAIC and NBIC values preferred. Furthermore, we test the null hypothesis $H_0$ that the data follow the PLS model (against the alternative that the data follow the LPPLS model) using the Wilks test, which is the most powerful test for comparing two competing models with one nested in another (here the PLS model is indeed nested in the LPPLS model). If the *p*-value for the Wilks test is below the prescribed significance level (e.g., 0.05), $H_0$ is rejected. Moreover, we perform two-sample Kolmogorov-Smirnov (KS) and Anderson-Darling (AD) tests of the null hypothesis that the probability distributions of the LPPLS and PLS residuals are identical. If the *p*-value is smaller than the significance level, the null is rejected, which may additionally indicate model superiority.

The first study case is the Veslemannen landslide in Norway. This instability complex, consisting of high-grade metamorphic rocks, has been continuously monitored since October 2014 by a ground-based interferometric synthetic-aperture radar system with an accuracy of 0.5 mm (*30*). This landslide primarily exhibited active movements during summer and autumn seasons, likely due to rainwater infiltration into the slope through the thawed upper frost zones. On September 5, 2019, a volume of ~54,000 m³ rock collapsed. Prior to the failure, this landslide shifted significantly, displacing several meters over the course of approximately 3 months, showing an accelerating and oscillating behavior (see Fig. 1A for the monitoring data at one of the seven radar points and Fig. S1 for the data of all radar points). It is evident that the LPPLS model gives a much better fit to the data, with both the acceleration and oscillation well captured, whereas the PLS model only depicts the general acceleration trend (Fig. 1A and Fig. S1). The LPPLS residuals are confined in a much narrower range compared to the PLS residuals (Fig. 1A and Fig. S1, insets). The LPPLS model shows significantly lower NRMSE, NAIC, and NBIC values, while the *p*-values for the Wilks, KS, and AD tests are all well below 0.05 (Table 1), suggesting that the LPPLS model outperforms the PLS model. The superior performance of the LPPLS model is further demonstrated by its estimated $t_c$ which is very close to the actual failure time (with a small discrepancy of 0.13 day; note that the data have a daily aggregation resolution), whereas the $t_c$ estimated by the PLS model is about 9 days after the actual failure (Table 2).

The second example is a rockburst event that occurred in an underground mine at New South Wales, Australia (*31*). This coal mine, located at a depth of about 250 m, was mined using the longwall method. An in-situ monitoring system using multipoint extensometers (with an accuracy of 0.5 mm) was installed to record the roof displacement of a gateroad, which is a rectangular tunnel with a width of 5.2 m and a height of 3 m providing access to the longwall face. The roof catastrophically failed on June 4, 2004, prior to which a precursor accelerating and oscillating behavior was observed (Fig. 1B). A visual inspection indicates that the LPPLS model well captures the superimposed acceleration-oscillation behavior of the roof, whereas the PLS model can only track the overall acceleration trend (Fig. 1B). This is consistent with the frequency and eCDF distributions of residuals (Fig. 1B inset), demonstrating that the LPPLS model more closely matches the data. The LPPLS model is also associated with smaller NRMSE, NAIC, and NBIC values than the PLS model, while the $p$-values for the Wilks, KS, and AD tests are all below 0.05 (Table 1). Furthermore, the estimated $t_c$ by the LPPLS model is closer to the actual failure time (Table 2). All these results suggest that the LPPLS model surpasses the PLS model.

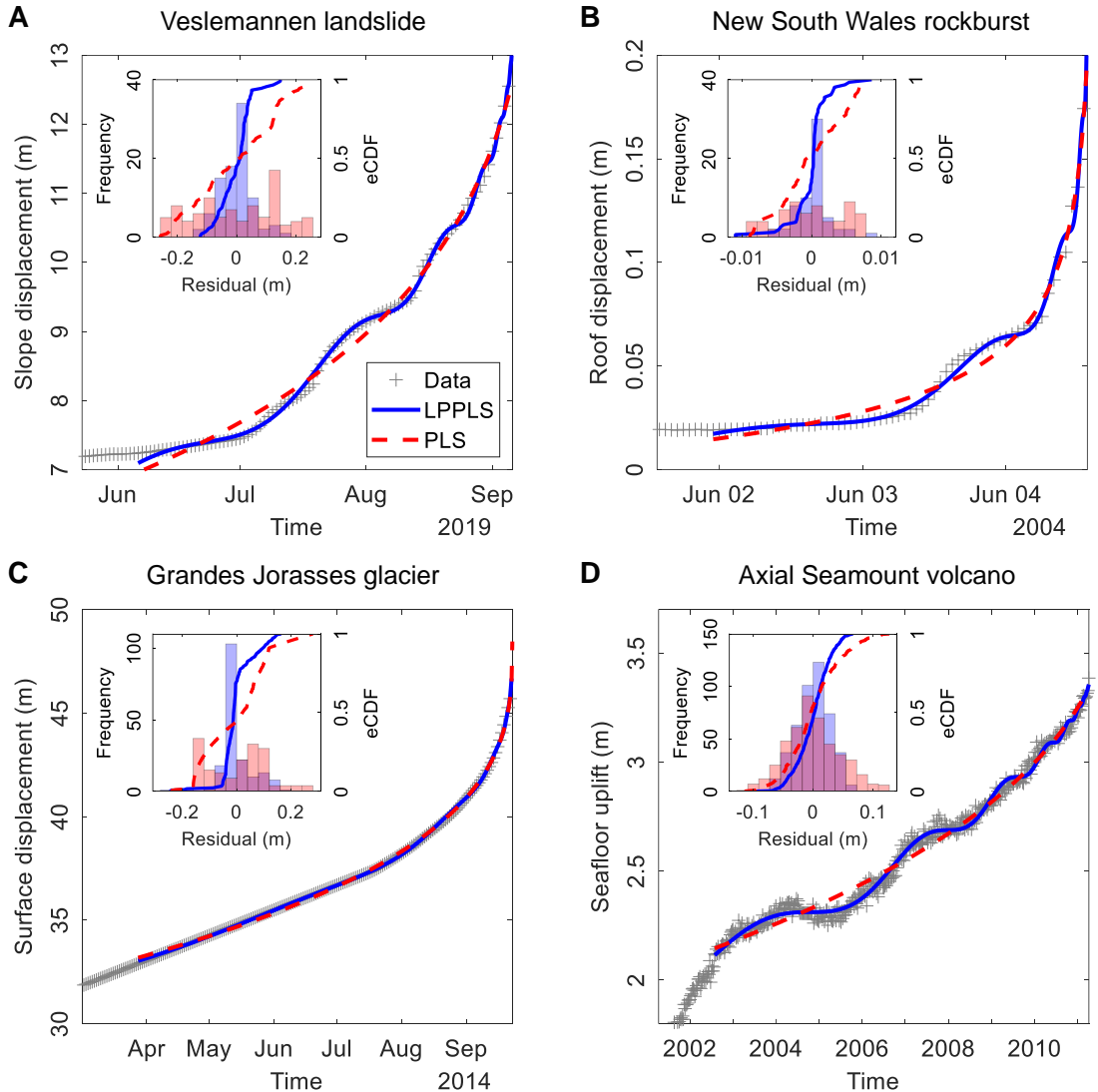

**Fig. 1. Comparison between the LPPLS and PLS models in fitting the monitoring data of various geohazard events.** (**A**) Slope surface displacement (aggregated on a daily basis) prior to a catastrophic landslide at Veslemannen, Norway. (**B**) Tunnel roof displacement (aggregated on an hourly basis) prior to a violent rockburst at New South Wales, Australia. (**C**) Glacier surface displacement (aggregated on a daily basis) prior to a rapid ice breakoff at Grandes Jorasses, Italy. (**D**) Seafloor uplift (aggregated on a weekly basis) prior to an explosive volcanic eruption at Axial Seamount, USA. Note that the raw measurement data for these events, originally recorded at higher frequencies, are aggregated on a lower frequency to expedite calibration and facilitate visualization.

The next case is a hanging cold glacier located on the south face of Grandes Jorasses, Italy and at an elevation of 3950 m above sea level (*13*). Surface displacements of this glacier, measured by a robotic total station with multiple reflectors, were evaluated with an accuracy of ~1 cm. A large volume of ~105,000 m$^3$ ice broke off in two consecutive events on September 23 and 29, 2014 (*32*). We focus on the first breakoff event. One can see that this glacier experienced a generally smooth acceleration behavior over several months, for which both the LPPLS and PLS models show a close match to the trend (Fig. 1C) and provide excellent estimates of the failure time (Table 2). However, the LPPLS residuals are much smaller than the PLS residuals (Fig. 1C inset). The NRMSE, NAIC, and NBIC values of the LPPLS model are considerably lower than those of the PLS model, while the *p*-values for the Wilks, KS, and AD tests are all well below 0.05 (Table 1). These metrics collectively demonstrate that the LPPLS model performs better than the PLS model.

**Table 1.** Comparison between the LPPLS and PLS models in fitting the monitoring data of the Veslemannen landslide, New South Wales rockburst, Grandes Jorasses glacier, and Axial Seamount volcano.

| Evaluation metrics | Veslemannen landslide | New South Wales rockburst | Grandes Jorasses glacier | Axial Seamount volcano |
|---|---|---|---|---|
| NRMSE$_{LPPLS}$ | 5.55×10$^{-4}$ | 4.34×10$^{-5}$ | 2.56×10$^{-4}$ | 6.02×10$^{-4}$ |
| NRMSE$_{PLS}$ | 3.73×10$^{-3}$ | 1.42×10$^{-4}$ | 1.67×10$^{-3}$ | 1.55×10$^{-3}$ |
| NAIC$_{LPPLS}$ | -2.84 | -8.85 | -2.80 | -4.31 |
| NAIC$_{PLS}$ | -1.00 | -7.76 | -1.46 | -3.37 |
| NBIC$_{LPPLS}$ | -2.64 | -8.61 | -2.68 | -4.24 |
| NBIC$_{PLS}$ | -0.89 | -7.62 | -1.39 | -3.34 |
| *p*-value (Wilks test) | 0.00 | 0.00 | 0.00 | 0.00 |
| *p*-value (KS test) | 0.00 | 0.03 | 0.00 | 0.00 |
| *p*-value (AD test) | 0.00 | 0.00 | 0.00 | 0.00 |

**Table 2.** Parameters of the LPPLS and PLS models fitted to the monitoring data of the Veslemannen landslide, New South Wales rockburst, Grandes Jorasses glacier, and Axial Seamount volcano.

| Model parameters | Veslemannen landslide | New South Wales rockburst | Grandes Jorasses glacier | Axial Seamount volcano |
|---|---|---|---|---|
| LPPLS: | | | | |
| $t_c$ | 0.13 | 0.02 | 0.00 | 6.00 |
| $m$ | 0.42 | -0.12 | 0.33 | 0.64 |
| $\omega$ | 7.49 | 4.94 | 4.94 | 9.23 |
| $\phi$ | -0.82 | -0.01 | 0.65 | 0.61 |
| $A$ | 13.64 | -0.24 | 48.76 | 3.38 |
| $B$ | -1.03 | 0.28 | -2.88 | -0.01 |
| $C$ | 4.09×10$^{-2}$ | 5.77×10$^{-3}$ | 3.56×10$^{-2}$ | 5.12×10$^{-4}$ |
| PLS: | | | | |
| $t_c$ | 9.23 | 0.05 | 0.00 | 323.10 |
| $m$ | 0.07 | -0.28 | 0.34 | 0.32 |
| $A$ | 43.93 | -0.08 | 48.46 | 4.38 |
| $B$ | -26.85 | 0.12 | -2.66 | -0.16 |

Note: $t_c$ is in day; $m$, $\omega$, and $\phi$ are dimensionless; $A$ is in meter; $B$ and C are in meter per day$^m$. The actual time of failure corresponds to time $t = 0$ day. The start of the calibration time window is detected by the Lagrange regularization approach (see Materials and Methods), while the calibration time window ends at the last available data point before the actual failure occurs.

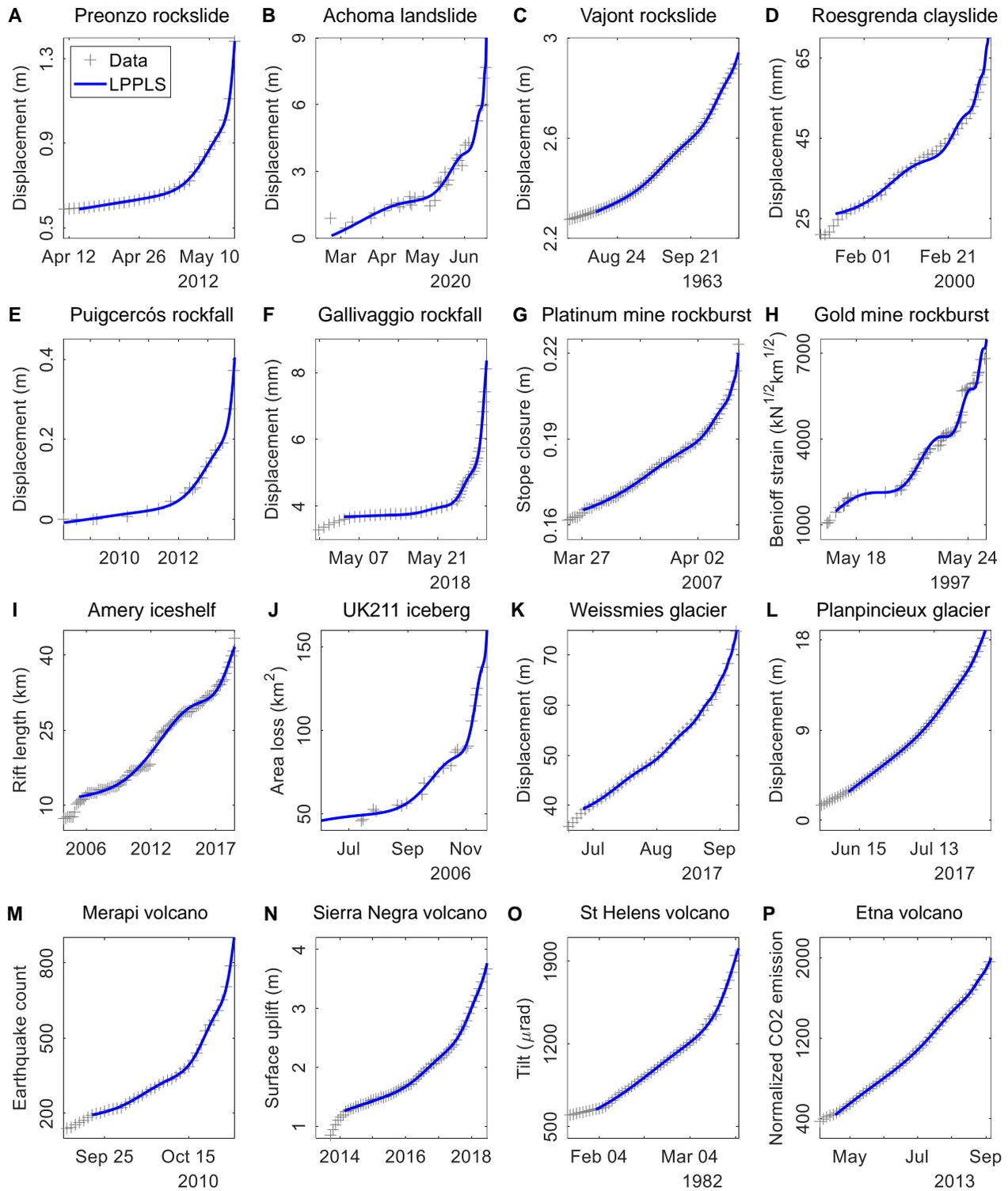

**Fig. 2. Application of the generalized failure law based on the LPPLS model to various geohazard events.** (**A**) Preonzo rockslide, Switzerland. (**B**) Achoma landslide, Peru. (**C**) Vajont rockslide, Italy. (**D**) Roesgrenda clayslide, Norway. (**E**) Puigcercós rockfall, Spain. (**F**) Gallivaggio rockfall, Italy. (**G**) Merensky rockburst, South Africa. (**H**) Goldmine rockburst, South Africa. (**I**) Amery iceshelf, Antarctica. (**J**) UK211 iceberg, Antarctica. (**K**) Weissmies glacier, Switzerland. (**L**) Planpincieux glacier, Italy. (**M**) Merapi volcano, Indonesia. (**N**) Sierra Negra volcano, Ecuador. (**O**) St. Helens volcano, USA. (**P**) Etna volcano, Italy.

The fourth example is Axial Seamount, an active submarine volcano with a summit caldera at ~1.5 km depth and a base at ~2.4 km, located ~500 km offshore Oregon, USA (*33*). This basaltic volcano with magma supplied from the Cobb hotspot has erupted three times over the past 26 years, in 1998, 2011, and 2015. It has been closely monitored by a cabled network of seafloor instruments since 1998, with the seafloor vertical deformation measured at a resolution of ~1 cm through bottom and mobile pressure recorders (*34*). We focus on the eruption event in April 2011, prior to which the volcano exhibited a series of inflation-deflation cycles over about 8 years (Fig. 1D). The LPPLS model effectively captures the oscillatory behavior of the seafloor deformation superimposed on an overall acceleration, whereas the PLS model only depicts the overall trend (Fig. 1D). Compared to the PLS model, the LPPLS model is associated with smaller residuals (Fig. 1D inset) and lower values for NRMSE, NAIC, and NBIC (Table 1). Additionally, the *p*-values for the Wilks, KS, and AD tests are all well below 0.05 (Table 1). The failure time estimated by the LPPLS model has a discrepancy of only 6 days compared to the actual eruption time (i.e., just one point of the weekly-aggregated data resolution), while that by the PLS model is almost 1 year after. All these results support the superior performance of the LPPLS model.

We further test the LPPLS model for describing the geomaterial failure behavior during various geohazard events (see Fig. 2 for some typical examples among 109 events analyzed in total; refer to Fig. S1-S18 and Tables S1-S16 for the complete inventory). More specifically, Fig. 2A shows the slope displacement data monitored by an extensometer at the Preonzo rockslide, Switzerland which collapsed on May 15, 2012 (*35*). Fig. 2B presents the displacement time series constructed from high-frequency optical satellite images for the Achoma landslide, Peru, which failed in June 2020 (*36*). Fig. 2C illustrates the displacement evolution of the Vajont rockslide, Italy, prior to a catastrophic failure on October 9, 1963 (*37*). Fig. 2D shows the acceleration behavior recorded by an extensometer at a quick clay slide at Roesgrenda, Norway (*38*). Fig. 2E displays the displacement data of a rock cliff acquired from a Terrestrial LiDAR instrument at Puigcercós, Spain, where a rockfall event occurred on December 3, 2013 (*39*). Fig. 2F depicts the displacement time series of a subvertical granitic slope monitored by a ground-based synthetic aperture radar at Gallivaggio, Italy, where a rockfall event occurred on May 29, 2018 (*40*). Fig. 2G shows the stope closure data measured by a closure meter prior to a violent rockburst in a deep platinum mine, South Africa (*41*). Fig. 2H gives the cumulative Benioff strain of seismic energy released before a large rockburst event in a deep gold mine, South Africa (*42*). Fig. 2I shows the multi-year rift propagation in the Amery Ice Shelf, Antarctica, preceding a large calving event on September 25, 2019 (*43*). Fig. 2J depicts the temporal evolution of area loss of the UK211 iceberg that rapidly disintegrated in 2006, tracked by the satellite-based Moderate Resolution Imaging Spectroradiometer sensor (*44*). Fig. 2K displays the surface displacement data of a polythermal glacier at Weissmies, Switzerland prior to a breakoff on September 10, 2017, recorded by a high-resolution camera (*45*). Fig. 2L gives the displacement data of a polythermal glacier at Planpincieux, Italy prior to a breakoff on September 16, 2015 (*46*). Fig. 2M illustrates the temporal evolution of earthquake count prior to an explosive eruption of the Merapi stratovolcano, Indonesia in October 2010 (*47*). Fig. 2N shows the ground uplift data recorded by a continuous Global Positioning System network during the 2018 eruption of the Sierra Negra shield volcano, Ecuador (*48*). Figs. 2O plots the tiltmeter measurements at the St. Helens stratovolcano during its 1982 eruption (*49*). Fig. 2P shows the temporal variation of cumulative normalized soil $CO_2$ efflux recorded by a geochemical monitoring network at the Etna stratovolcano that erupted in September 2013 (*50*).

The LPPLS model provides an excellent fit to the documented failure phenomena in all these cases, effectively capturing both the accelerating and oscillating behavior (a detailed comparison with the PLS model is given by Tables S9-S12). In total, we have tested 109 historical events (consisting of 160 time series data), including 49 landslides, 11 rockbursts, 17 glacier breakoffs, and 32 volcanic eruptions (Supplementary Materials). Fig. 3 illustrates the comparative model performance of LPPLS against PLS across the entire dataset (see Tables S9-S12 for more details). The NRMSE of the LPPLS model is consistently smaller than that of the PLS model across the entire dataset; notably, for 67.5% of the data, the NRMSE of the LPPLS model is less than half that of the PLS model. The NAIC (respectively NBIC) values of the LPPLS model are lower than those of the PLS model by 0.5 unit for 65% (respectively 70%) of the data and by 1 unit for 35% (respectively 40%). Note that a 0.5 to 1 unit decrease in NAIC or NBIC corresponds to a relative likelihood improvement of 1.65 to 2.72 times per data point. For 91% of the data (Tables S9-S12), the *p*-value from the Wilks test is below 0.05, providing strong evidence of the LPPLS model's superiority. The *p*-values of the KS and AD tests are below 0.05 for 42% and 47% of the data, respectively, aligning with the fact that these tests are less powerful than the Wilks test, particularly given the limited data for some events (Tables S13-16). However, the evidence remains strong, as only 8 cases with *p*-values below 0.05 would be expected if the *p*-values were random. All these results collectively indicate the superior performance of the LPPLS model over the PLS model, because the former can constrain the fit to the bursty oscillating structures in the data while the latter is often poorly defined, particularly in the presence of noise.

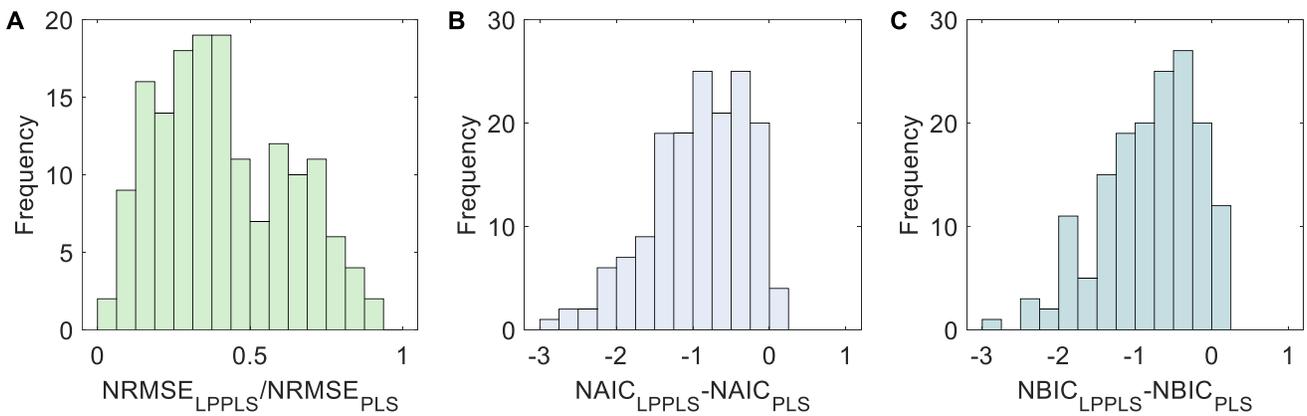

**Fig. 3. Histograms showing a comparison of the evaluation metrics between the LPPLS and PLS models in fitting 109 geohazard events (with 160 time series monitoring data in total).** (**A**) NRMSE ratio, (**B**) NAIC difference, and (**C**) NBIC difference between the LPPLS and PLS models.

**Discussion**

Our results reveal that the proposed generalized failure law can accurately describe the failure behavior of various geomaterials (rock, soil, and ice) under diverse contexts, ranging from landslides and rockbursts to glacier breakoffs and volcanic eruptions. These geohazard phenomena develop over a wide spectrum of time scales (from hours/days to months/years) and length scales (from meters to tens of kilometers). The broad applicability of our failure law highlights the commonality of different failure phenomena, which are mechanistically driven by the similar mechanisms involving interactions between many fractures, frictional processes, and final strain localization in heterogeneous materials. This failure law can be applied to various observables (e.g., surface displacement, tunnel closure, energy release, rift length, earthquake count, angular change, and gas emission) recorded by different instruments (e.g., extensometers, reflectors, tiltmeters, closuremeters, satellites, LiDAR, synthetic aperture radar, Global Positioning System, pressure recorders, and seismic/geochemical monitoring networks) (Tables S1-S4). These results indicate that our new failure law is general and robust, with significant potential to mitigate geohazard risks and enhance existing early warning systems. The LPPLS signatures observed in various geohazards imply the presence of characteristic time scales in the rupture of heterogeneous geomaterials, which could provide important insights into the underlying systems and/or physics as well as be very useful for prediction purposes (*21*). Our future work will focus on implementing this generalized failure law for prospective forecast of catastrophic events.

Supplementary Materials for

**Generalized failure law for landslides, rockbursts, glacier breakoffs, and volcanic eruptions**

Qinghua Lei[1]*, Didier Sornette[2]

[1] Department of Earth Sciences, Uppsala University; Uppsala, 752 36, Sweden.

[2] Institute of Risk Analysis, Prediction and Management, Academy for Advanced Interdisciplinary Studies, Southern University of Science and Technology; Shenzhen, 518055, China.

*Corresponding author. Email: qinghua.lei@geo.uu.se

**Materials and Methods**

Derivation of the power law singularity (PLS) model

An integration of equation (2) in the Main Text leads to:

$$\Omega(t) = \begin{cases} A - \dfrac{\kappa}{m}(t_c - t)^m, & m \neq 0 \\ A - \kappa \ln(t_c - t), & m = 0 \end{cases}, \quad (S1)$$

where $m = 1 - \xi < 1$ and $A$ can be determined from the initial condition of $\Omega(t = t_0) = \Omega_0$, so that:

$$\Omega(t) = \begin{cases} \Omega_0 + \dfrac{\kappa}{m}\left[(t_c - t_0)^m - (t_c - t)^m\right], & m \neq 0 \\ \Omega_0 - \kappa \ln\left(\dfrac{t_c - t}{t_c - t_0}\right), & m = 0 \end{cases}. \quad (S2)$$

Recall that $t_c$ is the time of failure determined from the other initial condition $\dot{\Omega}(t = t_0) = \dot{\Omega}_0$, giving $t_c = t_0 + (\xi/\eta)\dot{\Omega}_0^{-1/\xi}$.

For $m \to 0$, consider the Taylor expansion $(t_c - t_0)^m = e^{m\ln(t_c - t_0)} = 1 + m\ln(t_c - t_0) + O(m^2)$ and $(t_c - t)^m = e^{m\ln(t_c - t)} = 1 + m\ln(t_c - t) + O(m^2)$, we have:

$$\Omega(t) = \Omega_0 + \dfrac{\kappa}{m}\left[e^{m\ln(t_c - t_0)} - e^{m\ln(t_c - t)}\right] = \Omega_0 - \kappa \ln\left(\dfrac{t_c - t}{t_c - t_0}\right) + O(m). \quad (S3)$$

Thus, as $m \to 0$, the solution of $\Omega(t)$ for $m \neq 0$ converges to that for $m = 0$, so equation (2) in the Main Text with $B = -\kappa/m$ gives the general solution of $\Omega(t)$ for $m < 1$.

Derivation of the log-periodic power law singularity (LPPLS) model

Let us define:

$$\tilde{\Omega}(t) = \dfrac{\Omega(t) - A}{B} = (t_c - t)^m, \quad (S4)$$

where $B \neq 0$ ensures the presence of a finite-time singularity. The power law relation (S4) obeys the symmetry of scale invariance. This means that scaling $t_c - t$ by an arbitrary factor $\lambda$ leads to a corresponding scaling of the observable by factor $\mu(\lambda)$, while the $t_c - t$ dependence remains unchanged. Mathematically, this is expressed as:

$$\mu \tilde{\Omega}(t) = [\lambda(t_c - t)]^m. \tag{S5}$$

In other words, if we replace $t_c - t$ with $\lambda(t_c - t)$ and replace $\tilde{\Omega}$ with $\mu \tilde{\Omega}$ in equation (S4), the equality still holds, reflecting the scale invariance. From equations (S4) and (S5), we can derive the following equation that determines $\mu(\lambda)$:

$$\lambda^m / \mu = 1. \tag{S6}$$

It is often the case that, rather than deriving $\mu(\lambda)$ from $\lambda$ and $m$, first-principle considerations provide the values of $\lambda$ and $\mu$, from which one derives the value of the exponent as:

$$m = \frac{\ln \mu}{\ln \lambda}. \tag{S7}$$

In the presence of continuous scale invariance, the above derivation holds for arbitrary value of $\lambda$ with $\mu(\lambda)$ adjusting so that there is a single value of the exponent $m$ given by equation (S7). The general theoretical procedure is usually formulated with the Renormalization Group (*51*) by taking the limit $\lambda = 1 + \delta \to 1$ ($\delta \to 0$) and $\mu = 1 + \zeta \to 1$ ($\zeta \to 0$), so that $m = \delta / \zeta$ in the limit.

The continuous scale invariance discussed above can be partially broken into a discrete scale invariance. This means that the power law relationship (S4) holds under scaling $t_c - t$ by specific factors that are integer powers $\lambda^n$ of a specific scaling ratio $\lambda > 1$, where $n$ is an arbitrary integer indexing the hierarchy of scales (*22*). There are several derivations of the corresponding spectrum of power law exponents $m_n$. The simplest one is to start from the identity $1 = \exp(i2\pi n)$ and substitute it into the right-hand-side of equation (S6). Solving for $m_n$ yields the series of complex critical exponents:

$$m_n = m + in\omega, \tag{S8}$$

where $m$ is the real part given by (S7), $n$ is an arbitrary integer, and $\omega = 2\pi / \ln \lambda$. The general form of $\tilde{\Omega}$ can be expressed as an infinite sum of power laws $(t_c - t)^{m_n}$, similarly to a Fourier series (it is in fact a discrete Mellin transform generalization of discrete Fourier series, see e.g., (*52*)). The corresponding log-periodic power law is obtained by taking the real part of each power law $(t_c - t)^{m_n}$, since observables are real. This gives:

$$\tilde{\Omega}_n(t) = \mathrm{Re}\left[(t_c - t)^{m_n}\right] = (t_c - t)^m \cos[n\omega \ln(t_c - t)]. \tag{S9}$$

Thus, the most general solution of $\tilde{\Omega}$ is given by a superposition of all the components of this generalized Fourier series in logarithmic scale (*19*):

$$\tilde{\Omega}(t) = (t_c - t)^m \sum_{n=0}^{+\infty} a_n \cos[n\omega \ln(t_c - t) - \phi_n], \tag{S10}$$

where $a_n$ are the generalized Fourier coefficients and $\phi_n$ represent the phase shifts, with $a_0 = 1$ and $\phi_0 = 0$. The term for $n = 0$ defines a pure power law with fully continuous scale invariance governed by the real critical exponent $m$, while the terms for $n > 1$ introduce a partial breaking of continuous scale invariance into discrete scale invariance, resulting in a series of log-periodic oscillations decorating the pure power law (*22*). It has been proven that the amplitudes of the coefficients $a_n$ decay fast with $n$ (*52*), so that only the first correction term $n = 1$ is important in general (there are exceptions in which higher-order terms need to be considered but we do not consider this situation here). Keeping the terms $n=0$ and $n=1$ in (S10) recovers the LPPLS formula, i.e., equation (4) in the Main Text.

The local maxima of the log-periodic term $C\cos[\omega\ln(t_c-t)-\phi]$ in the LPPLS formula occur at times converging to $t_c$ according to a geometric time series $\{t_1,t_2,...,t_k,...\}$ with $t_c-t_k=\tilde{\tau}\lambda^{-k}$, where $k$ is an integer and $\tilde{\tau}=\exp(\phi/\omega)$ is a time constant determined from the initial conditions. The geometric time series corresponds to the times for which the argument of the cosine function is an integer multiple of $2\pi$.

Calibration of the LPPLS model

For the time series of $N$ measurements of the observable quantity $\mathbf{\Omega}=\{\Omega_1,\Omega_2,...,\Omega_N\}$ recorded at time $\mathbf{t}=\{t_1,t_2,...,t_N\}\in[\tau,T]$ ($\tau$ and $T$ respectively denote the start and end of the time window over which the fitting is performed), the LPPLS model is calibrated based on the following scheme (*29*).

The LPPLS model is written as:
$$\Omega(t)=A+\{B+C\cos[\omega\ln(t_c-t)-\phi]\}(t_c-t)^m, \tag{S11}$$

with a parameter set $\boldsymbol{\theta}_{\text{LPPLS}}=\{A,B,C,t_c,m,\omega,\phi\}$ including seven parameters with the former three being linear and the latter four being nonlinear. By introducing $C_1=C\cos\phi$ and $C_2=C\sin\phi$, we can rewrite the LPPLS formula as:
$$\Omega(t)=A+B(t_c-t)^m+C_1(t_c-t)^m\cos[\omega\ln(t_c-t)]+C_2(t_c-t)^m\sin[\omega\ln(t_c-t)], \tag{S12}$$

where the new parameter set $\boldsymbol{\theta}_{\text{LPPLS}}=\{A,B,C_1,C_2,t_c,m,\omega\}$ still has seven parameters but now with four linear and only three nonlinear parameters. To estimate all these parameters, we define the cost function as the sum of squared errors:
$$F(\boldsymbol{\theta}_{\text{LPPLS}};\mathbf{\Omega},\mathbf{t})=\sum_{i=1}^{N}\varepsilon_i^2, \tag{S13}$$

with each residual calculated as:
$$\varepsilon_i=\Omega_i-A-B(t_c-t_i)^m-C_1(t_c-t_i)^m\cos[\omega\ln(t_c-t_i)]-C_2(t_c-t_i)^m\sin[\omega\ln(t_c-t_i)]. \tag{S14}$$

The ordinary least squares method amounts to minimizing the cost function (S13) to obtain the estimates for the model parameters:
$$\hat{\boldsymbol{\theta}}_{\text{LPPLS}}=\arg\min_{\boldsymbol{\theta}_{\text{LPPLS}}}F(\boldsymbol{\theta}_{\text{LPPLS}};\mathbf{\Omega},\mathbf{t}). \tag{S15}$$

This is not a trivial task due to the strong nonlinearity of the cost function and the presence of multiple local minima.

To solve this minimization problem, we enslave the four linear parameters $\{A,B,C_1,C_2\}$ to the three nonlinear ones $\{t_c,m,\omega\}$ so as to reduce the minimization problem to:
$$\{\hat{t}_c,\hat{m},\hat{\omega}\}=\arg\min_{t_c,m,\omega}F_1(t_c,m,\omega), \tag{S16}$$

with the profiled cost function defined as:

$$F_1(t_c, m, \omega) = \min_{A,B,C_1,C_2} F(A, B, C_1, C_2, t_c, m, \omega) = F(t_c, m, \omega, \hat{A}, \hat{B}, \hat{C}_1, \hat{C}_2), \tag{S17}$$

and the estimates for the linear parameters $\{A, B, C_1, C_2\}$ obtained by solving the optimization problem for fixed values of the nonlinear parameters $\{t_c, m, \omega\}$:

$$\{\hat{A}, \hat{B}, \hat{C}_1, \hat{C}_2\} = \arg\min_{A,B,C_1,C_2} F(A, B, C_1, C_2, t_c, m, \omega), \tag{S18}$$

which has a unique solution analytically solved from the following system of linear equations:

$$\begin{bmatrix} N & \sum f_i & \sum g_i & \sum h_i \\ \sum f_i & \sum f_i^2 & \sum f_i g_i & \sum f_i h_i \\ \sum g_i & \sum f_i g_i & \sum g_i^2 & \sum g_i h_i \\ \sum h_i & \sum f_i h_i & \sum g_i h_i & \sum h_i^2 \end{bmatrix} \begin{bmatrix} \hat{A} \\ \hat{B} \\ \hat{C}_1 \\ \hat{C}_2 \end{bmatrix} = \begin{bmatrix} \sum \Omega_i \\ \sum \Omega_i f_i \\ \sum \Omega_i g_i \\ \sum \Omega_i h_i \end{bmatrix}, \tag{S19}$$

where $f_i = (t_c - t_i)^m$, $g_i = (t_c - t_i)^m \cos[\omega \ln(t_c - t_i)]$, and $h_i = (t_c - t_i)^m \sin[\omega \ln(t_c - t_i)]$.

The optimization problem (S16) can be further reformulated as:

$$\hat{t}_c = \arg\min_{t_c} F_2(t_c), \tag{S20}$$

with the cost function given by:

$$F_2(t_c) = \min_{m,\omega} F_1(t_c, m, \omega) = F_1(t_c, \hat{m}, \hat{\omega}), \tag{S21}$$

and the estimates for parameters $\{m, \omega\}$ obtained by solving the optimization problem:

$$\{\hat{m}, \hat{\omega}\} = \arg\min_{m,\omega} F_1(t_c, m, \omega). \tag{S22}$$

Here, a constraint of $4.94 \leq \omega \leq 15$ is imposed with the lower bound defined to prevent chaotic scenarios (*53*) and the upper bound defined to avoid spurious oscillations (*29*), so that the scaling ratio $\lambda = \exp(2\pi/\omega)$ is at the order of 2 (*22*), as suggested by general theoretical arguments (*54*).

Calibration of the PLS model

For the time series of $N$ measurements of the observable quantity $\mathbf{\Omega} = \{\Omega_1, \Omega_2, ..., \Omega_N\}$ recorded at time $\mathbf{t} = \{t_1, t_2, ..., t_N\} \in [\tau, T]$ ($\tau$ and $T$ respectively denote the start and end of the time window over which the fitting is performed), the PLS model is calibrated based on a scheme similar to the one for the LPPLS model.

The PLS model is written as:

$$\Omega(t) = A + B(t_c - t)^m, \tag{S23}$$

with a parameter set $\mathbf{\theta}_{\text{PLS}} = \{A, B, t_c, m\}$ including four parameters with the first two being linear and the last two being nonlinear. To estimate all these parameters, we define the cost function as the sum of squared errors:

$$F(\mathbf{\theta}_{\text{PLS}}; \mathbf{\Omega}, \mathbf{t}) = \sum_{i=1}^{N} \varepsilon_i^2, \tag{S24}$$

with each residual calculated as:

$$\varepsilon_i = \Omega_i - A - B(t_c - t_i)^m. \tag{S25}$$

The ordinary least squares method amounts to minimizing the cost function (S24) to obtain the estimates for the model parameters:

$$\hat{\boldsymbol{\theta}}_{PLS} = \arg\min_{\boldsymbol{\theta}_{PLS}} F(\boldsymbol{\theta}_{PLS}). \tag{S26}$$

To do so, we enslave the two linear parameters $\{A, B\}$ to the two nonlinear ones $\{t_c, m\}$ to obtain the nonlinear optimization problem:

$$\{\hat{t}_c, \hat{m}\} = \arg\min_{t_c, m} F_1(t_c, m), \tag{S27}$$

with the cost function defined as:

$$F_1(t_c, m) = \min_{A,B} F(t_c, m, A, B) = F(t_c, m, \hat{A}, \hat{B}). \tag{S28}$$

The estimates for parameters $\{A, B\}$ are obtained by solving the optimization problem:

$$\{\hat{A}, \hat{B}\} = \arg\min_{A,B} F(t_c, m, A, B), \tag{S29}$$

which has a unique solution analytically solved from the following system of linear equations:

$$\begin{bmatrix} N & \sum f_i \\ \sum f_i & \sum f_i^2 \end{bmatrix} \begin{bmatrix} \hat{A} \\ \hat{B} \end{bmatrix} = \begin{bmatrix} \sum \Omega_i \\ \sum \Omega_i f_i \end{bmatrix}, \tag{S30}$$

where $f_i = (t_c - t_i)^m$.

The optimization problem (S27) can be further reformulated as:

$$\hat{t}_c = \arg\min_{t_c} F_2(t_c), \tag{S31}$$

with the cost function given by:

$$F_2(t_c) = \min_m F_1(t_c, m) = F_1(t_c, \hat{m}), \tag{S32}$$

and the estimate for parameter $m$ is obtained by solving the optimization problem:

$$\hat{m} = \arg\min_m F_1(t_c, m). \tag{S33}$$

Lagrange regularization for determining the onset of the fitting time window

For a fixed end time $T$, the optimal start time $\tau$ of the time window for calibrating the LPPLS or PLS model can be endogenously detected using the Lagrange regularization approach (*28*) with the following cost function to minimize:

$$\tilde{F}'(\tau) = \tilde{F}(\tau) - \chi N(\tau), \tag{S34}$$

where $\chi$ is the Lagrange parameter, $N$ is the number of observations within the time window $[\tau, T]$, and $\tilde{F}(\tau)$ is the normalized sum of squared residuals given by:

$$\tilde{F}(\tau) = \frac{F}{N(\tau) - n}, \tag{S35}$$

where $F$ is the sum of squared errors given by equation (S13) for the LPPLS fitting or by equation (S24) for the PLS fitting and $n$ is the number of degrees of freedom of the model (i.e., 7 and 4 for the LPPLS and PLS model, respectively). The Lagrange parameter $\chi$ can be heuristically approximated by the slope in the linear regression model of $\tilde{F}(\tau)$ with respect to $\tau$.

When comparing the LPPLS and PLS models, we estimate $\tau$ based on the LPPLS model to determine the time window for comparison. Alternatively, we can estimate $\tau$ based on the PLS model or as the maximum of the $\tau$ values from the LPPLS and PLS models, which do not affect our conclusions about the model's superiority.

Goodness-of-fit tests for comparing the LPPLS and PLS models

We employ a set of evaluation metrics to compare the LPPLS and PLS models, including normalized root mean square error (NRMSE), normalized Akaike information criterion (NAIC) and normalized Bayesian information criterion (NBIC) as well as $p$-values derived from the Wilks likelihood-ratio test, the two-sample Kolmogorov-Smirnov (KS) test, and the two-sample Anderson-Darling (AD) test.

For the time series of $N$ measurements of the observable quantity $\boldsymbol{\Omega} = \{\Omega_1, \Omega_2, ..., \Omega_N\}$ recorded at time $\mathbf{t} = \{t_1, t_2, ..., t_N\} \in [\tau, T]$, we compute NRMSE as:

$$\text{NRMSE} = \frac{1}{\Omega_N - \Omega_1} \sqrt{\frac{1}{N} F(\boldsymbol{\theta}; \boldsymbol{\Omega}, \mathbf{t})}, \tag{S36}$$

where $F$ is the sum of squared errors given by equation (S13) for the LPPLS fitting or by equation (S24) for the PLS fitting.

In the context of ordinary least squares, if the model is well-specified (i.e., it represents the true generative process of the data, with the noise being independent and identically distributed), then the error term $\varepsilon(\boldsymbol{\theta}; \boldsymbol{\Omega}, \mathbf{t})$ obeys a zero-mean Gaussian distribution:

$$f(\varepsilon; \sigma^2) = \frac{1}{\sqrt{2\pi\sigma^2}} \exp\left(-\frac{\varepsilon^2}{2\sigma^2}\right), \tag{S37}$$

where $\sigma^2$ is the variance. We construct the likelihood function as:

$$L(\boldsymbol{\theta}, \sigma^2; \boldsymbol{\Omega}, \mathbf{t}) = \prod_{i=1}^{N} \left[ \frac{1}{\sqrt{2\pi\sigma^2}} \exp\left(-\frac{\varepsilon_i^2}{2\sigma^2}\right) \right] = (2\pi\sigma^2)^{-N/2} \exp\left[-\frac{F(\boldsymbol{\theta}; \boldsymbol{\Omega}, \mathbf{t})}{2\sigma^2}\right]. \tag{S38}$$

The corresponding log-likelihood function is:

$$\ln L(\boldsymbol{\theta}, \sigma^2; \boldsymbol{\Omega}, \mathbf{t}) = -\frac{N}{2} \ln(2\pi\sigma^2) - \frac{F(\boldsymbol{\theta}; \boldsymbol{\Omega}, \mathbf{t})}{2\sigma^2}, \tag{S39}$$

with the maximum likelihood estimate for $\sigma^2$ given as:

$$\hat{\sigma}^2 = \frac{1}{N} F(\hat{\boldsymbol{\theta}}; \boldsymbol{\Omega}, \mathbf{t}), \tag{S40}$$

so that we obtain:

$$\ln L(\hat{\boldsymbol{\theta}}; \boldsymbol{\Omega}, \mathbf{t}) = -\frac{N}{2} \left[ \ln F(\hat{\boldsymbol{\theta}}; \boldsymbol{\Omega}, \mathbf{t}) + \ln\left(\frac{2\pi}{N}\right) + 1 \right]. \tag{S41}$$

We test the null hypothesis $H_0$ that the observable quantity $\boldsymbol{\Omega}(\mathbf{t})$ follows the PLS model against the alternative that it follows the LPPLS model. Note that the PLS model is a special case of (or a model nested in) the LPPLS model (i.e., the latter reduces to the former if $C = 0$ or in other words $\boldsymbol{\theta}_{\text{PLS}}$ is a subset of $\boldsymbol{\theta}_{\text{LPPLS}}$), Thus, we use the Wilks likelihood ratio test which is powerful and widely applicable, especially for nested models, because it relies on comparing the likelihoods of the two models, which is a robust way of assessing model fit. It has desirable properties under certain conditions, such as asymptotically following a chi-squared distribution under the null hypothesis, which makes it convenient for deriving *p*-values and making decisions about model selection. The corresponding likelihood-ratio test statistic is defined as:

$$\Lambda = -2\left[\ln L(\hat{\boldsymbol{\theta}}_{\text{PLS}}; \boldsymbol{\Omega}, \mathbf{t}) - \ln L(\hat{\boldsymbol{\theta}}_{\text{LPPLS}}; \boldsymbol{\Omega}, \mathbf{t})\right] = N \ln \frac{F(\hat{\boldsymbol{\theta}}_{\text{PLS}}; \boldsymbol{\Omega}, \mathbf{t})}{F(\hat{\boldsymbol{\theta}}_{\text{LPPLS}}; \boldsymbol{\Omega}, \mathbf{t})}, \quad \text{(S42)}$$

which converges asymptotically to a chi-squared distribution (with $\kappa = 3$ degrees of freedom, which is the difference in the number of parameters between the LPPLS and PLS models), if $H_0$ happens to be true. For the finite sample here with *N* number of data points, the distribution of the likelihood-ratio test statistic is unknown. Here, we perform Monte Carlo simulations (1000 runs) to estimate the *p*-value of the null hypothesis as the fraction of exceedances of the test statistic of simulated data compared to that of the actual data. If the *p*-value is smaller than the prescribed significance level (e.g., 0.05), the null hypothesis is rejected.

We also perform a two-sample KS test and AD test on the null hypothesis that the probability distributions of the LPPLS and PLS residuals do not differ. If the *p*-value is smaller than the prescribed significance level, the null hypothesis is rejected, meaning that the residuals of the LPPLS and PLS models are not from the same distribution.

Furthermore, we estimate the relative model quality based on the NAIC and NBIC values calculated as (*55*):

$$\text{NAIC} = \frac{2\kappa - 2\ln L(\hat{\boldsymbol{\theta}}, \hat{\sigma}^2; \boldsymbol{\Omega}, \mathbf{t})}{N} = \frac{2\kappa}{N} + \ln F(\hat{\boldsymbol{\theta}}; \boldsymbol{\Omega}, \mathbf{t}) + \ln\left(\frac{2\pi}{N}\right) + 1, \quad \text{(S43)}$$

$$\text{NBIC} = \frac{\kappa \ln N - 2\ln L(\hat{\boldsymbol{\theta}}, \hat{\sigma}^2; \boldsymbol{\Omega}, \mathbf{t})}{N} = \frac{\kappa \ln N}{N} + \ln F(\hat{\boldsymbol{\theta}}; \boldsymbol{\Omega}, \mathbf{t}) + \ln\left(\frac{2\pi}{N}\right) + 1, \quad \text{(S44)}$$

which reward the goodness-of-fit while introducing a penalty term for the number of parameters.

Data acquisition approach

In this study, we have compiled a large dataset of >100 geohazard events of landslides, rockbursts, glacier breakoffs, and volcanic eruptions (see Tables S1-S4). These data were retrieved through two major ways: (1) exported directly from the monitoring system and obtained from either published dataset/database or from the authors (indicated as "Original"), and (2) digitized from figures in published literature using digitization software (indicated as "Digitized").

The data of 10 cases (Agoyama, Arvigo, Galterengraben, Grabengufer, Hogarth, Kagemori, La Saxe, Nevis Bluff, Vajont, and Weissmiess) were extracted from the published dataset (https://mediatum.ub.tum.de/1688868) (*15*), where the data of 4 cases are "Original" and the data of other 6 cases are "Digitized". Most volcano data are downloaded from the WOVOdat platform (https://www.wovodat.org), which is a publicly accessible database of volcanic unrest (*56*). For most of digitized data in our dataset, we employ the software PlotDigitizer Pro (https://plotdigitizer.com/) to retrieve the data from the published literature. The references for all the data are indicated in Tables S1-S4.

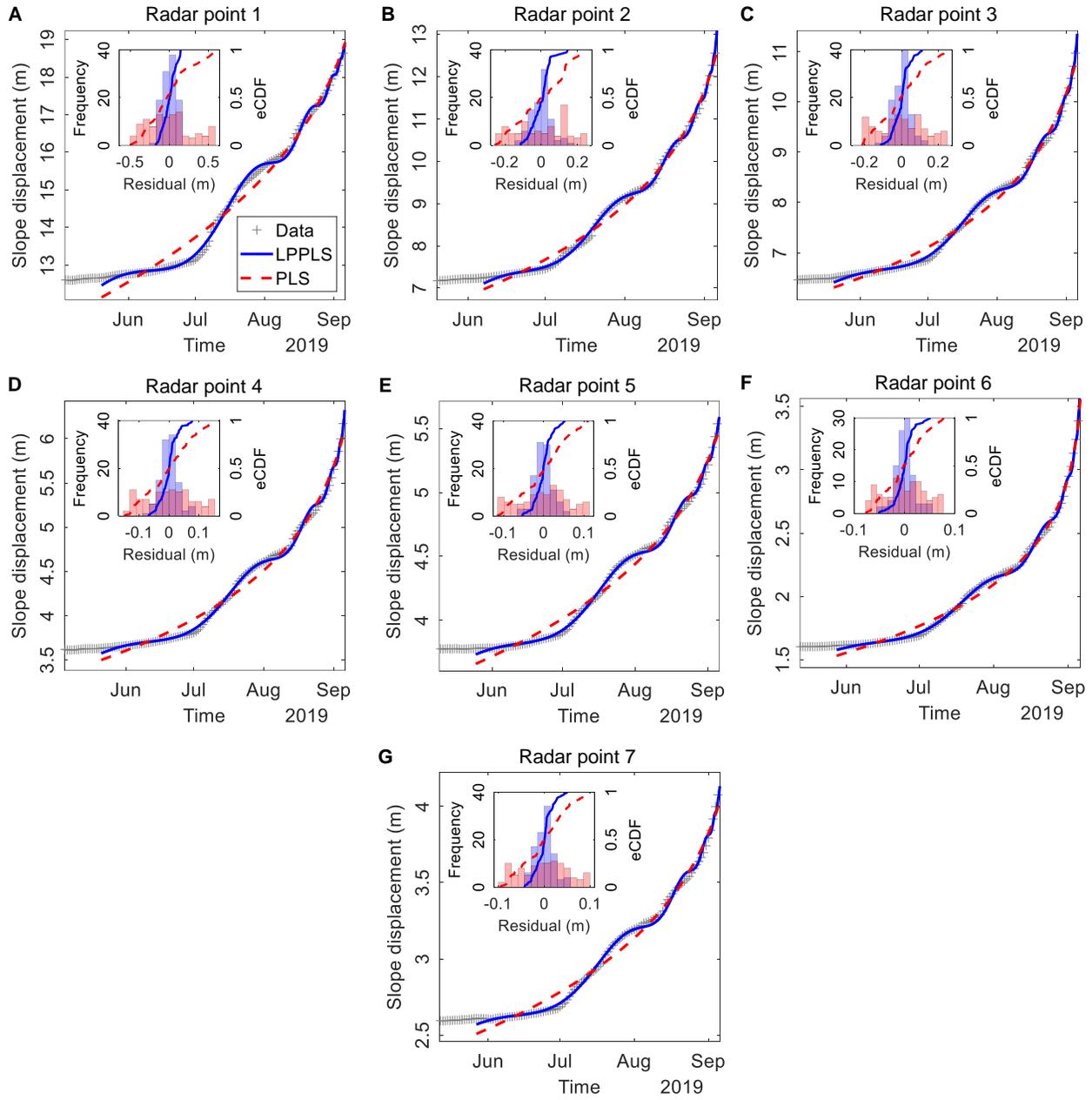

**Fig. S1.** Comparison of the LPPLS versus PLS models in fitting the data of the Veslemannen landslide.

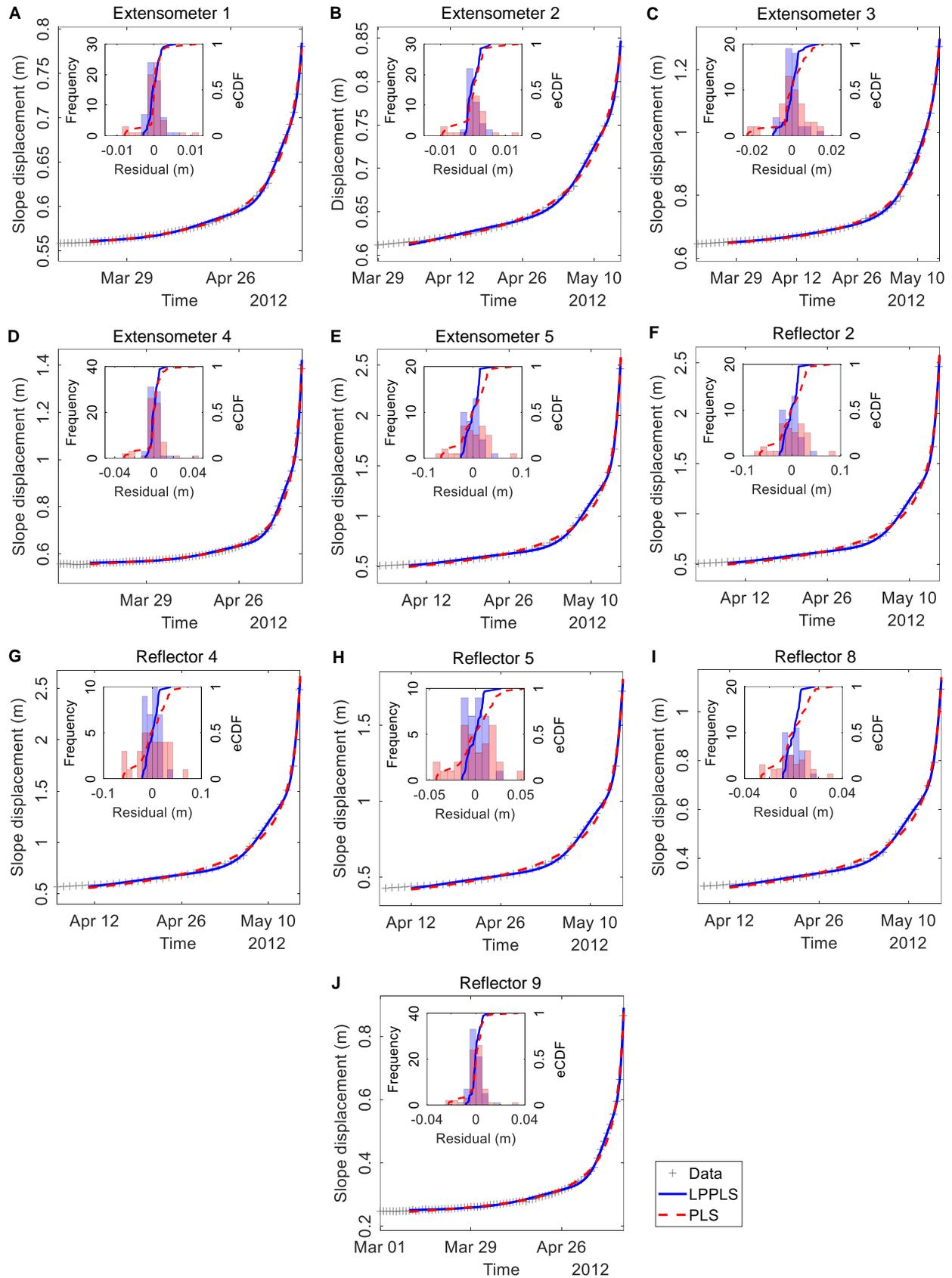

**Fig. S2.** Comparison of the LPPLS versus PLS models in fitting the data of the Preonzo landslide.

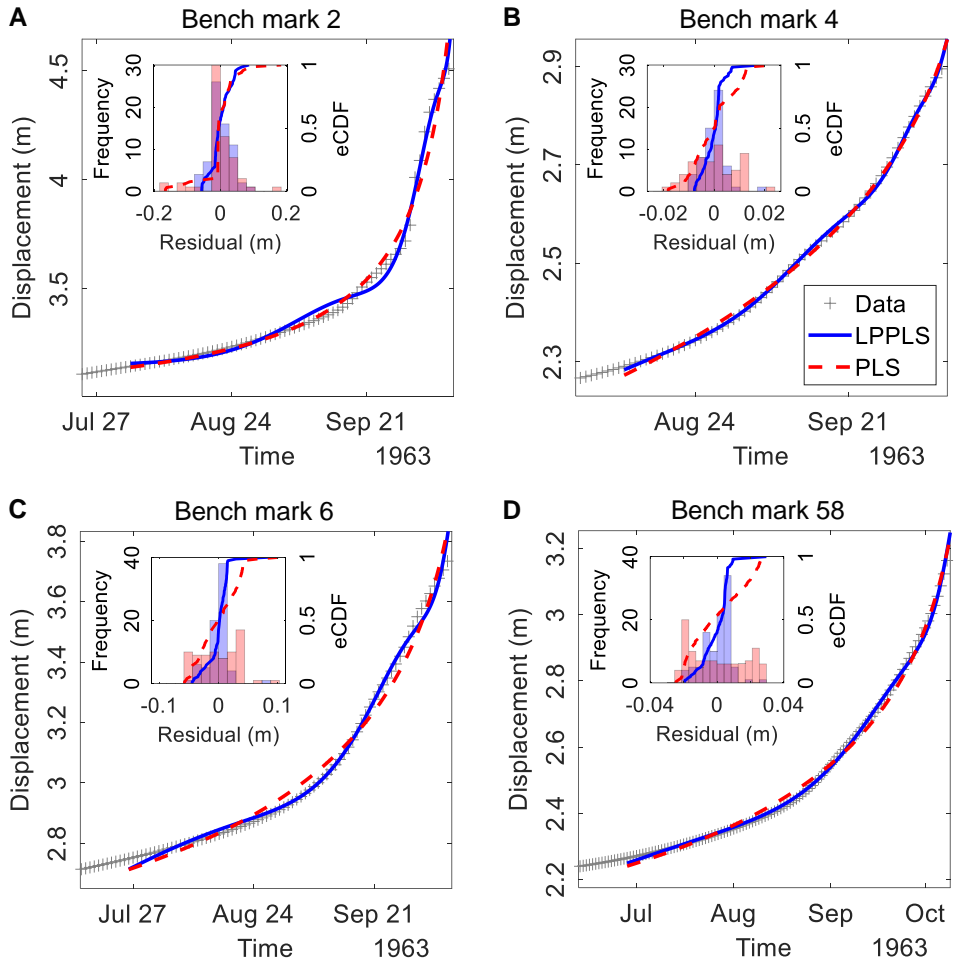

**Fig. S3.** Comparison of the LPPLS versus PLS models in fitting the data of the Vajont landslide.

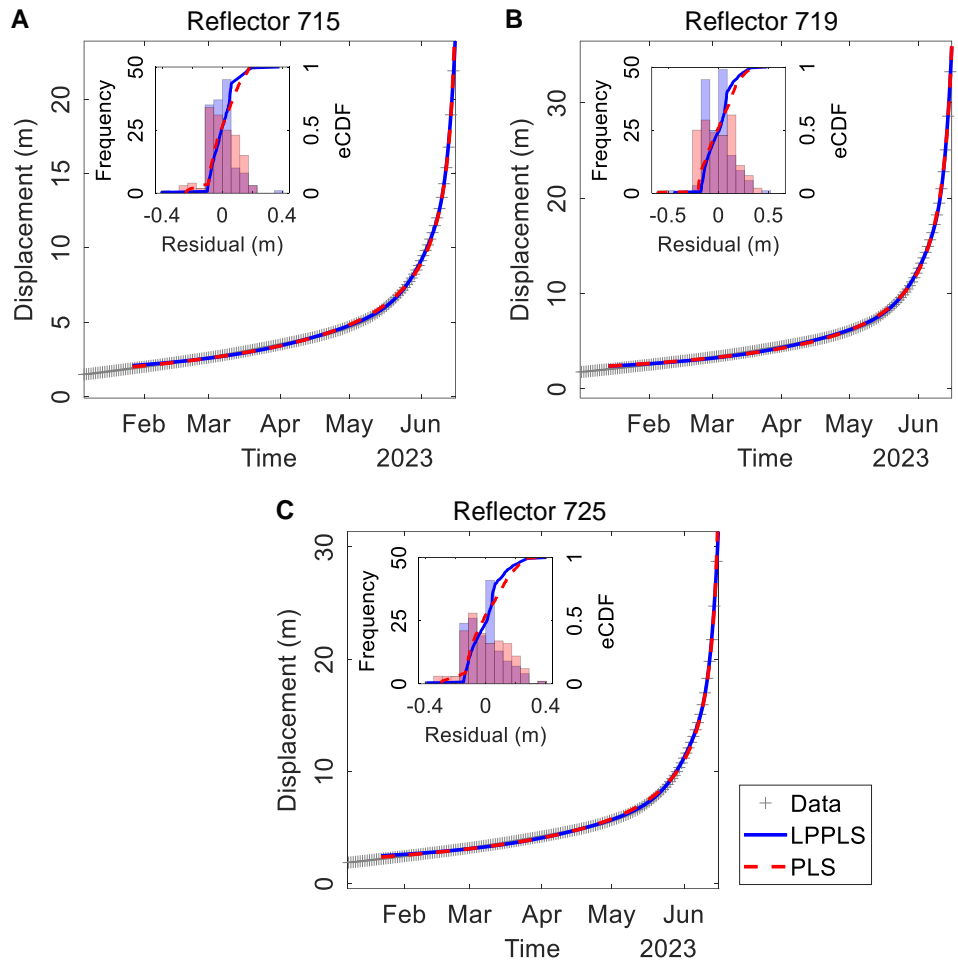

**Fig. S4.** Comparison of the LPPLS versus PLS models in fitting the reflector monitoring data of the Brienz/Brinzauls landslide.

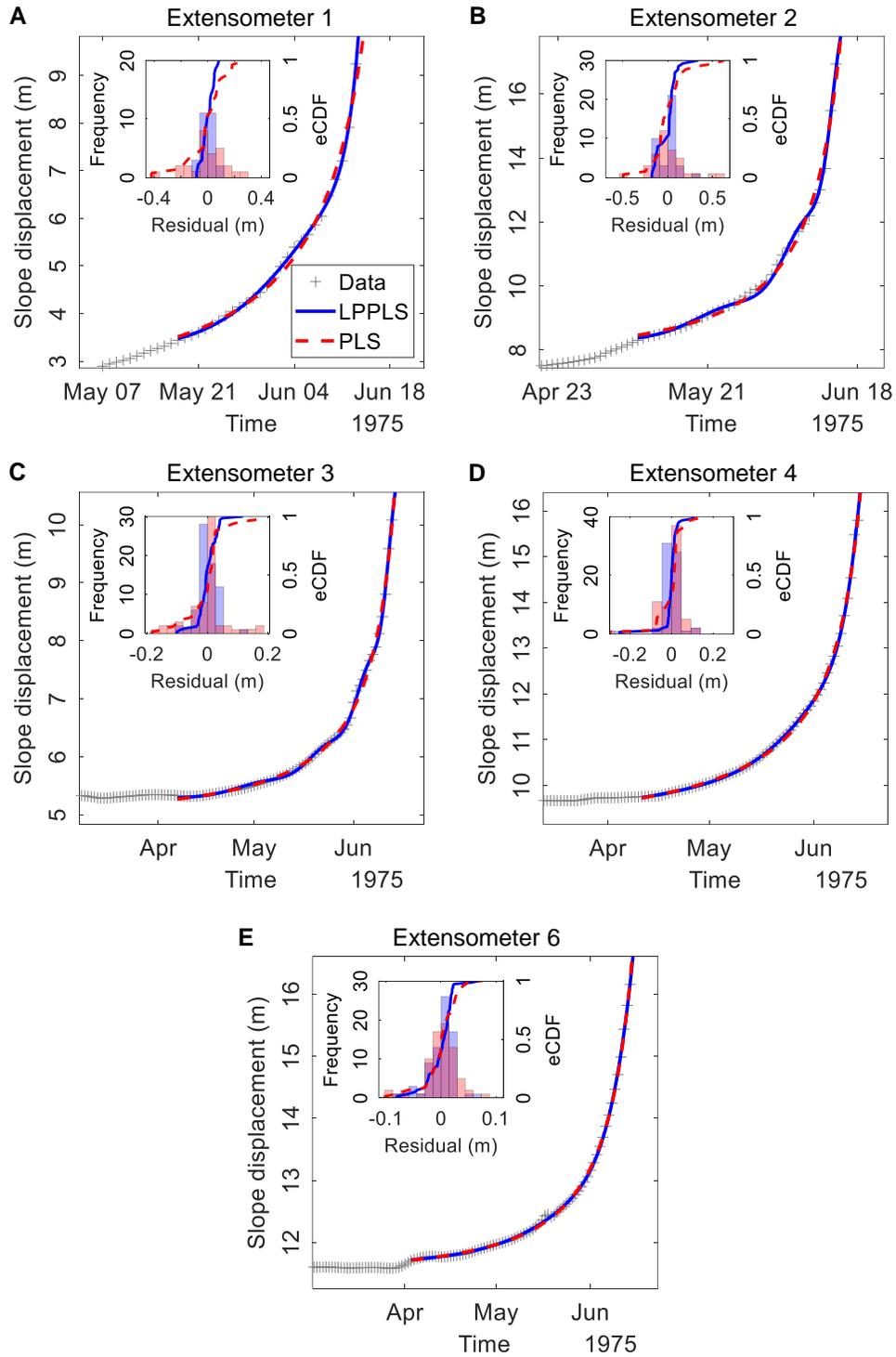

**Fig. S5.** Comparison of the LPPLS versus PLS models in fitting the data of the Hogarth landslide.

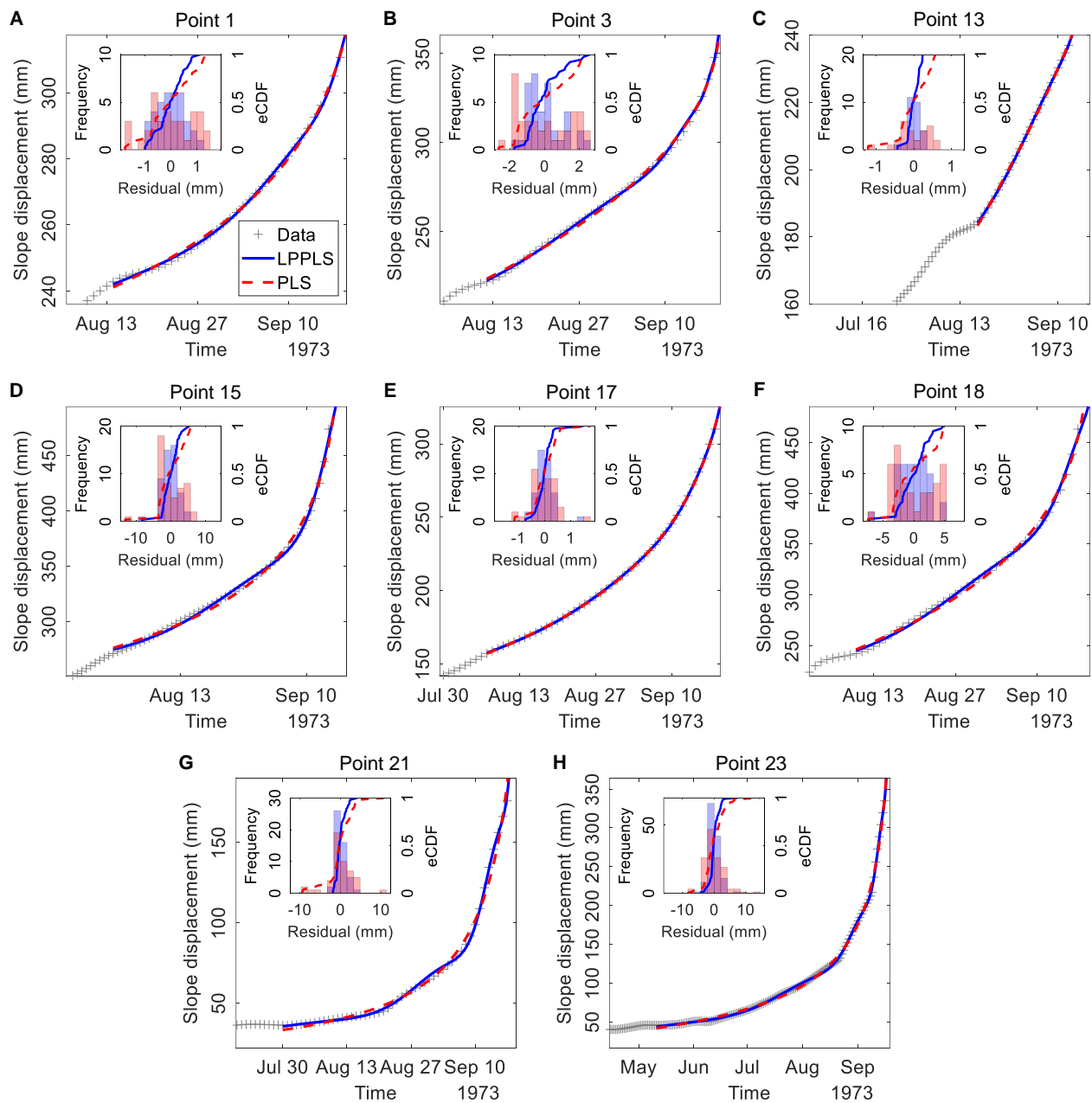

**Fig. S6.** Comparison of the LPPLS versus PLS models in fitting the data of the Kagemori landslide.

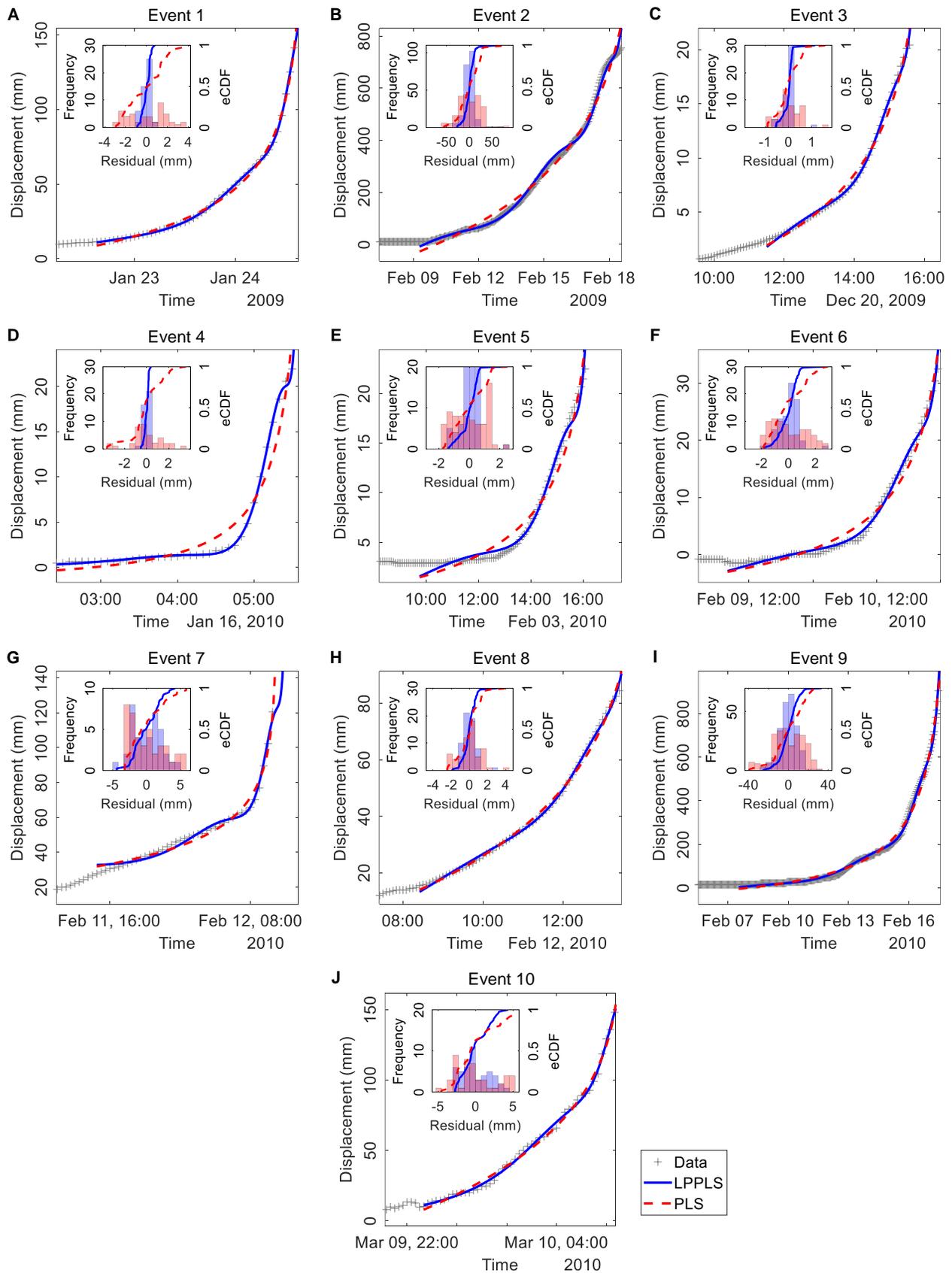

**Fig. S7.** Comparison of the LPPLS versus PLS models in fitting the monitoring data of a road slope.

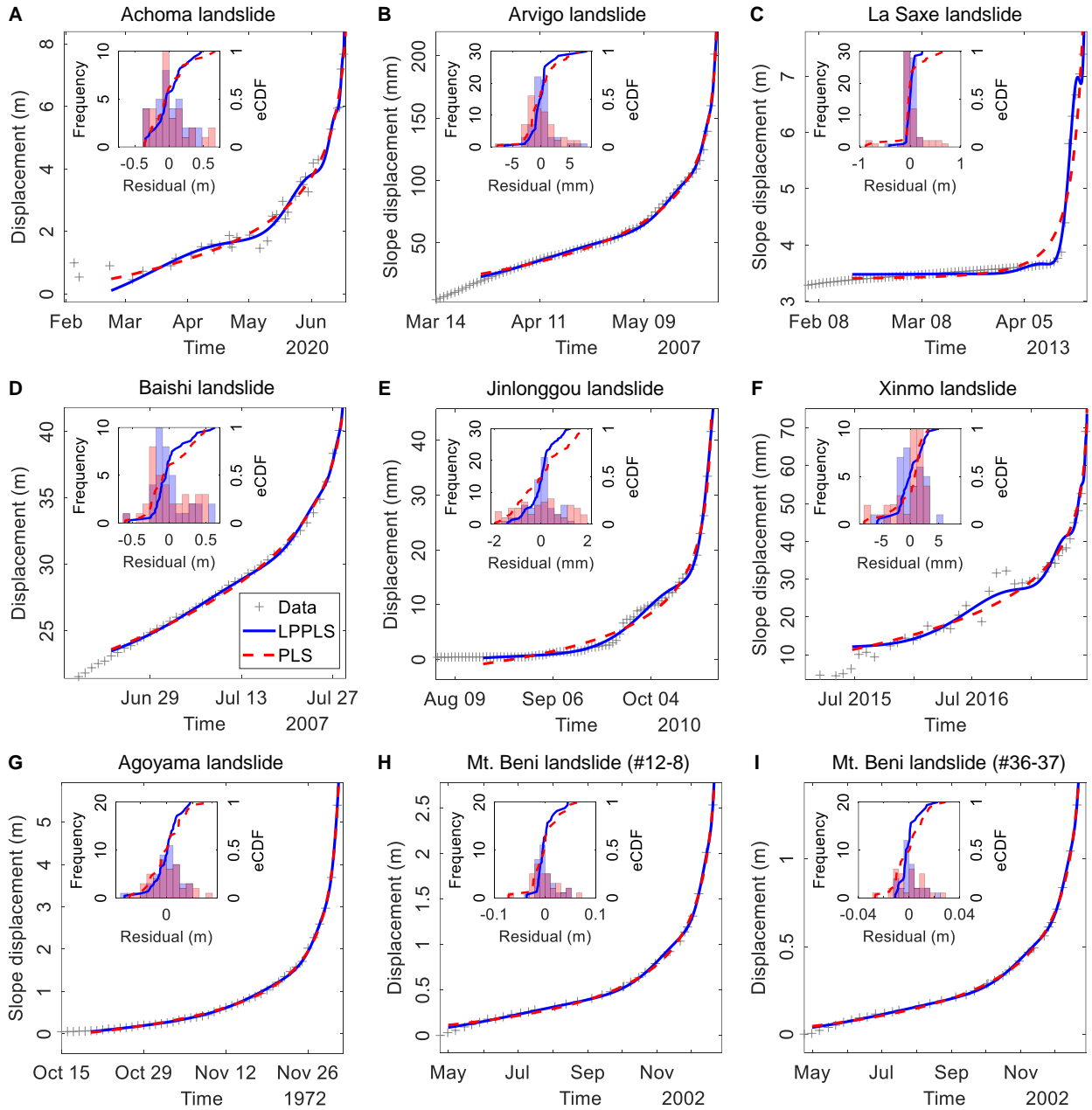

**Fig. S8.** Comparison of the LPPLS versus PLS models in fitting the data of various landslides in rock.

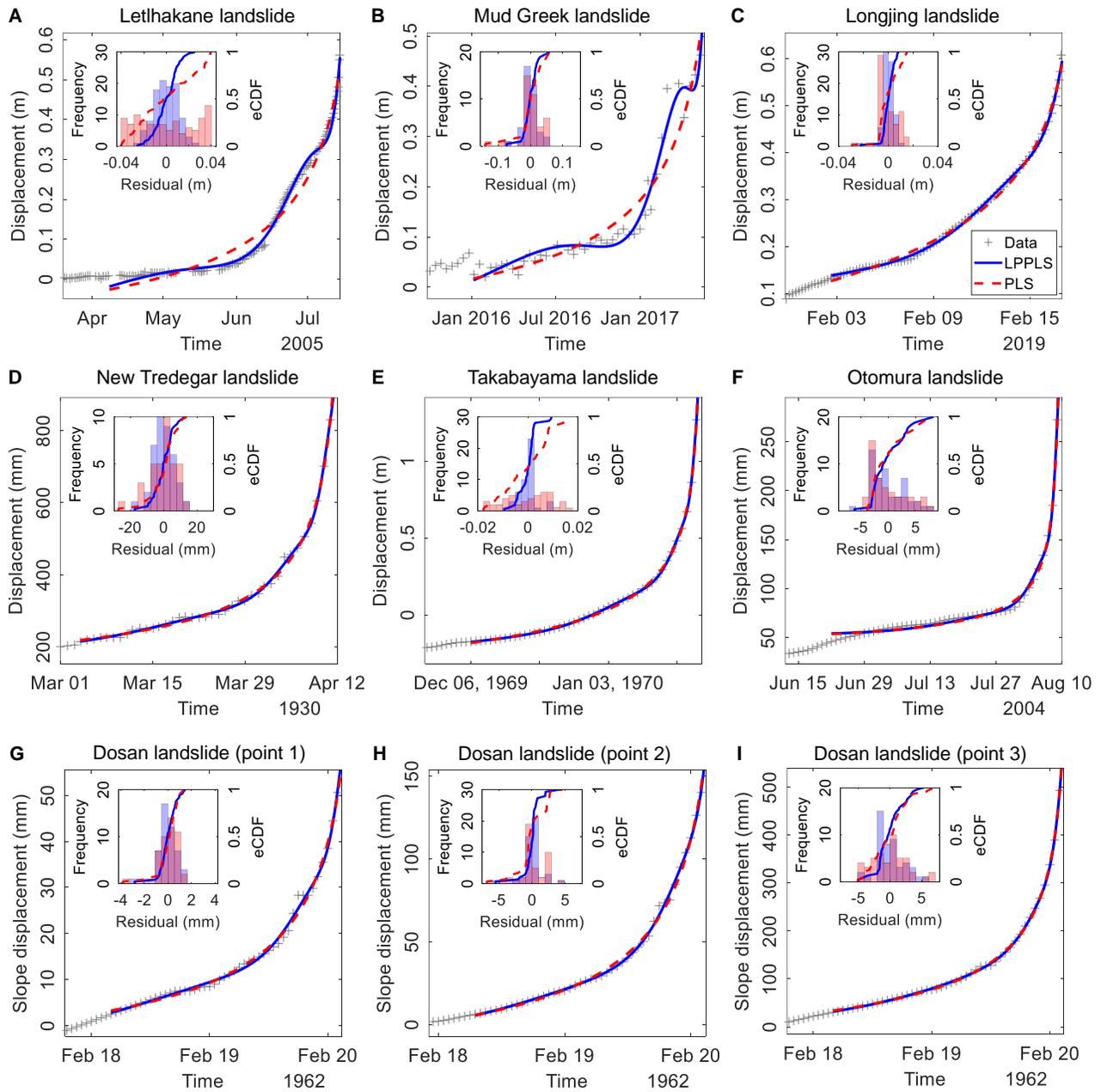

**Fig. S9.** Comparison of the LPPLS versus PLS models in fitting the data of various landslides in rock.

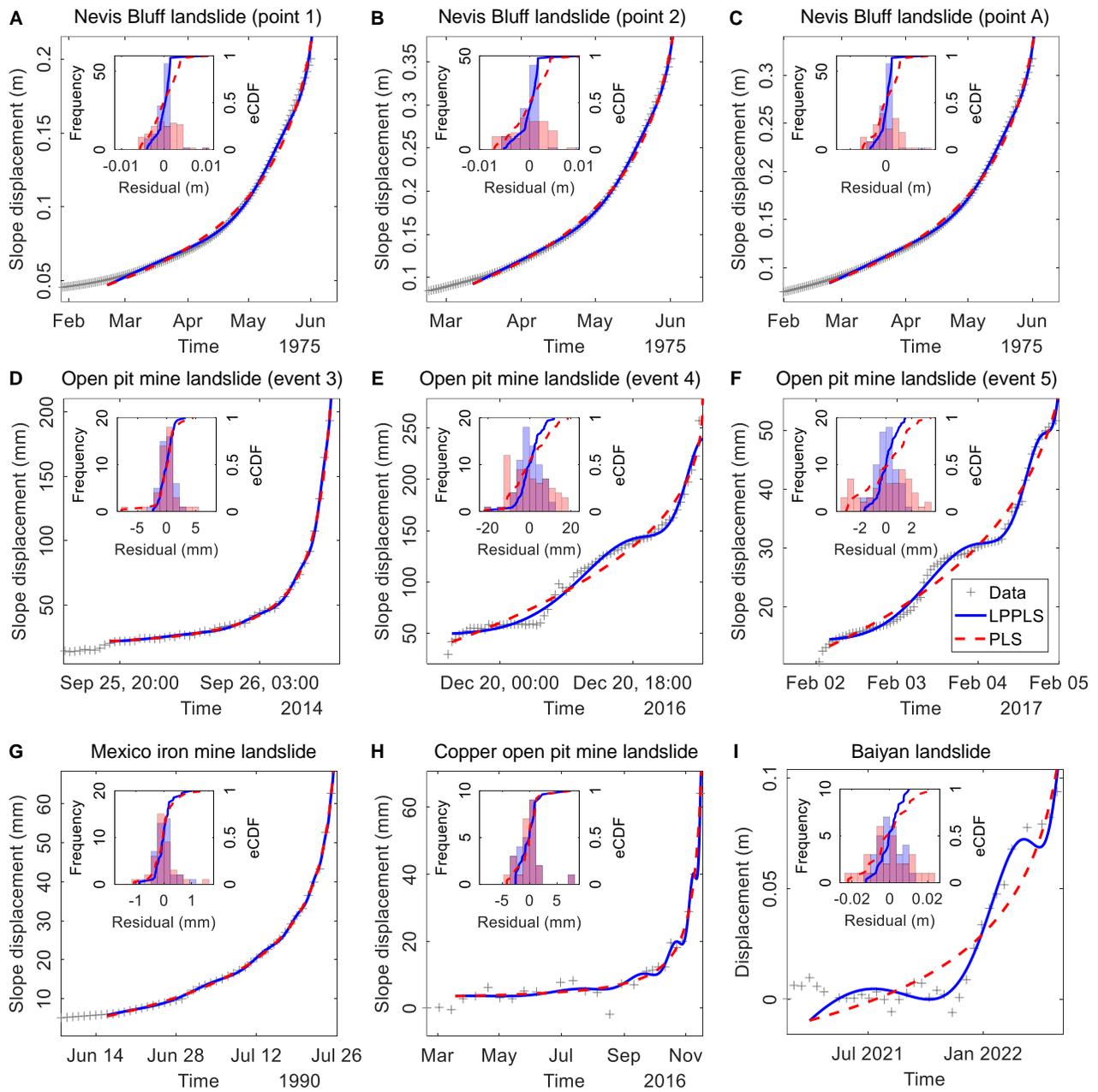

**Fig. S10.** Comparison of the LPPLS versus PLS models in fitting the data of various landslides in rock.

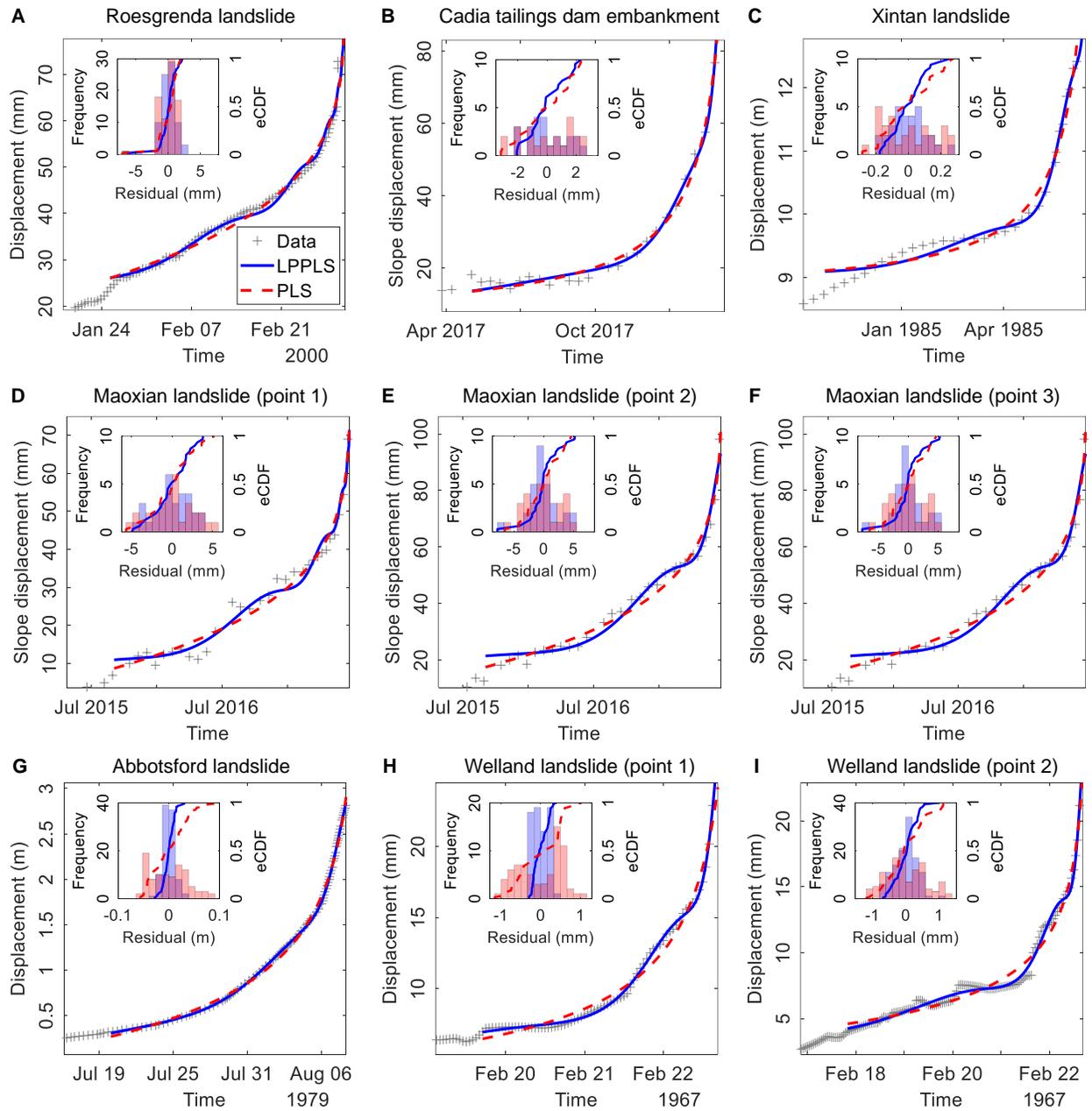

**Fig. S11.** Comparison of the LPPLS versus PLS models in fitting the data of various landslides in soil.

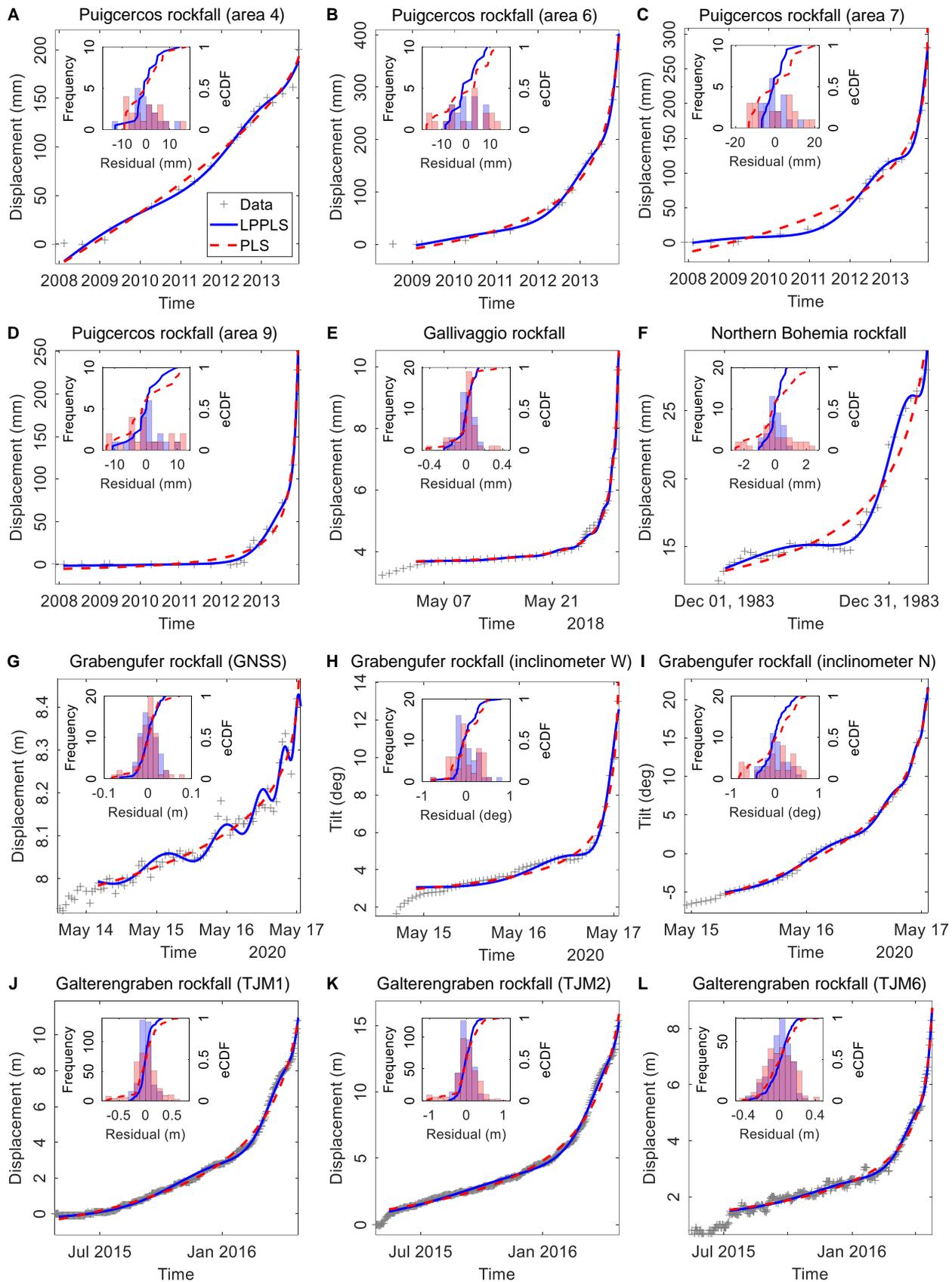

**Fig. S12.** Comparison of the LPPLS versus PLS models in fitting the data of various rockfalls.

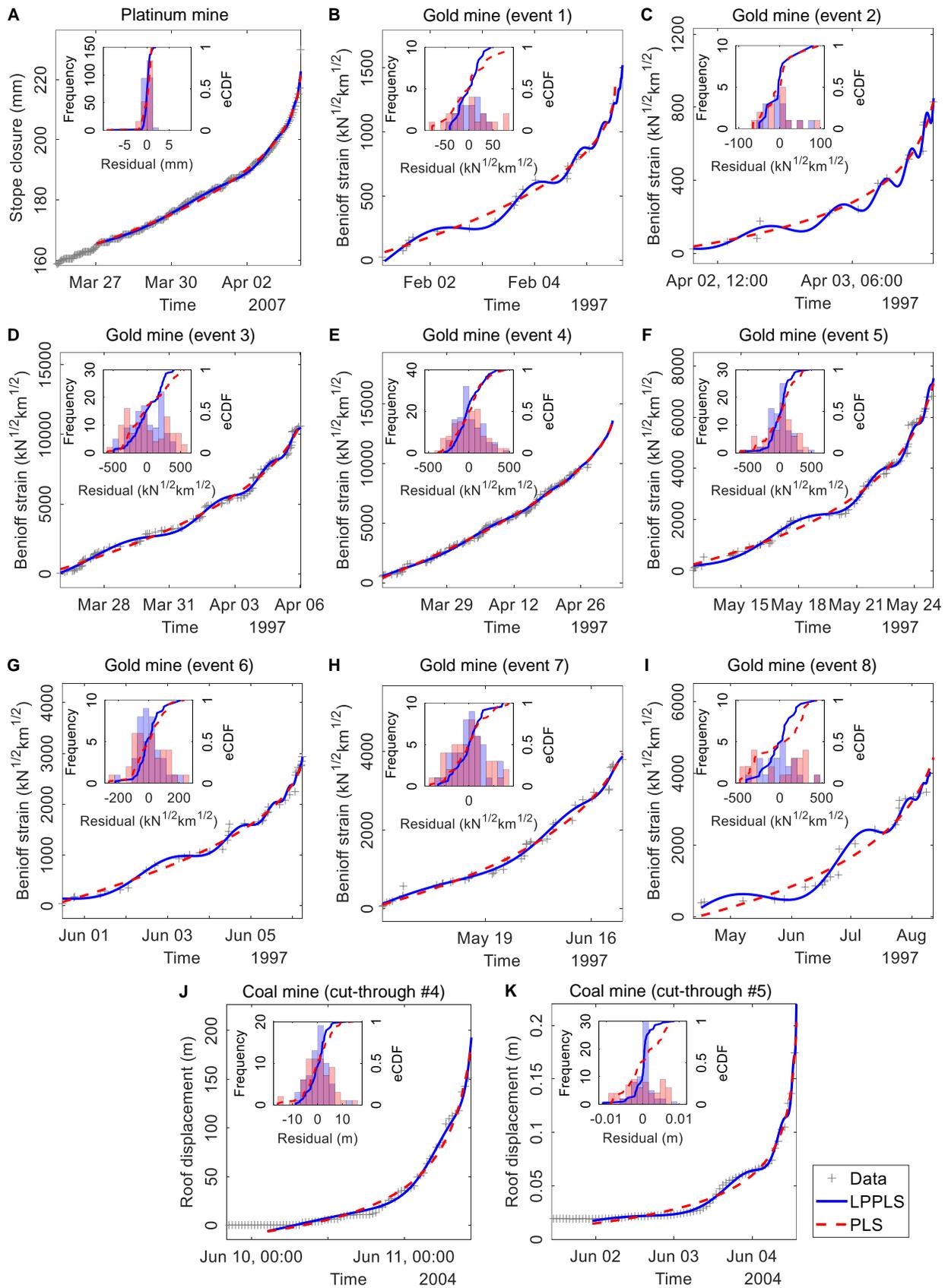

**Fig. S13.** Comparison of the LPPLS versus PLS models in fitting the data of various rockbursts.

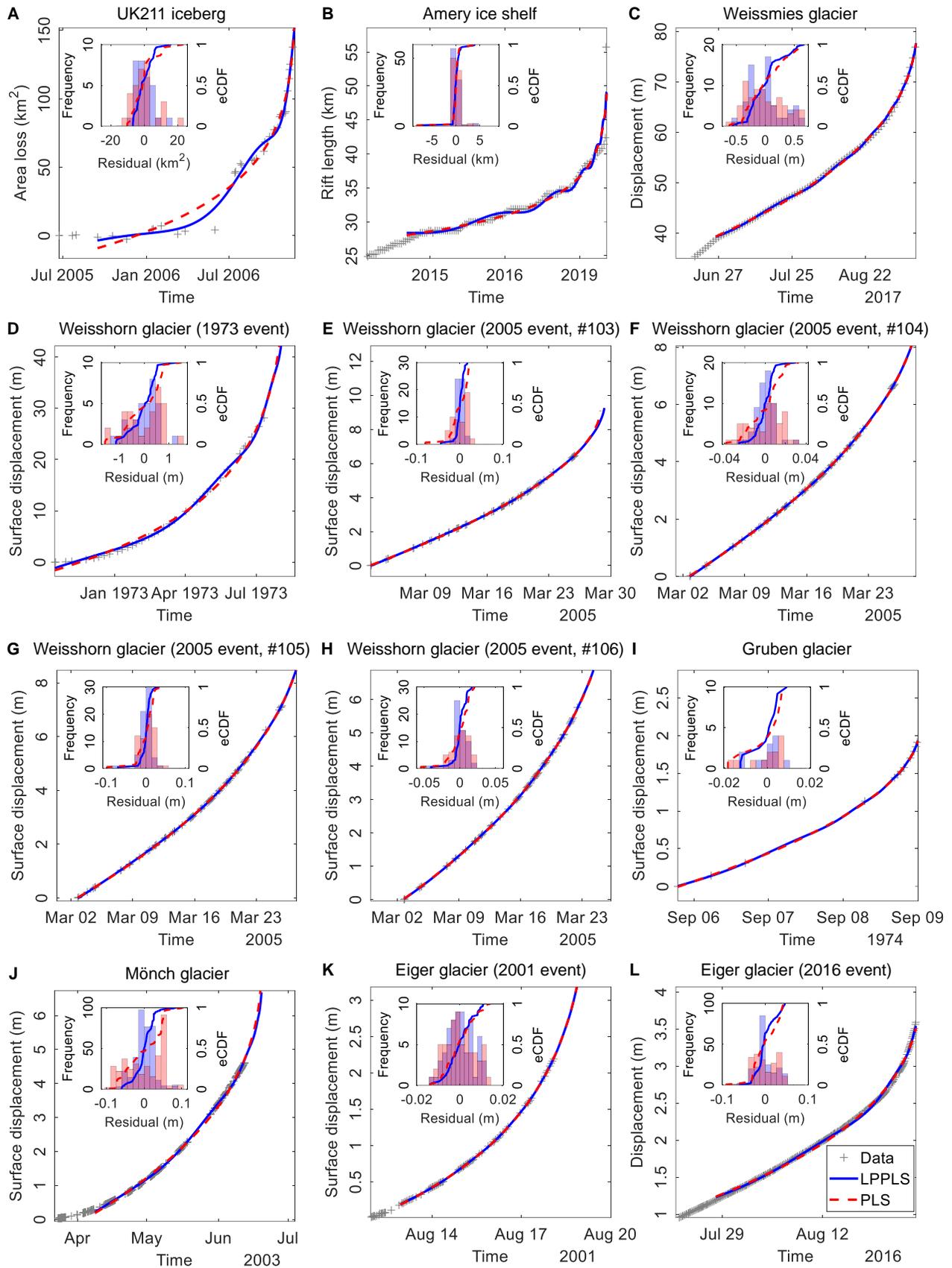

**Fig. S14.** Comparison of the LPPLS versus PLS models in fitting the data of various glaciers.

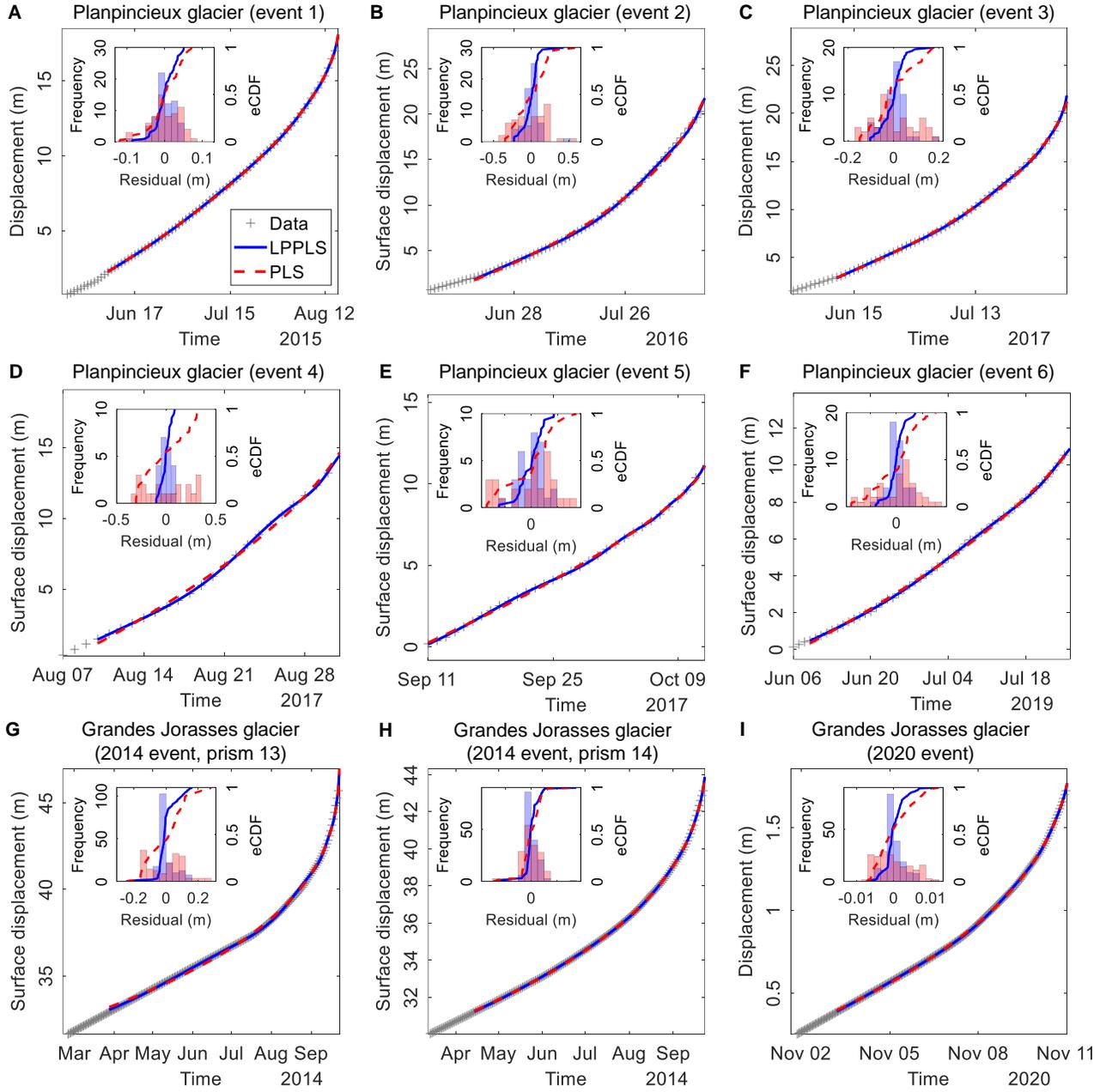

**Fig. S15.** Comparison of the LPPLS versus PLS models in fitting the data of multiple breakoff events of the Planpincieux glacier and Grandes Jorasses glacier.

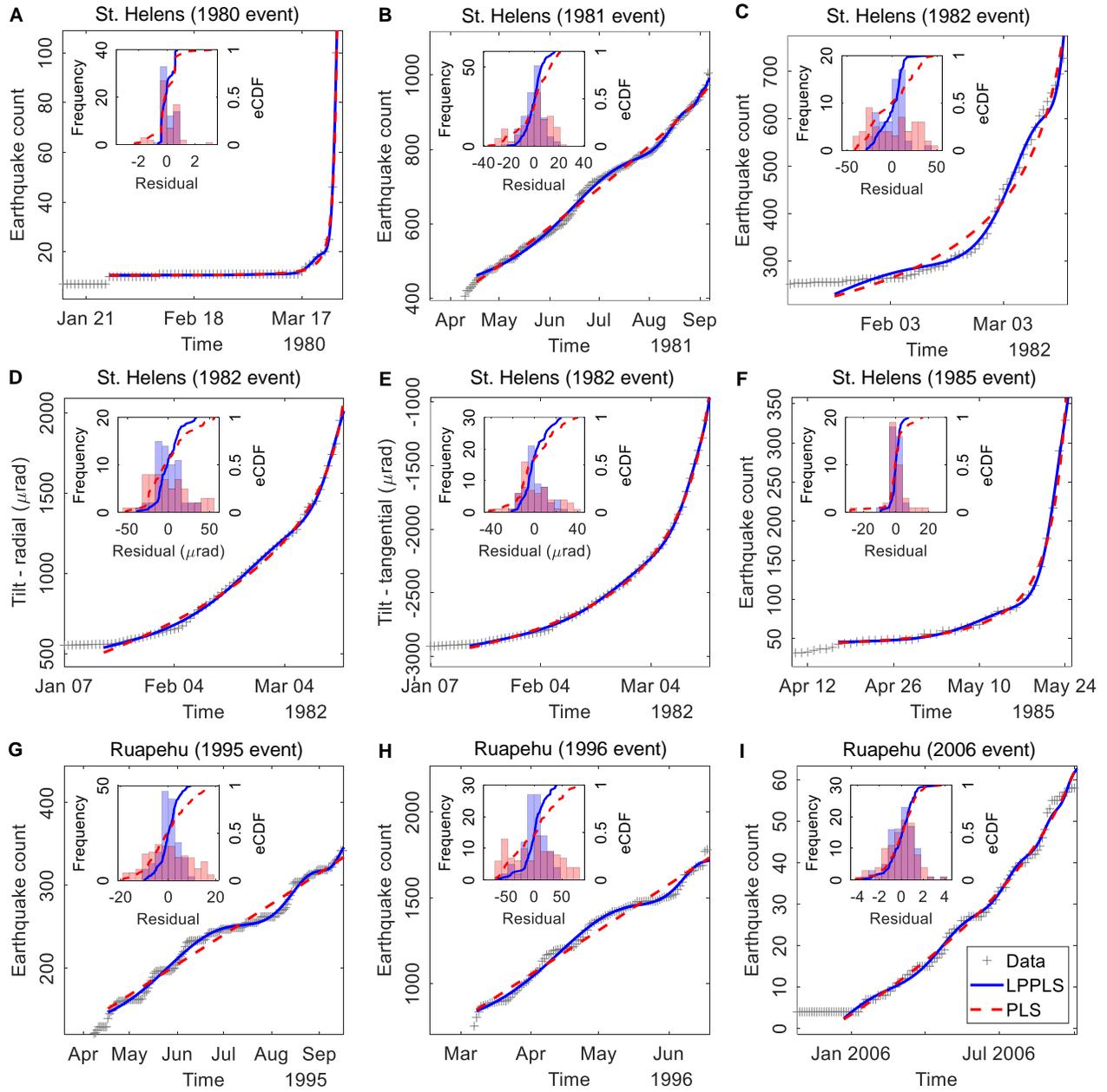

**Fig. S16.** Comparison of the LPPLS versus PLS models in fitting the data of multiple eruption events at the Mount St. Helen volcano and the Ruapehu volcano.

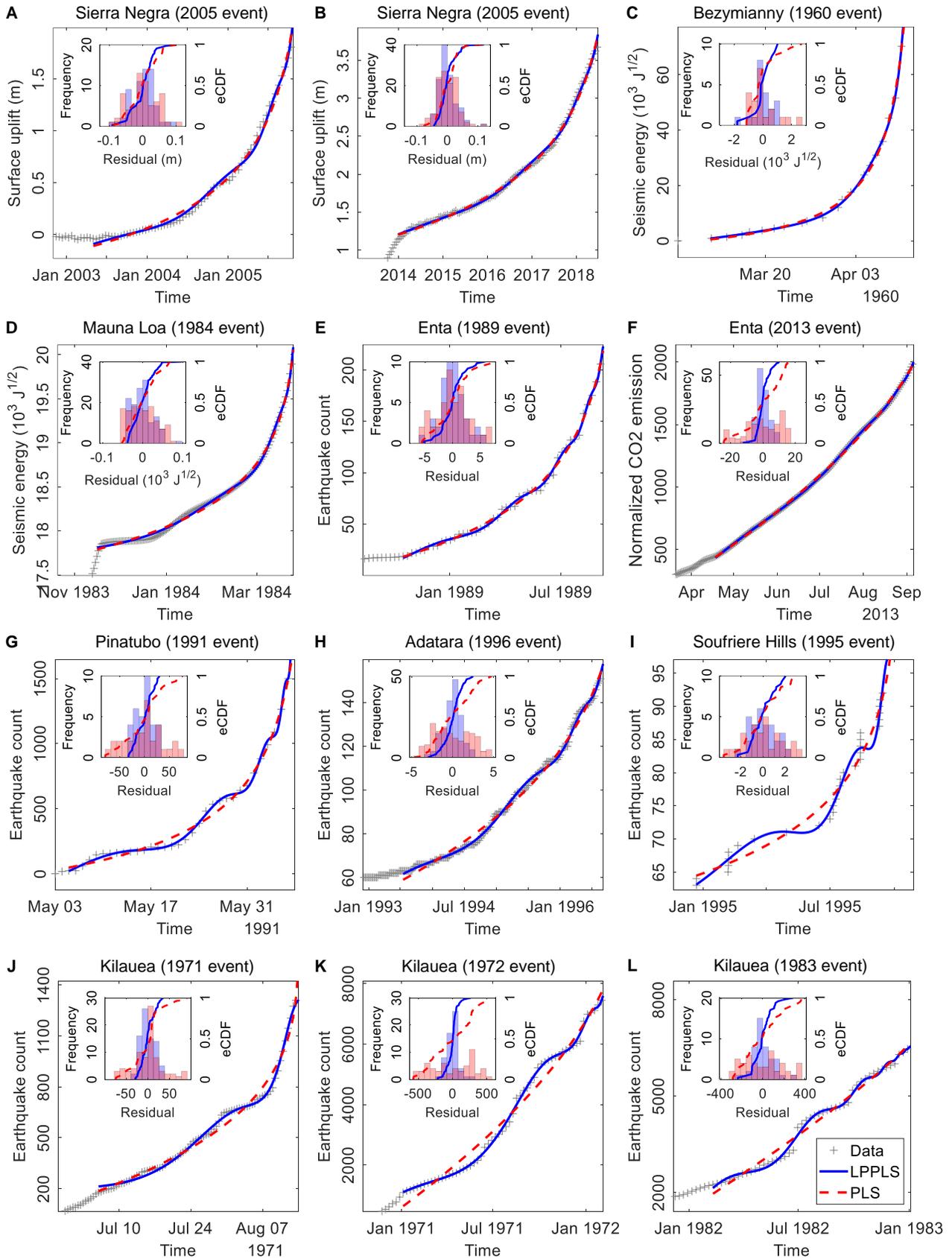

**Fig. S17.** Comparison of the LPPLS versus PLS models in fitting the data of various volcanoes.

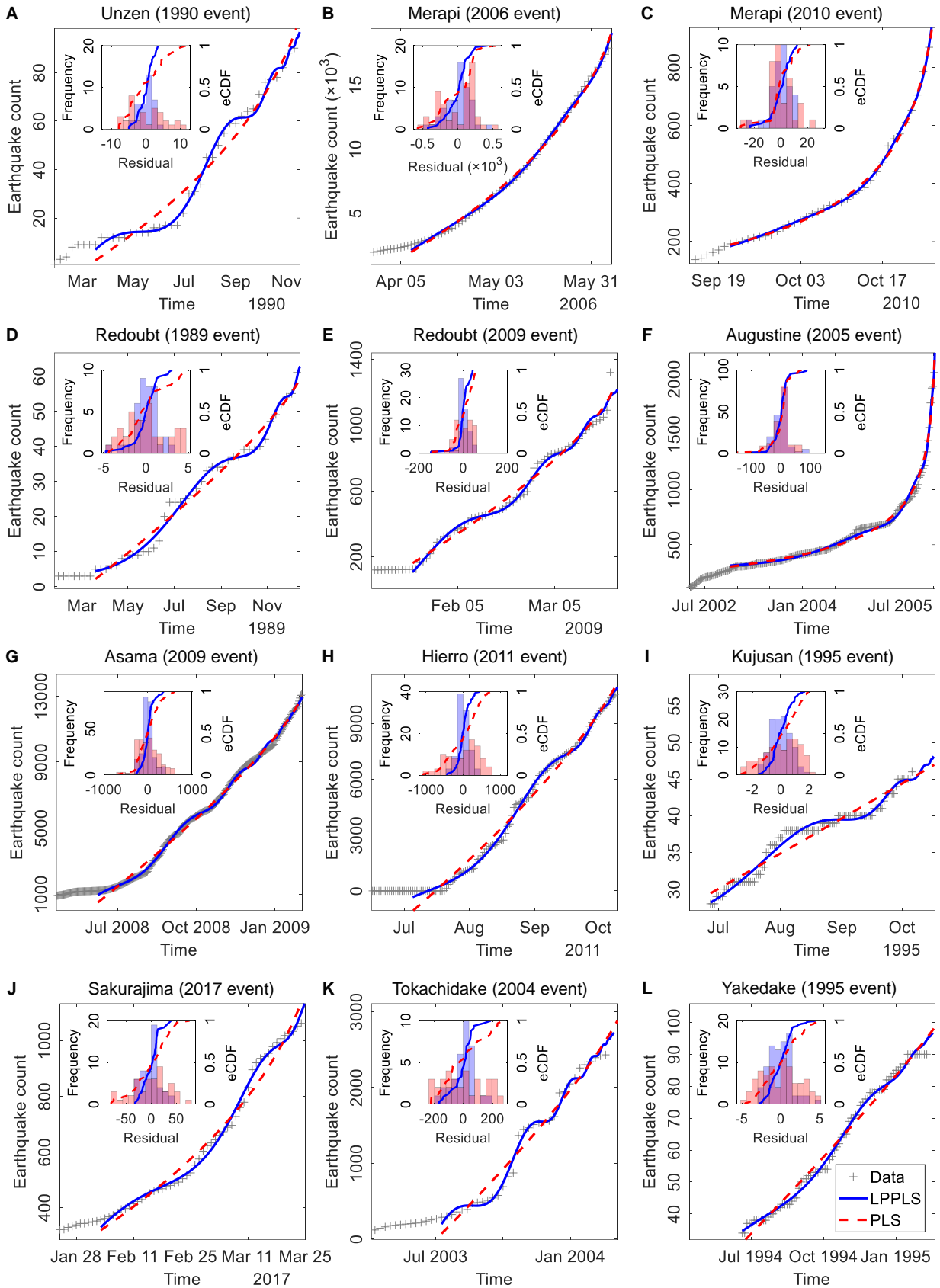

**Fig. S18.** Comparison of the LPPLS versus PLS models in fitting the data of various volcanoes.

**Table S1. Landslide cases (49 events in total).**

| Site | Location | Type | Material | Failure time | Volume (m$^3$) | Monitoring method | Data source | Reference |
|---|---|---|---|---|---|---|---|---|
| Abbotsford | New Zealand | Soilslide | Clay | 1979-08-08 | $5 \times 10^6$ | Survey lines | Digitized | (57) |
| Achoma | Peru | Rock-soilslide | Lacustrine sediments | 2020-06-16 | $5.4 \times 10^6$ | Optical satellites | Digitized | (36) |
| Agoyama | Japan | Rockslide | Tuffaceous sandstone | 1972-12-02 | Unknown | Geodetic bench marks | Digitized | (58) |
| Arvigo | Switzerland | Topple / rockslide | Gneiss | 2007-05-28 | $2 \times 10^5$ | Telejointmeter | Original | (15) |
| Baishi | China | Rockslide | Phyllite | 2007-07-28 | $2 \times 10^6$ | Total station with reflectors | Digitized | (59) |
| Baiyan | China | Rockslide | Limestone | 2022-05-08 | $2.5 \times 10^4$ | Satellite-based InSAR | Digitized | (60) |
| Brienz / Brinzauls | Switzerland | Rockslide | Flysch, schists, dolomite | 2023-06-15 | $1.2 \times 10^6$ | Total station with reflectors | Original | (61) |
| Cadia | Australia | Soilslide | Earthfill materials | 2018-03-09 | $7.2 \times 10^4$ | Satellite-based InSAR | Digitized | (62) |
| Copper open pit | Undisclosed | Rockslide | Limestone, spilite | 2016-11-17 | $6.4 \times 10^5$ | Satellite-based InSAR | Digitized | (62) |
| Dosan | Japan | Rockslide | Schist | 1962-02-20 | $6 \times 10^4$ | Crack meter | Digitized | (63) |
| Gallivaggio | Italy | Rockfall | Granite | 2018-05-29 | $5 \times 10^3$ | Ground-based InSAR | Digitized | (40) |
| Galterengraben | Switzerland | Rockfall | Sandstone | 2016-04-24 | $2.5 \times 10^3$ | Telejointmeter | Original | (15) |
| Grabengufer | Switzerland | Rockfall | Rock & ice | 2020-05-17 | $5 \times 10^2$ | GNSS & inclinometer | Original | (64) |
| Hogarth | Canada | Topple | Diorite | 1975-06-23 | $2 \times 10^5$ | Extensometers | Digitized | (65) |
| Iron mine | Mexico | Rockslide | Rock | 1990-07-26 | Unknown | Total station with reflectors | Digitized | (66) |
| Jinlonggou | China | Rockslide | Syenite & basalt | 2010-10-23 | $2 \times 10^5$ | Extensometer | Digitized | (67) |
| Kagemori | Japan | Rockslide | Limestone | 1973-09-20 | $3 \times 10^5$ - $4 \times 10^5$ | Measuring tapes | Digitized | (68) |
| La Saxe | Italy | Rockslide | Meta-sedimentary sequences | 2013-04-21 | $5 \times 10^2$ - $1 \times 10^3$ | Total station with reflectors | Original | (69) |
| Letlhakane diamond mine | Botswana | Rockslide | Sandstone | 2005-07-14 | $2.3 \times 10^5$ | Total station with reflectors | Digitized | (70) |
| Longjing | China | Rockslide | Dolomite & limestone | 2019-02-17 | $1.4 \times 10^6$ | Extensometers | Digitized | (71) |

**Table S1 (continued). Landslide cases (49 events in total).**

| Site | Location | Type | Material | Failure time | Volume (m$^3$) | Monitoring method | Data source | Reference |
|---|---|---|---|---|---|---|---|---|
| Maoxian | China | Soilslide | Soil | 2017-06-24 | $1.5\times10^7$ | Satellite-based InSAR | Digitized | (*72*) |
| Mt. Beni | Italy | Rockslide | Basalt & limestone | 2002-12-28 | $5.0\times10^5$ | Distometric benchmarks | Digitized | (*73*) |
| Mud Greek | USA | Rockslide | Shale, sandstone, sediments | 2017-05-20 | $3\times10^6$ | Satellite-based InSAR | Digitized | (*74*) |
| Nevis Bluff | New Zealand | Flexural topple / rockslide | Schist | 1975-06-14 | $3.2\times10^4$ | Survey markers | Digitized | (*75*) |
| New Tredegar | UK | Rockslide | Sandstone | 1930-04-12 | Unknown | Unspecified | Digitized | (*76*) |
| Northern Bohemia | Czech Republic | Rockfall | Sandstone | 1984-01-07 | $1.4\times10^3$ | Extensometers | Digitized | (*77*) |
| Open pit mine (event 3) | Undisclosed | Rockslide | Anorthosite | 2014-09-26 | $6\times10^2$ | Ground-based InSAR | Digitized | (*78*) |
| Open pit mine (event 4) | Undisclosed | Rockslide | Anorthosite | 2014-2017 | $3\times10^3$ | Ground-based InSAR | Digitized | (*78*) |
| Open pit mine (event 5) | Undisclosed | Topple | Anorthosite | 2017-02-05 | $4\times10^3$ | Ground-based InSAR | Digitized | (*78*) |
| Otomura | Japan | Rockslide | Sandstone & shale | 2004-08-10 | $2\times10^5$ | Extensometers | Digitized | (*79*) |
| Preonzo | Switzerland | Rockslide | Gneiss | 2012-05-15 | $2.1\times10^5$ | Extensometers & total station with reflectors | Original | (*35*) |
| Puigcercós | Spain | Rockfall | Marl, silt, sandstone, limestone | 2013-12-03 | $1\times10^3$ | LiDAR | Digitized | (*39*) |
| Road slope (event 1) | Undisclosed | Rockslide | Mobilized gneiss | 2009-01-24 | $3\times10^2$ | Ground-based InSAR | Digitized | (*80*) |
| Road slope (event 2) | Undisclosed | Flow | Colluvium | 2009-02-18 | $1.4\times10^1$ | Ground-based InSAR | Digitized | (*80*) |
| Road slope (event 3) | Undisclosed | Soilslide | Colluvium & beton | 2009-12-20 | $1.6\times10^2$ | Ground-based InSAR | Digitized | (*80*) |
| Road slope (event 4) | Undisclosed | Soilslide | Colluvium & beton | 2010-01-16 | $2\times10^2$ | Ground-based InSAR | Digitized | (*80*) |
| Road slope (event 5) | Undisclosed | Soilslide | Colluvium & beton | 2009-02-03 | $8\times10^1$ | Ground-based InSAR | Digitized | (*80*) |
| Road slope (event 6) | Undisclosed | Soilslide | Colluvium & beton | 2010-02-11 | $5\times10^2$ | Ground-based InSAR | Digitized | (*80*) |
| Road slope (event 7) | Undisclosed | Flow | Mobilized & altered gneiss | 2010-02-12 | $2\times10^2$ | Ground-based InSAR | Digitized | (*80*) |
| Road slope (event 8) | Undisclosed | Rockslide | Colluvium & beton | 2010-02-12 | $3\times10^2$ | Ground-based InSAR | Digitized | (*80*) |

**Table S1 (continued). Landslide cases (49 events in total).**

| Site | Location | Type | Material | Failure time | Volume (m$^3$) | Monitoring method | Data source | Reference |
|---|---|---|---|---|---|---|---|---|
| Road slope (event 9) | Undisclosed | Rockslide | Mobilized & altered gneiss | 2010-02-17 | $8\times10^1$ | Ground-based InSAR | Digitized | (*80*) |
| Road slope (event 10) | Undisclosed | Rockslide | Mobilized & altered gneiss | 2010-03-10 | $1.5\times10^2$ | Ground-based InSAR | Digitized | (*80*) |
| Roesgrenda | Norway | Soilslide | Quick clay | 2000-03-02 | $2\times10^3$ | Extensometers | Digitized | (*38*) |
| Takabayama | Japan | Rockslide | Mudstone, sandstone | 1970-01-22 | $5\times10^3$ | Extensometers | Digitized | (*81*) |
| Vajont | Italy | Rockslide | Limestone | 1963-10-09 | $2.7\times10^8$ | Geodetic bench marks | Digitized | (*37*) |
| Veslemannen | Norway | Rockslide | Gneiss | 2019-09-05 | $5.4\times10^4$ | Ground-based InSAR | Original | (*30*) |
| Welland | Canada | Soilslide | Clay | 1967-02-22 | $5\times10^2$ | Extensometers | Digitized | (*82*) |
| Xinmo | China | Rockslide | Sandstone, phyllite | 2017-06-24 | $1.3\times10^7$ | Satellite-based InSAR | Digitized | (*62*) |
| Xintan | China | Rockslide | Sediments | 1985-06-12 | $3\times10^7$ | Geodetic bench marks | Digitized | (*83*) |

Note: InSAR - Interferometric Synthetic Aperture Radar; LiDAR - Light Detection and Ranging; GNSS - Global navigation satellite system.

**Table S2. Rockburst cases (11 events in total).**

| Site | Location | Material | Failure time | Monitoring method | Data source | Reference |
|---|---|---|---|---|---|---|
| Coal mine (cut-through #4) | Australia | Coal | 2004-06-11 | Extensometer | Digitized | (*31*) |
| Coal mine (cut-through #5) | Australia | Coal | 2004-06-04 | Extensometer | Digitized | (*31*) |
| Gold mine (event 1) | South Africa | Metamorphic rock | 1997-02-05 | Seismic stations | Digitized | (*42*) |
| Gold mine (event 2) | South Africa | Metamorphic rock | 1997-04-03 | Seismic stations | Digitized | (*42*) |
| Gold mine (event 3) | South Africa | Metamorphic rock | 1997-04-06 | Seismic stations | Digitized | (*42*) |
| Gold mine (event 4) | South Africa | Metamorphic rock | 1997-05-04 | Seismic stations | Digitized | (*42*) |
| Gold mine (event 5) | South Africa | Metamorphic rock | 1997-05-26 | Seismic stations | Digitized | (*42*) |
| Gold mine (event 6) | South Africa | Metamorphic rock | 1997-06-06 | Seismic stations | Digitized | (*42*) |
| Gold mine (event 7) | South Africa | Metamorphic rock | 1997-06-23 | Seismic stations | Digitized | (*42*) |
| Gold mine (event 8) | South Africa | Metamorphic rock | 1997-08-15 | Seismic stations | Digitized | (*42*) |
| Platinum mine | South Africa | Merensky reef | 2007-04-04 | Closure meters | Digitized | (*41*) |

**Table S3. Glacier cases (17 events in total).**

| Site | Location | Type | Failure time | Size | Monitoring method | Data source | Reference |
|---|---|---|---|---|---|---|---|
| Amery | Antarctica | Ice shelf | 2019-09-25 | 1,636 km$^2$ | Satellite images | Digitized | (*43*) |
| Eiger glacier (2001 event) | Switzerland | Polythermal glacier | 2001-08-20 | 2.7×10$^5$ m$^3$ | Theodolite with reflectors | Digitized | (*84*) |
| Eiger glacier (2016 event) | Switzerland | Polythermal glacier | 2016-08-25 | 1.5×10$^4$ m$^3$ | Ground-based InSAR | Digitized | (*85*) |
| Grandes Jorasses (2014 event) | Italy | Cold glacier | 2014-09-23 & 2014-09-29 | 1.5×10$^5$ m$^3$ | Total station with reflectors | Original | (*32*) |
| Grandes Jorasses (2020 event) | Italy | Cold glacier | 2020-11-11 | 2.0×10$^4$ m$^3$ | Total station with reflectors | Digitized | (*86*) |
| Gruben | Switzerland | Temperate glacier | 1974-09-09 | Unknown | Dial gauge with wire | Digitized | (*84*) |
| Mönch | Switzerland | Temperate glacier | 2003-07-04 | 6×10$^5$ m$^3$ | Theodolite with reflectors | Original | (*84*) |
| Planpincieux (event 1) | Italy | Polythermal glacier | 2015-08-16 | 1.4×10$^4$ m$^3$ | Time-lapse camera | Original | (*46, 86*) |
| Planpincieux (event 2) | Italy | Polythermal glacier | 2016-08-15 | 2.5×10$^4$ m$^3$ | Time-lapse camera | Original | (*46, 86*) |
| Planpincieux (event 3) | Italy | Polythermal glacier | 2017-08-03 | 2.0×10$^4$ m$^3$ | Time-lapse camera | Original | (*46, 86*) |
| Planpincieux (event 4) | Italy | Polythermal glacier | 2017-08-31 | 5.4×10$^4$ m$^3$ | Time-lapse camera | Original | (*46, 86*) |
| Planpincieux (event 5) | Italy | Polythermal glacier | 2017-10-12 | 1.5×10$^4$ m$^3$ | Time-lapse camera | Original | (*46, 86*) |
| Planpincieux (event 6) | Italy | Polythermal glacier | 2019-07-26 | 2.2×10$^4$ m$^3$ | Time-lapse camera | Original | (*46, 86*) |
| UK211 | Antarctica | Iceberg | 2006-11-23 | 2,400 km$^2$ | Satellite images | Digitized | (*44*) |
| Weisshorn (1973 event) | Switzerland | Cold glacier | 1972-10-17 | 5×10$^5$ m$^3$ | Theodolite with reflectors | Original | (*87*) |
| Weisshorn (2005 event) | Switzerland | Cold glacier | 2005-03-30 | 5×10$^5$ m$^3$ | Total station with reflectors | Original | (*87, 88*) |
| Weissmies | Switzerland | Polythermal glacier | 2017-09-10 | 2.5×10$^5$ m$^3$ | Time-lapse camera | Original | (*15, 45*) |

**Table S4. Volcano cases (32 events in total).**

| Site | Location | Type | Eruption time | Erupted volume (m$^3$) | Monitoring method | Data source | Reference |
|---|---|---|---|---|---|---|---|
| Adatara | Japan | Stratovolcano | 1996-09-01 | $1.1 \times 10^6$ | JMA seismic network | Original | (*56*) |
| Asama | Japan | Complex volcano | 2009-02-02 | $\sim 10^4$ | JMA seismic network | Original | (*56*) |
| Augustine | USA | Lava dome | 2006-01-11 | $7.3 \times 10^7$ | USGS seismic network-AVO | Original | (*56*) |
| Axial Seamount | USA | Submarine fissure volcano | 2011-04-06 | $9.9 \times 10^7$ | Bottom pressure recorders & mobile pressure recorders | Original | (*89*) |
| Bezymianny | Russia | Stratovolcano | 1960-04-10 | $\sim 10^6$ | Unspecified | Digitized | (*6*) |
| Etna (1989 event) | Italy | Stratovolcano | 1989-09-08 | $\sim 10^7$ | ISCSN | Original | (*56*) |
| Etna (2013 event) | Italy | Stratovolcano | 2013-09-05 | $\sim 10^6$ | Geochemical monitoring stations | Original | (*50, 56*) |
| Hierro | Spain | Submarine shield volcano | 2011-10-10 | $3.3 \times 10^7$ | El Hierro seismic network | Original | (*56, 90*) |
| Kilauea (1971 event) | USA | Shield volcano | 1971-08-14 | $9.1 \times 10^6$ | NCDC-ANSS seismic network | Original | (*56*) |
| Kilauea (1972 event) | USA | Shield volcano | 1972-02-04 | $1.2 \times 10^8$ | NCDC-ANSS seismic network | Original | (*56*) |
| Kilauea (1983 event) | USA | Shield volcano | 1983-01-03 | $4 \times 10^6$ | NCDC-ANSS seismic network | Original | (*56*) |
| Kujusan | Japan | Stratovolcano | 1995-10-11 | $2 \times 10^5$ | JMA seismic network | Original | (*56, 91*) |
| Mauna Loa | USA | Shield volcano | 1984-03-25 | $\sim 10^8$ | Unspecified | Digitized | (*92*) |
| Merapi (2006 event) | Indonesia | Stratovolcano | 2006-06-06 | $5.3 \times 10^6$ | Plawangan Observatory | Original | (*56, 93*) |
| Merapi (2010 event) | Indonesia | Stratovolcano | 2010-10-26 | $\sim 10^7$ | Plawangan Observatory | Original | (*47, 56*) |
| Pinatubo | Philippines | Stratovolcano | 1991-06-07 | $5 \times 10^9$ | Seismic network | Original | (*56*) |
| Redoubt (1989 event) | USA | Stratovolcano | 1989-12-14 | $\sim 10^8$ | ANSS seismic network | Original | (*56, 94*) |
| Redoubt (2009 event) | USA | Stratovolcano | 2009-03-22 | $\sim 10^8$ | ANSS seismic network | Original | (*56, 95*) |
| Ruapehu (1995 event) | New Zealand | Stratovolcano | 1995-09-18 | $2 \times 10^5$ | New Zealand seismic network | Original | (*56*) |
| Ruapehu (1996 event) | New Zealand | Stratovolcano | 1996-06-19 | $\sim 10^7$ | New Zealand seismic network | Original | (*56*) |
| Ruapehu (2006 event) | New Zealand | Stratovolcano | 2006-10-04 | $\sim 10^5$ | New Zealand seismic network | Original | (*56*) |

**Table S4 (continued). Volcano cases (32 events in total).**

| Site | Location | Type | Eruption time | Erupted volume ($m^3$) | Monitoring method | Data source | Reference |
|---|---|---|---|---|---|---|---|
| Sakurajima | Japan | Stratovolcano | 2017-03-25 | $\sim10^6$ | JMA seismic network | Original | (*56*) |
| Sierra Negra (2005 event) | Ecuador | Shield volcano | 2005-10-22 | $1.5\times10^8$ | Continuous GPS network | Original | (*96*) |
| Sierra Negra (2018 event) | Ecuador | Shield volcano | 2018-06-26 | $1.4\times10^8$ | Continuous GPS network | Original | (*48*) |
| St. Helens (1980 event) | USA | Stratovolcano | 1980-03-27 | $\sim10^5$ | PNSN | Original | (*56, 97*) |
| St. Helens (1981 event) | USA | Stratovolcano | 1981-09-06 | $\sim10^6$ | PNSN | Original | (*56, 97*) |
| St. Helens (1982 event) | USA | Stratovolcano | 1982-03-19 | $\sim10^6$ | PNSN, tiltmeter | Original | (*56, 97*) |
| St. Helens (1985 event) | USA | Stratovolcano | 1985-05-25 | $\sim10^6$ | PNSN | Original | (*56, 98*) |
| Soufriere Hills | UK | Stratovolcano | 1995-11-15 | $7\times10^7$ | ISCSN | Original | (*56, 99*) |
| Tokachidake | Japan | Stratovolcano | 2004-02-25 | $\sim10^5$ | JMA seismic network | Original | (*56*) |
| Unzen | Japan | Complex volcano | 1990-11-17 | $2.1\times10^8$ | ISCSN | Original | (*56, 100*) |
| Yakedake | Japan | Stratovolcano | 1995-02-11 | $6\times10^3$ | JMA seismic network | Original | (*56*) |

Note: PNSN - Pacific Northwest Seismic Network; ISCSN - International Seismological Centre Seismographic Network; JMASN - Japan Meteorological Agency Seismic Network; NCDC - National Climatic Data Center; ANSS - Advanced National Seismic System; GPS - Global Positioning System; USGS - United States Geological Survey; AVO - Alaska Volcano Observatory.

**Table S5. Parameters of the LPPLS and PLS models fitted to landslide data (49 events and 94 time series monitoring data in total).**

| Data | LPPLS | | | | | | | PLS | | | |
|---|---|---|---|---|---|---|---|---|---|---|---|
| | $t_c$ | $m$ | $\omega$ | $\phi$ | $A$ | $B$ | $C$ | $t_c$ | $m$ | $A$ | $B$ |
| Abbotsford | 1.40 | -0.24 | 4.94 | 1.56 | -2.60e+00 | 5.90e+00 | 8.03e-02 | 1.19 | -0.20 | -3.12e+00 | 6.24e+00 |
| Achoma | -0.06 | 0.076 | 5.01 | -1.21 | 2.59e+01 | -1.79e+01 | 1.84e-01 | 2.49 | -0.13 | -1.16e+01 | 2.27e+01 |
| Agoyama | 0.00 | -0.35 | 5.27 | 0.29 | -1.87e+00 | 7.23e+00 | 7.87e-02 | 0.00 | -0.36 | -1.84e+00 | 7.20e+00 |
| Arvigo | 0.17 | -0.063 | 5.04 | -1.28 | -5.37e-01 | 7.29e-01 | 2.74e-03 | 0.17 | -0.07 | -4.40e-01 | 6.34e-01 |
| Baishi | 0.02 | 0.19 | 5.73 | 0.93 | 5.68e+01 | -1.67e+01 | 1.23e-01 | 0.02 | 0.22 | 5.36e+01 | -1.36e+01 |
| Baiyan | 26.78 | 0.33 | 4.94 | 0.49 | 1.96e-01 | -3.10e-02 | 2.86e-03 | 31.17 | -0.29 | -1.25e-01 | 6.70e-01 |
| Brienz/Brinzauls (reflector 715) | 3.11 | -0.53 | 4.94 | -0.03 | -1.40e+00 | 4.76e+01 | 5.79e-01 | 1.57 | -0.39 | -3.41e+00 | 3.64e+01 |
| Brienz/Brinzauls (reflector 719) | 3.40 | -0.63 | 4.94 | -0.20 | -1.11e+00 | 8.53e+01 | 1.23e+00 | 3.40 | -0.64 | -1.09e+00 | 8.62e+01 |
| Brienz/Brinzauls (reflector 725) | 3.22 | -0.61 | 4.94 | -0.10 | -8.45e-01 | 6.89e+01 | 9.67e-01 | 1.62 | -0.45 | -2.79e+00 | 4.91e+01 |
| Cadia | 16.16 | -0.8 | 4.94 | -0.43 | 2.78e-03 | 1.09e+00 | 6.15e-02 | 19.48 | -0.87 | 4.31e-03 | 1.44e+00 |
| Copper open pit | 10.28 | -1.2 | 9.88 | -0.16 | 1.82e-03 | 1.15e+00 | 1.88e-01 | 10.28 | -1.28 | 2.30e-03 | 1.55e+00 |
| Dosan (point 1) | 0.06 | -0.15 | 9.41 | 0.24 | -7.73e-02 | 8.90e-02 | 8.68e-04 | 0.10 | -0.27 | -4.28e-02 | 5.50e-02 |
| Dosan (point 2) | 0.16 | -0.52 | 4.94 | -0.43 | -5.63e-02 | 8.71e-02 | 2.06e-03 | 0.18 | -0.57 | -4.86e-02 | 8.07e-02 |
| Dosan (point 3) | 0.05 | -0.38 | 4.94 | 0.94 | -1.63e-01 | 2.55e-01 | 2.12e-03 | 0.05 | -0.38 | -1.59e-01 | 2.51e-01 |
| Gallivaggio | 0.85 | -1.2 | 11.63 | -1.46 | 3.54e-03 | 5.78e-03 | 4.16e-04 | 0.85 | -1.20 | 3.56e-03 | 5.89e-03 |
| Galterengraben (TJM1) | 32.07 | -0.49 | 5.99 | -1.51 | -4.97e+00 | 8.83e+01 | 3.47e+00 | 36.07 | -0.42 | -6.65e+00 | 7.95e+01 |
| Galterengraben (TJM2) | 30.83 | -0.53 | 4.94 | -0.19 | -4.31e+00 | 1.29e+02 | 5.47e+00 | 34.67 | -0.53 | -4.49e+00 | 1.35e+02 |
| Galterengraben (TJM6) | 12.96 | -0.51 | 4.94 | -1.56 | -1.22e-01 | 3.12e+01 | 1.38e+00 | 6.52 | -0.31 | -1.66e+00 | 1.85e+01 |
| Grabengufer (GNSS) | 0.24 | -0.35 | 12.42 | 1.16 | 7.71e+00 | 4.28e-01 | 2.75e-02 | -0.01 | 0.13 | 8.95e+00 | -8.36e-01 |
| Grabengufer (inclinometer W) | 0.23 | -1.2 | 4.94 | -1.39 | 2.60e+00 | 1.44e+00 | 3.06e-01 | 0.03 | -0.46 | 1.02e+00 | 2.77e+00 |
| Grabengufer (inclinometer N) | 0.03 | 0.1 | 6.70 | -0.84 | 7.55e+01 | -7.59e+01 | 5.56e-01 | 0.03 | 0.13 | 5.83e+01 | -5.88e+01 |
| Hogarth (extensometer 1) | 3.11 | -2.7 | 4.94 | 1.05 | 3.09e-03 | 8.73e+00 | 2.23e+00 | 0.00 | -1.20 | 2.06e-03 | 1.07e-01 |
| Hogarth (extensometer 2) | 0.00 | -1.3 | 9.86 | -1.22 | 7.35e-03 | 1.48e-01 | 9.85e-03 | 0.00 | -1.37 | 7.50e-03 | 1.81e-01 |
| Hogarth (extensometer 3) | 6.76 | -1.8 | 14.45 | -1.43 | 5.01e-03 | 7.06e-01 | 3.98e-02 | 0.00 | -1.19 | 4.81e-03 | 8.05e-02 |
| Hogarth (extensometer 4) | 0.00 | -0.87 | 5.39 | -0.81 | 8.56e-03 | 4.85e-02 | 1.16e-03 | 0.00 | -0.89 | 8.61e-03 | 5.12e-02 |
| Hogarth (extensometer 6) | 7.11 | -1.6 | 6.98 | -0.01 | 1.14e-02 | 4.47e-01 | 9.90e-03 | 5.33 | -1.53 | 1.14e-02 | 2.83e-01 |

**Table S5 (continued). Parameters of the LPPLS and PLS models fitted to landslide data (49 events and 94 time series monitoring data in total).**

| Data | LPPLS | | | | | | | PLS | | | |
|---|---|---|---|---|---|---|---|---|---|---|---|
| | $t_c$ | $m$ | $\omega$ | $\phi$ | $A$ | $B$ | $C$ | $t_c$ | $m$ | $A$ | $B$ |
| Iron mine | 0.43 | -0.13 | 11.46 | 0.21 | -9.44e-01 | 1.64e+00 | 5.84e-03 | 0.43 | -0.15 | -8.46e-01 | 1.55e+00 |
| Jinlonggou | 4.78 | -0.87 | 4.94 | -1.36 | -5.56e-03 | 2.19e-01 | 2.52e-02 | -1.00 | -0.22 | -2.89e-02 | 7.09e-02 |
| Kagemori (point 1) | 3.11 | 0.14 | 4.94 | 1.52 | 5.04e-01 | -1.59e-01 | 7.03e-04 | 3.50 | 0.09 | 6.36e-01 | -2.83e-01 |
| Kagemori (point 3) | 12.44 | 0.22 | 4.94 | 1.21 | 6.91e-01 | -2.00e-01 | 3.32e-03 | 14.00 | 0.12 | 1.12e+00 | -5.58e-01 |
| Kagemori (point 13) | 2.84 | 0.57 | 4.94 | 0.45 | 2.82e-01 | -1.29e-02 | 2.17e-04 | 3.20 | 0.71 | 2.72e-01 | -7.01e-03 |
| Kagemori (point 15) | 4.53 | -0.45 | 4.94 | 0.63 | 1.48e-01 | 8.12e-01 | 2.42e-02 | 0.00 | -0.10 | -3.87e-01 | 9.67e-01 |
| Kagemori (point 17) | 3.73 | 0.012 | 6.19 | 1.31 | 5.86e+00 | -5.44e+00 | 5.51e-04 | 4.20 | -0.01 | -8.55e+00 | 8.98e+00 |
| Kagemori (point 18) | 0.00 | 0.21 | 4.94 | -0.95 | 6.95e-01 | -2.05e-01 | 1.70e-03 | 0.00 | 0.24 | 6.49e-01 | -1.63e-01 |
| Kagemori (point 21) | 4.44 | -0.95 | 7.20 | 0.72 | 1.03e-02 | 1.07e+00 | 7.21e-02 | 5.00 | -0.93 | 7.49e-03 | 1.11e+00 |
| Kagemori (point 23) | 2.89 | -0.4 | 6.67 | -0.21 | -7.85e-02 | 8.65e-01 | 1.77e-02 | 4.33 | -0.48 | -5.52e-02 | 1.04e+00 |
| La Saxe | 5.64 | -3.6 | 7.72 | -0.46 | 3.47e+00 | 6.57e+03 | 3.09e+03 | 6.34 | -2.43 | 3.39e+00 | 5.83e+02 |
| Letlhakane diamond mine | 9.53 | -0.37 | 4.94 | 1.20 | -4.44e-01 | 2.29e+00 | 1.32e-01 | 10.61 | -0.50 | -2.81e-01 | 2.61e+00 |
| Longjing | 2.22 | -0.45 | 4.94 | -1.24 | -1.66e-01 | 1.08e+00 | 2.84e-02 | 0.95 | -0.04 | -4.19e+00 | 4.77e+00 |
| Maoxian (point 1) | 2.86 | 0.17 | 5.31 | 0.53 | 1.12e-01 | -3.42e-02 | 8.37e-04 | 2.86 | 0.18 | 1.09e-01 | -3.14e-02 |
| Maoxian (point 2) | 2.86 | 0.15 | 5.32 | 0.43 | 1.69e-01 | -5.89e-02 | 6.38e-04 | 2.86 | 0.15 | 1.65e-01 | -5.47e-02 |
| Maoxian (point 3) | 2.86 | 0.14 | 5.29 | 0.69 | 1.75e-01 | -6.48e-02 | 8.57e-04 | 2.86 | 0.14 | 1.73e-01 | -6.19e-02 |
| Mt. Beni | -0.86 | -0.33 | 5.21 | 0.15 | -1.02e+00 | 6.91e+00 | 1.23e-01 | 1.73 | -0.45 | -6.44e-01 | 8.76e+00 |
| Mud Greek | 42.13 | -0.46 | 4.94 | 0.32 | -2.88e-01 | 5.01e+00 | 5.57e-01 | 47.38 | -0.52 | -1.98e-01 | 5.74e+00 |
| Nevis Bluff (point 1) | -3.11 | -0.11 | 4.94 | 1.11 | -5.28e-01 | 9.42e-01 | 5.57e-03 | -2.00 | -0.21 | -2.20e-01 | 7.06e-01 |
| Nevis Bluff (point 2) | -4.80 | -0.051 | 4.94 | -0.95 | -2.07e+00 | 2.71e+00 | 5.33e-03 | -3.90 | -0.13 | -6.87e-01 | 1.41e+00 |
| Nevis Bluff (point A) | -3.38 | -0.077 | 4.94 | 1.42 | -1.14e+00 | 1.75e+00 | 6.00e-03 | -2.30 | -0.16 | -4.58e-01 | 1.16e+00 |
| New Tredegar | 1.63 | -0.58 | 5.93 | 0.96 | 6.11e-02 | 1.35e+00 | 3.52e-02 | 1.21 | -0.52 | 4.21e-02 | 1.21e+00 |
| Northern Bohemia | 3.94 | -0.68 | 4.94 | 0.12 | 7.62e-03 | 6.59e-02 | 9.97e-03 | 4.35 | -0.64 | 7.64e-03 | 5.98e-02 |
| Open pit mine (event 3) | 0.03 | -1.4 | 10.94 | -0.14 | 1.50e-02 | 2.64e-03 | 1.08e-04 | 0.04 | -1.49 | 1.51e-02 | 2.49e-03 |

**Table S5 (continued). Parameters of the LPPLS and PLS models fitted to landslide data (49 events and 94 time series monitoring data in total).**

| Data | LPPLS | | | | | | | PLS | | | |
|---|---|---|---|---|---|---|---|---|---|---|---|
| | $t_c$ | $m$ | $\omega$ | $\phi$ | $A$ | $B$ | $C$ | $t_c$ | $m$ | $A$ | $B$ |
| Open pit mine (event 4) | 0.13 | 0.041 | 4.94 | 0.81 | 1.92e+00 | -1.83e+00 | 1.27e-02 | 0.00 | 0.38 | 2.85e-01 | -2.14e-01 |
| Open pit mine (event 5) | 0.25 | -0.1 | 5.97 | -0.94 | -1.33e-01 | 1.67e-01 | 2.54e-03 | 0.25 | 0.13 | 1.63e-01 | -1.29e-01 |
| Otomura | 0.53 | -0.81 | 4.94 | -1.07 | 4.17e-02 | 3.12e-01 | 1.67e-02 | 0.00 | -0.65 | 3.48e-02 | 2.37e-01 |
| Preonzo (extensometer 1) | 5.95 | -1.1 | 6.47 | 1.15 | 5.44e-01 | 1.84e+00 | 1.11e-01 | 6.59 | -1.14 | 5.42e-01 | 2.01e+00 |
| Preonzo (extensometer 2) | 3.62 | -0.65 | 4.94 | -0.15 | 5.54e-01 | 6.72e-01 | 3.24e-02 | 4.99 | -0.85 | 5.74e-01 | 1.04e+00 |
| Preonzo (extensometer 3) | 5.51 | -1.4 | 5.16 | -1.44 | 6.19e-01 | 7.31e+00 | 4.50e-01 | 6.09 | -1.35 | 6.15e-01 | 7.86e+00 |
| Preonzo (extensometer 4) | 5.42 | -1.5 | 6.63 | 1.27 | 7.32e-01 | 1.21e+01 | 7.97e-01 | 4.85 | -1.30 | 7.20e-01 | 7.89e+00 |
| Preonzo (extensometer 5) | 4.44 | -1.3 | 5.97 | 0.20 | 5.39e-01 | 6.21e+00 | 5.58e-01 | 3.02 | -1.01 | 5.19e-01 | 2.66e+00 |
| Preonzo (reflector 2) | 2.49 | -0.93 | 4.94 | 0.77 | 3.45e-01 | 4.80e+00 | 4.18e-01 | 0.89 | -0.50 | 1.21e-01 | 2.32e+00 |
| Preonzo (reflector 4) | 2.40 | -0.87 | 4.94 | 0.89 | 3.69e-01 | 4.43e+00 | 3.30e-01 | 0.89 | -0.47 | 1.18e-01 | 2.36e+00 |
| Preonzo (reflector 5) | 2.35 | -0.79 | 4.94 | 0.93 | 2.56e-01 | 2.83e+00 | 1.97e-01 | 0.89 | -0.43 | 5.77e-02 | 1.65e+00 |
| Preonzo (reflector 8) | 2.35 | -0.76 | 4.94 | 0.94 | 1.67e-01 | 1.74e+00 | 1.14e-01 | 0.89 | -0.41 | 2.97e-02 | 1.06e+00 |
| Preonzo (reflector 9) | 3.82 | -1.1 | 5.61 | 1.47 | 2.24e-01 | 2.81e+00 | 2.12e-01 | 2.35 | -0.79 | 1.98e-01 | 1.33e+00 |
| Puigcercós (area 4) | 191.03 | 0.63 | 4.94 | -1.48 | 2.42e-01 | -2.01e-03 | 8.77e-05 | 2.87 | 0.69 | 1.96e-01 | -1.06e-03 |
| Puigcercós (area 6) | 151.36 | -0.67 | 4.94 | -0.52 | -9.30e-02 | 1.35e+01 | 7.96e-01 | 73.42 | -0.36 | -1.92e-01 | 2.76e+00 |
| Puigcercós (area 7) | 189.80 | -0.5 | 4.94 | 1.18 | -1.16e-01 | 5.03e+00 | 4.20e-01 | 1.35 | 0.14 | 5.20e-01 | -1.85e-01 |
| Puigcercós (area 9) | 130.55 | -1.7 | 9.16 | -1.44 | -7.31e-03 | 9.23e+02 | 2.53e+02 | 33.62 | -0.83 | -1.81e-02 | 5.80e+00 |
| Road slope (event 1) | 0.13 | -0.41 | 4.94 | 0.03 | -6.02e-02 | 9.56e-02 | 2.58e-03 | 0.07 | -0.25 | -1.12e-01 | 1.45e-01 |
| Road slope (event 2) | 0.86 | 0.031 | 6.89 | 1.45 | 1.18e+01 | -1.10e+01 | 2.80e-02 | 0.96 | -0.05 | -7.71e+00 | 8.54e+00 |
| Road slope (event 3) | -0.05 | 0.0043 | 4.94 | 1.29 | 2.10e+00 | -2.12e+00 | 6.94e-04 | -0.05 | -0.12 | -6.23e-02 | 5.25e-02 |
| Road slope (event 4) | 0.01 | -0.51 | 4.94 | 1.34 | -1.75e-02 | 4.66e-03 | 4.90e-04 | 0.01 | -0.51 | -1.68e-02 | 4.72e-03 |
| Road slope (event 5) | -0.03 | -0.4 | 4.94 | -0.92 | -1.46e-02 | 9.38e-03 | 5.20e-04 | -0.03 | -0.54 | -8.09e-03 | 4.92e-03 |

**Table S5 (continued). Parameters of the LPPLS and PLS models fitted to landslide data (49 events and 94 time series monitoring data in total).**

| Data | LPPLS | | | | | | | PLS | | | |
|---|---|---|---|---|---|---|---|---|---|---|---|
| | $t_c$ | $m$ | $\omega$ | $\phi$ | $A$ | $B$ | $C$ | $t_c$ | $m$ | $A$ | $B$ |
| Road slope (event 6) | 0.14 | -0.46 | 4.94 | -0.40 | -2.37e-02 | 2.58e-02 | 1.28e-03 | 0.16 | -0.58 | -1.70e-02 | 1.94e-02 |
| Road slope (event 7) | -0.03 | -0.64 | 4.94 | -0.11 | 1.12e-02 | 2.18e-02 | 2.06e-03 | -0.10 | -0.13 | -1.13e-01 | 1.42e-01 |
| Road slope (event 8) | 0.02 | 0.19 | 4.94 | 0.42 | 2.27e-01 | -2.83e-01 | 2.47e-03 | 0.02 | 0.09 | 4.13e-01 | -4.56e-01 |
| Road slope (event 9) | 0.82 | -0.92 | 6.87 | 0.53 | -1.12e-01 | 9.80e-01 | 5.31e-02 | 0.93 | -0.96 | -1.11e-01 | 1.06e+00 |
| Road slope (event 10) | 0.07 | -0.75 | 4.94 | 0.64 | -3.52e-02 | 2.58e-02 | 1.79e-03 | 0.07 | -0.82 | -2.87e-02 | 2.15e-02 |
| Roesgrenda | -0.15 | 0.2 | 6.28 | -1.23 | 1.05e-01 | -3.84e-02 | 6.89e-04 | -0.15 | 0.25 | 9.35e-02 | -2.76e-02 |
| Takabayama | -0.06 | -0.28 | 4.94 | -0.01 | -9.23e-01 | 2.19e+00 | 2.84e-02 | -0.06 | -0.27 | -9.75e-01 | 2.23e+00 |
| Vajont (bench mark 2) | 6.13 | -0.98 | 5.59 | 0.33 | 2.98e+00 | 1.24e+01 | 1.30e+00 | 6.90 | -0.89 | 2.90e+00 | 1.13e+01 |
| Vajont (bench mark 4) | 5.42 | 0.21 | 6.89 | 0.48 | 3.99e+00 | -7.26e-01 | 5.85e-03 | 6.10 | 0.16 | 4.42e+00 | -1.09e+00 |
| Vajont (bench mark 6) | 6.40 | 0.02 | 4.94 | -0.59 | 2.76e+01 | -2.28e+01 | 4.17e-02 | 7.20 | -0.15 | -1.05e-01 | 5.36e+00 |
| Vajont (bench mark 58) | 8.71 | -0.025 | 4.94 | 0.52 | -1.27e+01 | 1.68e+01 | 2.39e-02 | 9.80 | -0.13 | -3.72e-01 | 4.88e+00 |
| Veslemannen (radar point 1) | 0.13 | 0.65 | 6.52 | -0.75 | 1.90e+01 | -3.39e-01 | 3.06e-02 | 10.93 | 0.29 | 2.55e+01 | -3.28e+00 |
| Veslemannen (radar point 2) | 0.13 | 0.42 | 7.49 | -0.82 | 1.36e+01 | -1.03e+00 | 4.09e-02 | 9.23 | 0.07 | 4.39e+01 | -2.68e+01 |
| Veslemannen (radar point 3) | 0.13 | 0.37 | 6.46 | -0.42 | 1.19e+01 | -1.00e+00 | 4.16e-02 | 10.93 | -0.04 | -4.08e+01 | 5.68e+01 |
| Veslemannen (radar point 4) | 0.13 | 0.39 | 6.42 | -0.27 | 6.61e+00 | -5.03e-01 | 2.36e-02 | 9.64 | 0.03 | 4.06e+01 | -3.22e+01 |
| Veslemannen (radar point 5) | 0.13 | 0.49 | 6.44 | -0.31 | 5.70e+00 | -2.13e-01 | 1.25e-02 | 10.53 | 0.15 | 9.63e+00 | -2.87e+00 |
| Veslemannen (radar point 6) | 0.13 | 0.21 | 6.51 | -0.48 | 4.54e+00 | -1.15e+00 | 2.45e-02 | 1.26 | 0.11 | 7.00e+00 | -3.35e+00 |
| Veslemannen (radar point 7) | 0.13 | 0.47 | 6.57 | -0.81 | 4.23e+00 | -2.03e-01 | 1.12e-02 | 10.23 | 0.14 | 7.85e+00 | -2.74e+00 |
| Welland (point 1) | 0.28 | -0.5 | 4.94 | -1.41 | -1.22e-03 | 1.37e-02 | 8.72e-04 | 0.31 | -0.54 | -3.01e-04 | 1.30e-02 |
| Welland (point 2) | 0.23 | -0.47 | 4.96 | -0.82 | -1.02e-03 | 1.21e-02 | 9.11e-04 | 0.12 | -0.31 | -4.56e-03 | 1.50e-02 |
| Xinmo | 2.62 | 0.049 | 4.94 | -1.35 | 2.68e-01 | -1.85e-01 | 1.46e-03 | 2.62 | 0.06 | 2.25e-01 | -1.42e-01 |
| Xintan | 12.89 | -0.96 | 4.94 | 0.24 | 8.78e+00 | 7.04e+01 | 8.00e+00 | 15.36 | -0.97 | 8.71e+00 | 8.45e+01 |

Note: $t_c$ is in day; $m$, $\omega$, and $\phi$ are dimensionless; $A$ is in deg for tilt (Grabengufer inclinometer) and in meter for displacement (all other cases); $B$ and $C$ are in the unit of $A$ per day$^m$. The actual failure corresponds to time $t = 0$ day. Parameters $A$, $B$, and $C$ are displayed using scientific exponential notation, where 1.00e-02 represents $1.00 \times 10^{-2}$.

**Table S6. Parameters of the LPPLS and PLS models fitted to rockburst data (11 events and 11 time series monitoring data in total).**

| Data | LPPLS | | | | | | | PLS | | | |
|---|---|---|---|---|---|---|---|---|---|---|---|
| | $t_c$ | $m$ | $\omega$ | $\phi$ | $A$ | $B$ | $C$ | $t_c$ | $m$ | $A$ | $B$ |
| Coal mine (cut-through #4) | 0.12 | -0.58 | 4.94 | 0.87 | -7.14e+01 | 7.81e+01 | 3.86e+00 | 0.13 | -0.62 | -6.29e+01 | 7.15e+01 |
| Coal mine (cut-through #5) | 0.02 | -0.12 | 4.94 | -0.01 | -2.43e-01 | 2.85e-01 | 5.77e-03 | 0.05 | -0.28 | -7.64e-02 | 1.19e-01 |
| Gold mine (event 1) | -0.14 | 0.5 | 7.26 | 1.56 | 1.34e+03 | -6.42e+02 | 4.43e+01 | -0.14 | 0.43 | 1.40e+03 | -7.14e+02 |
| Gold mine (event 2) | 0.08 | -0.32 | 11.94 | 1.21 | -4.91e+02 | 6.06e+02 | 3.61e+01 | 0.07 | -0.09 | -2.68e+03 | 2.80e+03 |
| Gold mine (event 3) | 1.01 | 0.24 | 9.59 | -0.79 | 2.33e+04 | -1.26e+04 | 2.11e+02 | 0.28 | 0.40 | 1.39e+04 | -5.10e+03 |
| Gold mine (event 4) | -2.46 | 0.59 | 9.14 | 0.36 | 1.36e+04 | -1.34e+03 | 1.85e+01 | -1.98 | 0.58 | 1.40e+04 | -1.45e+03 |
| Gold mine (event 5) | 0.42 | 0.11 | 7.75 | -1.40 | 2.31e+04 | -1.70e+04 | 2.10e+02 | 0.97 | 0.01 | 2.12e+05 | -2.05e+05 |
| Gold mine (event 6) | 0.03 | 0.42 | 8.37 | 1.04 | 3.28e+03 | -1.51e+03 | 7.88e+01 | 0.03 | 0.44 | 3.22e+03 | -1.45e+03 |
| Gold mine (event 7) | 3.75 | 0.43 | 4.98 | -0.18 | 5.47e+03 | -9.12e+02 | 3.47e+01 | 7.26 | 0.13 | 1.52e+04 | -8.75e+03 |
| Gold mine (event 8) | -5.42 | 0.64 | 5.65 | 0.26 | 4.14e+03 | -2.12e+02 | 3.33e+01 | 4.74 | 0.09 | 2.06e+04 | -1.33e+04 |
| Platinum mine | 0.04 | 0.31 | 4.94 | 1.38 | 2.35e-01 | -3.56e-02 | 6.93e-04 | 0.04 | 0.36 | 2.31e-01 | -3.07e-02 |

Note: $t_c$ is in day; $m$, $\omega$, and $\phi$ are dimensionless; $A$ is in meter for displacement (Coal and Platinum mines) and in kN$^{1/2}$km$^{1/2}$ for Benioff strain (Gold mine); $B$ and C are in the unit of $A$ per day$^m$. The actual failure corresponds to time $t = 0$ day. Parameters $A$, $B$, and $C$ are displayed using scientific exponential notation, where 1.00e-02 represents 1.00×10$^{-2}$.

**Table S7. Parameters of the LPPLS and PLS models fitted to glacier data (17 events and 21 time series monitoring data in total).**

| Data | LPPLS | | | | | | | PLS | | | |
|---|---|---|---|---|---|---|---|---|---|---|---|
| | $t_c$ | $m$ | $\omega$ | $\phi$ | $A$ | $B$ | $C$ | $t_c$ | $m$ | $A$ | $B$ |
| Amery | 2.69 | 0.14 | 11.47 | 1.35 | 6.64e+01 | -1.39e+01 | 1.26e-01 | 2.69 | 0.14 | 6.47e+01 | -1.25e+01 |
| Eiger glacier (2001 event) | -0.46 | 0.027 | 14.15 | -0.06 | 5.10e+01 | -4.83e+01 | 3.96e-03 | -0.40 | 0.00 | 3.12e+02 | -3.09e+02 |
| Eiger glacier (2016 event) | 0.00 | 0.51 | 4.94 | -0.44 | 3.61e+00 | -4.41e-01 | 3.43e-03 | 0.00 | 0.53 | 3.56e+00 | -4.03e-01 |
| Grandes Jorasses (2014 event, prism 13) | 0.00 | 0.33 | 4.94 | 0.65 | 4.88e+01 | -2.88e+00 | 3.56e-02 | 0.00 | 0.34 | 4.85e+01 | -2.66e+00 |
| Grandes Jorasses (2014 event, prism 14) | 1.79 | 0.36 | 4.94 | 1.07 | 4.70e+01 | -2.57e+00 | 6.86e-03 | 1.79 | 0.36 | 4.69e+01 | -2.50e+00 |
| Grandes Jorasses (2020 event) | 0.26 | 0.36 | 5.80 | 0.62 | 2.34e+00 | -9.28e-01 | 4.06e-03 | 0.26 | 0.35 | 2.36e+00 | -9.52e-01 |
| Gruben | 0.03 | 0.5 | 8.11 | 0.16 | 2.14e+00 | -1.20e+00 | 9.30e-03 | 0.03 | 0.50 | 2.14e+00 | -1.20e+00 |
| Mönch | -14.13 | 0.45 | 4.94 | -1.10 | 7.27e+00 | -1.04e+00 | 1.39e-02 | -13.40 | 0.37 | 8.33e+00 | -1.63e+00 |
| Planpincieux (event 1) | 0.00 | 0.58 | 4.94 | 0.41 | 1.85e+01 | -1.38e+00 | 7.70e-03 | 0.00 | 0.60 | 1.83e+01 | -1.31e+00 |
| Planpincieux (event 2) | 5.07 | 0.36 | 4.94 | 0.22 | 3.56e+01 | -7.83e+00 | 8.69e-02 | 5.70 | 0.27 | 4.41e+01 | -1.40e+01 |
| Planpincieux (event 3) | 0.00 | 0.53 | 4.94 | 1.19 | 2.26e+01 | -2.42e+00 | 2.66e-02 | 1.73 | 0.43 | 2.67e+01 | -4.34e+00 |
| Planpincieux (event 4) | 1.78 | 0.69 | 4.94 | 1.21 | 1.71e+01 | -1.89e+00 | 6.52e-02 | 2.00 | 0.56 | 1.92e+01 | -3.07e+00 |
| Planpincieux (event 5) | 0.00 | 0.7 | 7.41 | 1.28 | 1.13e+01 | -1.00e+00 | 1.65e-02 | 0.00 | 0.72 | 1.11e+01 | -9.04e-01 |
| Planpincieux (event 6) | 4.09 | 0.62 | 4.94 | -0.50 | 1.36e+01 | -1.15e+00 | 1.71e-02 | 4.60 | 0.62 | 1.39e+01 | -1.16e+00 |
| UK211 | 40.47 | -0.36 | 4.94 | -0.68 | -1.12e+02 | 9.71e+02 | 4.90e+01 | 2.39 | 0.10 | 4.22e+02 | -2.35e+02 |
| Weisshorn (1973 event) | -9.84 | 0.21 | 4.94 | -1.07 | 8.06e+01 | -2.53e+01 | 2.40e-01 | -7.03 | 0.11 | 1.42e+02 | -7.66e+01 |
| Weisshorn (2005 event, #103) | -0.92 | 0.56 | 4.94 | -0.44 | 9.32e+00 | -1.48e+00 | 7.17e-03 | -1.18 | 0.57 | 9.11e+00 | -1.44e+00 |
| Weisshorn (2005 event, #104) | -0.92 | 0.53 | 4.94 | -0.59 | 9.98e+00 | -1.79e+00 | 6.59e-03 | -0.92 | 0.52 | 1.01e+01 | -1.86e+00 |
| Weisshorn (2005 event, #105) | -0.92 | 0.44 | 4.94 | -0.82 | 1.19e+01 | -2.86e+00 | 1.11e-02 | -0.66 | 0.41 | 1.27e+01 | -3.37e+00 |
| Weisshorn (2005 event, #106) | -2.99 | 0.36 | 9.24 | 1.19 | 1.28e+01 | -4.11e+00 | 6.46e-03 | -2.99 | 0.37 | 1.25e+01 | -3.87e+00 |
| Weissmies | 0.75 | 0.39 | 11.20 | 1.09 | 8.52e+01 | -8.41e+00 | 5.63e-02 | 0.75 | 0.40 | 8.48e+01 | -8.12e+00 |

Note: $t_c$ is in day; $m$, $\omega$, and $\phi$ are dimensionless; $A$ is in km$^2$ for area loss (UK211), in km for rift length (Amery), and in meter for displacement (all other cases); $B$ and $C$ are in the unit of $A$ per day$^m$. The actual failure corresponds to time $t = 0$ day. Parameters $A$, $B$, and $C$ are displayed using scientific exponential notation, where 1.00e-02 represents $1.00\times10^{-2}$.

**Table S8. Parameters of the LPPLS and PLS models fitted to volcano data (32 events and 34 time series monitoring data in total).**

| Data | LPPLS | | | | | | | PLS | | | |
|---|---|---|---|---|---|---|---|---|---|---|---|
| | $t_c$ | $m$ | $\omega$ | $\phi$ | $A$ | $B$ | $C$ | $t_c$ | $m$ | $A$ | $B$ |
| Adatara | 93.18 | 0.36 | 7.12 | 0.90 | 2.20e+02 | -1.22e+01 | 2.76e-01 | 105.70 | 0.29 | 2.49e+02 | -2.33e+01 |
| Asama | 0.00 | 0.8 | 10.41 | 0.70 | 1.31e+04 | -1.59e+02 | 6.40e+00 | 0.00 | 0.78 | 1.31e+04 | -1.73e+02 |
| Augustine | 15.68 | -0.31 | 4.96 | -0.65 | -4.19e+02 | 6.43e+03 | 1.69e+02 | 15.68 | -0.28 | -5.73e+02 | 6.12e+03 |
| Axial Seamount | 6.00 | 0.64 | 9.23 | 0.61 | 3.38e+00 | -7.50e-03 | 5.12e-04 | 323.10 | 0.32 | 4.38e+00 | -1.62e-01 |
| Bezymianny | 1.78 | -1 | 4.94 | 0.41 | -8.13e+00 | 2.93e+02 | 1.28e+01 | 2.11 | -1.02 | -8.36e+00 | 3.15e+02 |
| Etna (1989 event) | 31.62 | -0.15 | 10.84 | -0.91 | -4.28e+02 | 1.10e+03 | 6.93e+00 | 35.20 | -0.21 | -3.02e+02 | 1.09e+03 |
| Etna (2013 event) | 0.35 | 0.76 | 4.94 | -0.54 | 2.02e+03 | -3.74e+01 | 8.23e-01 | 14.35 | 0.59 | 2.49e+03 | -1.03e+02 |
| Hierro | 2.49 | 0.72 | 4.94 | -0.41 | 1.20e+04 | -4.56e+02 | 5.12e+01 | 11.20 | 0.38 | 2.03e+04 | -3.40e+03 |
| Kilauea (1971 event) | 3.02 | -0.12 | 4.94 | 0.33 | -2.56e+03 | 4.33e+03 | 7.93e+01 | 0.00 | 0.30 | 1.64e+03 | -4.80e+02 |
| Kilauea (1972 event) | 12.71 | 0.81 | 4.94 | -0.48 | 8.03e+03 | -5.86e+01 | 6.29e+00 | 43.20 | 0.55 | 1.06e+04 | -3.46e+02 |
| Kilauea (1983 event) | 2.00 | 0.99 | 9.05 | -0.29 | 6.58e+03 | -1.56e+01 | 1.34e+00 | 34.20 | 0.82 | 7.46e+03 | -4.50e+01 |
| Kujusan | 4.98 | 0.75 | 4.94 | 0.19 | 4.80e+01 | -5.10e-01 | 7.93e-02 | 6.10 | 0.99 | 4.70e+01 | -1.64e-01 |
| Mauna Loa | 5.73 | -0.15 | 5.05 | 0.02 | 1.41e+01 | 7.75e+00 | 6.31e-02 | 4.30 | -0.05 | 4.60e+00 | 1.65e+01 |
| Merapi (2006 event) | 5.51 | 0.39 | 4.94 | -0.57 | 2.98e+04 | -5.57e+03 | 8.98e+01 | 6.20 | 0.28 | 3.76e+04 | -1.09e+04 |
| Merapi (2010 event) | 3.02 | -0.35 | 5.44 | -1.00 | -3.90e+02 | 2.08e+03 | 2.74e+01 | 3.40 | -0.41 | -2.91e+02 | 2.14e+03 |
| Pinatubo | 0.36 | -0.016 | 6.94 | 0.39 | -2.90e+04 | 3.06e+04 | 6.06e+01 | 3.20 | -0.50 | -7.04e+02 | 4.50e+03 |
| Redoubt (1989 event) | 8.91 | 0.52 | 4.94 | -0.16 | 7.26e+01 | -3.40e+00 | 2.93e-01 | 3.00 | 0.79 | 6.10e+01 | -6.93e-01 |
| Redoubt (2009 event) | 3.43 | 0.7 | 8.43 | 0.10 | 1.39e+03 | -6.89e+01 | 3.93e+00 | -1.00 | 0.72 | 1.24e+03 | -5.90e+01 |
| Ruapehu (1995 event) | 13.60 | 0.67 | 6.40 | 0.32 | 3.87e+02 | -7.26e+00 | 6.23e-01 | 0.00 | 0.99 | 3.36e+02 | -1.26e+00 |
| Ruapehu (1996 event) | 9.07 | 0.83 | 4.94 | 0.75 | 1.84e+03 | -1.79e+01 | 2.11e+00 | 10.20 | 0.99 | 1.83e+03 | -9.04e+00 |
| Ruapehu (2006 event) | 25.07 | 0.45 | 11.66 | 1.50 | 9.22e+01 | -6.70e+00 | 1.01e-01 | 28.20 | 0.40 | 1.02e+02 | -1.02e+01 |
| Sakurajima | 8.18 | -0.055 | 4.94 | -0.88 | -7.37e+03 | 9.66e+03 | 5.01e+01 | 9.20 | -0.13 | -2.67e+03 | 5.22e+03 |
| Sierra Negra (2005 event) | 85.89 | -0.31 | 7.21 | -0.92 | -2.06e+00 | 1.63e+01 | 2.43e-01 | 96.00 | -0.37 | -1.69e+00 | 1.99e+01 |
| Sierra Negra (2018 event) | 154.60 | -0.1 | 7.36 | 0.55 | -8.32e+00 | 2.02e+01 | 4.79e-02 | 172.80 | -0.15 | -4.83e+00 | 1.92e+01 |
| St. Helens (1980 event) | 3.41 | -3.4 | 7.42 | -0.63 | 1.06e+01 | 1.64e+04 | 5.55e+03 | 0.13 | -2.04 | 1.05e+01 | 2.51e+02 |

**Table S8 (continued). Parameters of the LPPLS and PLS models fitted to volcano data (32 events and 34 time series monitoring data in total).**

| Data | LPPLS | | | | | | | PLS | | | |
|---|---|---|---|---|---|---|---|---|---|---|---|
| | $t_c$ | $m$ | $\omega$ | $\phi$ | $A$ | $B$ | $C$ | $t_c$ | $m$ | $A$ | $B$ |
| St. Helens (1981 event) | 0.20 | 0.73 | 4.94 | 0.14 | 9.98e+02 | -1.42e+01 | 9.96e-01 | 0.20 | 0.86 | 9.76e+02 | -7.42e+00 |
| St. Helens (1982 event, seismic data) | 4.62 | -0.099 | 4.94 | 0.83 | -1.97e+03 | 3.25e+03 | 4.32e+01 | 5.20 | -0.31 | -3.28e+02 | 1.92e+03 |
| St. Helens (1982 event, radial tilt) | 0.00 | 0.32 | 6.39 | 0.09 | 2.49e+03 | -5.44e+02 | 9.76e+00 | 0.64 | 0.26 | 2.85e+03 | -8.00e+02 |
| St. Helens (1982 event, tangential tilt) | 5.24 | -0.4 | 6.14 | 0.25 | -4.10e+03 | 6.24e+03 | 7.49e+01 | 5.90 | -0.44 | -4.00e+03 | 6.76e+03 |
| St. Helens (1985 event) | 3.29 | -1.5 | 5.62 | 1.06 | 3.65e+01 | 2.88e+03 | 3.84e+02 | 3.70 | -1.63 | 3.60e+01 | 3.86e+03 |
| Soufriere Hills | -37.48 | 0.0039 | 5.50 | 0.60 | 2.92e+03 | -2.79e+03 | 2.20e+00 | -34.53 | -0.04 | -2.04e+02 | 3.38e+02 |
| Tokachidake | 4.98 | 0.99 | 9.38 | 0.95 | 2.73e+03 | -1.26e+01 | 1.37e+00 | 22.40 | 0.80 | 3.18e+03 | -3.78e+01 |
| Unzen | 0.00 | 0.82 | 6.11 | -0.52 | 9.72e+01 | -1.16e+00 | 1.49e-01 | 23.80 | 0.37 | 1.71e+02 | -2.12e+01 |
| Yakedake | 6.20 | 0.99 | 4.94 | -1.18 | 9.68e+01 | -2.89e-01 | 2.44e-02 | 24.40 | 0.91 | 1.04e+02 | -4.76e-01 |

Note: $t_c$ is in day; $m$, $\omega$, and $\phi$ are dimensionless; $A$ is dimensionless for earthquake count, in meter for displacement, and in $\mu$rad for tilt; $B$ and C are in the unit of $A$ per day$^m$. The actual failure corresponds to time $t = 0$ day. Parameters $A$, $B$, and $C$ are displayed using scientific exponential notation, where 1.00e-02 represents $1.00 \times 10^{-2}$.

**Table S9. Comparison between the LPPLS and PLS models for landslide data (49 events and 94 time series monitoring data in total).**

| Data | LPPLS | | | PLS | | | $p$-value | | |
|---|---|---|---|---|---|---|---|---|---|
| | NRMSE | NAIC | NBIC | NRMSE | NAIC | NBIC | Wilks | KS | AD |
| Abbotsford | 4.98e-05 | -6.01 | -5.83 | 5.25e-04 | -3.72 | -3.61 | 0.00 | 0.00 | 0.00 |
| Achoma | 6.53e-03 | 0.19 | 0.51 | 9.20e-03 | 0.36 | 0.54 | 0.14 | 0.96 | 0.72 |
| Agoyama | 1.13e-04 | -4.25 | -3.97 | 1.97e-04 | -3.84 | -3.67 | 0.00 | 0.77 | 0.48 |
| Arvigo | 2.28e-05 | -9.34 | -9.10 | 3.53e-05 | -9.00 | -8.86 | 0.00 | 0.05 | 0.10 |
| Baishi | 2.36e-03 | -0.01 | 0.31 | 3.51e-03 | 0.20 | 0.38 | 0.05 | 0.38 | 0.46 |
| Baiyan | 3.59e-04 | -6.93 | -6.60 | 1.16e-03 | -5.96 | -5.78 | 0.00 | 0.34 | 0.17 |
| Brienz/Brinzauls (reflector 715) | 3.56e-04 | -2.01 | -1.86 | 4.51e-04 | -1.81 | -1.73 | 0.00 | 0.04 | 0.17 |
| Brienz/Brinzauls (reflector 719) | 5.77e-04 | -1.09 | -0.95 | 8.25e-04 | -0.77 | -0.69 | 0.00 | 0.06 | 0.02 |
| Brienz/Brinzauls (reflector 725) | 4.80e-04 | -1.43 | -1.28 | 6.70e-04 | -1.14 | -1.06 | 0.00 | 0.12 | 0.17 |
| Cadia | 3.41e-05 | -9.61 | -9.27 | 4.75e-05 | -9.54 | -9.35 | 0.42 | 0.84 | 0.71 |
| Copper open pit | 6.52e-05 | -8.94 | -8.59 | 7.88e-05 | -9.00 | -8.80 | 0.52 | 0.99 | 0.73 |
| Dosan (point 1) | 9.47e-06 | -11.45 | -11.19 | 1.60e-05 | -11.04 | -10.89 | 0.00 | 0.18 | 0.14 |
| Dosan (point 2) | 1.57e-05 | -9.91 | -9.62 | 3.27e-05 | -9.32 | -9.16 | 0.00 | 0.03 | 0.09 |
| Dosan (point 3) | 9.96e-06 | -9.15 | -8.88 | 1.67e-05 | -8.76 | -8.61 | 0.00 | 0.21 | 0.27 |
| Gallivaggio | 1.12e-06 | -15.61 | -15.33 | 2.48e-06 | -14.95 | -14.79 | 0.00 | 0.61 | 0.60 |
| Galterengraben (TJM1) | 1.20e-03 | -1.46 | -1.38 | 4.50e-03 | -0.15 | -0.11 | 0.00 | 0.00 | 0.00 |
| Galterengraben (TJM2) | 2.26e-03 | -0.53 | -0.45 | 6.89e-03 | 0.57 | 0.61 | 0.00 | 0.01 | 0.00 |
| Galterengraben (TJM6) | 2.09e-03 | -1.34 | -1.26 | 3.32e-03 | -0.90 | -0.85 | 0.00 | 0.04 | 0.01 |
| Grabengufer (GNSS) | 8.41e-04 | -4.82 | -4.59 | 1.41e-03 | -4.40 | -4.26 | 0.00 | 0.69 | 0.68 |
| Grabengufer (inclinometer W) | 7.02e-03 | 0.49 | 0.75 | 8.07e-03 | 0.52 | 0.67 | 0.34 | 0.55 | 0.37 |
| Grabengufer (inclinometer N) | 2.16e-03 | 0.26 | 0.54 | 6.84e-03 | 1.27 | 1.43 | 0.00 | 0.27 | 0.08 |
| Hogarth (extensometer 1) | 3.50e-07 | -16.66 | -16.32 | 3.26e-06 | -14.65 | -14.46 | 0.00 | 0.28 | 0.17 |
| Hogarth (extensometer 2) | 1.29e-06 | -15.11 | -14.80 | 4.37e-06 | -14.04 | -13.87 | 0.00 | 0.12 | 0.21 |
| Hogarth (extensometer 3) | 2.24e-07 | -17.61 | -17.39 | 8.74e-07 | -16.34 | -16.21 | 0.00 | 0.15 | 0.08 |
| Hogarth (extensometer 4) | 2.66e-07 | -17.20 | -16.96 | 6.17e-07 | -16.45 | -16.31 | 0.00 | 0.00 | 0.00 |
| Hogarth (extensometer 6) | 1.31e-07 | -18.23 | -18.01 | 1.97e-07 | -17.91 | -17.78 | 0.00 | 0.18 | 0.18 |

**Table S9 (continued). Comparison between the LPPLS and PLS models for landslide data (49 events and 94 time series monitoring data in total).**

| Data | LPPLS | | | PLS | | | p-value | | |
|---|---|---|---|---|---|---|---|---|---|
| | NRMSE | NAIC | NBIC | NRMSE | NAIC | NBIC | Wilks | KS | AD |
| Iron mine | 1.73e-05 | -8.35 | -8.05 | 3.20e-05 | -7.88 | -7.71 | 0.00 | 0.98 | 0.67 |
| Jinlonggou | 8.70e-06 | -11.79 | -11.56 | 2.55e-05 | -10.81 | -10.68 | 0.00 | 0.09 | 0.01 |
| Kagemori (point 1) | 4.66e-06 | -11.74 | -11.43 | 1.08e-05 | -11.07 | -10.89 | 0.00 | 0.46 | 0.21 |
| Kagemori (point 3) | 5.33e-05 | -8.30 | -8.16 | 1.69e-04 | -7.19 | -7.11 | 0.00 | 0.00 | 0.00 |
| Kagemori (point 13) | 8.75e-07 | -13.54 | -13.21 | 5.30e-06 | -11.96 | -11.77 | 0.00 | 0.09 | 0.03 |
| Kagemori (point 15) | 2.88e-05 | -8.93 | -8.66 | 6.70e-05 | -8.21 | -8.05 | 0.00 | 0.09 | 0.02 |
| Kagemori (point 17) | 1.05e-06 | -12.48 | -12.19 | 1.58e-06 | -12.21 | -12.04 | 0.00 | 0.77 | 0.42 |
| Kagemori (point 18) | 2.84e-05 | -8.78 | -8.48 | 4.79e-05 | -8.41 | -8.24 | 0.00 | 0.04 | 0.06 |
| Kagemori (point 21) | 1.33e-05 | -10.08 | -9.81 | 7.73e-05 | -8.44 | -8.29 | 0.00 | 0.36 | 0.13 |
| Kagemori (point 23) | 7.97e-06 | -10.03 | -9.88 | 3.27e-05 | -8.67 | -8.58 | 0.00 | 0.00 | 0.00 |
| La Saxe | 1.95e-03 | -1.79 | -1.55 | 1.50e-02 | 0.16 | 0.29 | 0.00 | 0.12 | 0.07 |
| Letlhakane diamond mine | 2.51e-04 | -5.91 | -5.73 | 1.18e-03 | -4.42 | -4.32 | 0.00 | 0.00 | 0.00 |
| Longjing | 4.27e-05 | -7.78 | -7.57 | 1.24e-04 | -6.80 | -6.68 | 0.00 | 0.00 | 0.00 |
| Maoxian (point 1) | 8.45e-05 | -8.91 | -8.59 | 1.12e-04 | -8.82 | -8.64 | 0.35 | 0.94 | 0.76 |
| Maoxian (point 2) | 8.87e-05 | -8.57 | -8.24 | 9.80e-05 | -8.66 | -8.47 | 0.95 | 0.94 | 0.77 |
| Maoxian (point 3) | 8.80e-05 | -8.57 | -8.25 | 1.04e-04 | -8.60 | -8.41 | 0.74 | 0.94 | 0.75 |
| Mt. Beni | 1.25e-04 | -4.81 | -4.49 | 2.86e-04 | -4.16 | -3.98 | 0.00 | 0.23 | 0.24 |
| Mud Greek | 1.21e-03 | -4.33 | -4.03 | 3.25e-03 | -3.50 | -3.33 | 0.00 | 0.69 | 0.60 |
| Nevis Bluff (point 1) | 3.08e-05 | -9.31 | -9.13 | 6.23e-05 | -8.67 | -8.56 | 0.00 | 0.00 | 0.00 |
| Nevis Bluff (point 2) | 2.78e-05 | -8.85 | -8.64 | 5.22e-05 | -8.29 | -8.17 | 0.00 | 0.00 | 0.01 |
| Nevis Bluff (point A) | 2.48e-05 | -9.10 | -8.91 | 5.31e-05 | -8.40 | -8.29 | 0.00 | 0.00 | 0.00 |
| New Tredegar | 5.61e-05 | -7.08 | -6.78 | 9.85e-05 | -6.67 | -6.50 | 0.00 | 0.71 | 0.64 |
| Northern Bohemia | 1.78e-05 | -11.95 | -11.64 | 1.01e-04 | -10.39 | -10.21 | 0.00 | 0.18 | 0.05 |
| Open pit mine (event 3) | 7.27e-06 | -10.44 | -10.15 | 1.88e-05 | -9.62 | -9.46 | 0.00 | 0.92 | 0.75 |
| Open pit mine (event 4) | 1.60e-04 | -7.23 | -7.00 | 3.59e-04 | -6.51 | -6.38 | 0.00 | 0.06 | 0.02 |

**Table S9 (continued). Comparison between the LPPLS and PLS models for landslide data (49 events and 94 time series monitoring data in total).**

| Data | LPPLS | | | PLS | | | p-value | | |
|---|---|---|---|---|---|---|---|---|---|
| | NRMSE | NAIC | NBIC | NRMSE | NAIC | NBIC | Wilks | KS | AD |
| Open pit mine (event 5) | 1.71e-05 | -11.21 | -10.98 | 7.90e-05 | -9.77 | -9.64 | 0.00 | 0.01 | 0.00 |
| Otomura | 4.16e-05 | -8.45 | -8.18 | 4.76e-05 | -8.44 | -8.28 | 0.44 | 0.95 | 0.72 |
| Preonzo (extensometer 1) | 1.14e-05 | -9.81 | -9.56 | 3.90e-05 | -8.69 | -8.55 | 0.00 | 0.09 | 0.12 |
| Preonzo (extensometer 2) | 1.56e-05 | -9.38 | -9.09 | 7.09e-05 | -8.01 | -7.85 | 0.00 | 0.55 | 0.23 |
| Preonzo (extensometer 3) | 2.71e-05 | -7.88 | -7.62 | 8.45e-05 | -6.86 | -6.71 | 0.00 | 0.17 | 0.15 |
| Preonzo (extensometer 4) | 1.27e-05 | -8.21 | -7.94 | 1.10e-04 | -6.16 | -6.01 | 0.00 | 0.37 | 0.13 |
| Preonzo (extensometer 5) | 1.90e-05 | -8.01 | -7.78 | 1.29e-04 | -6.19 | -6.05 | 0.00 | 0.28 | 0.30 |
| Preonzo (reflector 2) | 1.09e-04 | -5.24 | -4.93 | 4.72e-04 | -3.94 | -3.76 | 0.00 | 0.03 | 0.06 |
| Preonzo (reflector 4) | 8.14e-05 | -5.53 | -5.21 | 4.54e-04 | -3.98 | -3.80 | 0.00 | 0.16 | 0.05 |
| Preonzo (reflector 5) | 6.35e-05 | -6.15 | -5.84 | 3.38e-04 | -4.66 | -4.48 | 0.00 | 0.08 | 0.05 |
| Preonzo (reflector 8) | 3.87e-05 | -7.12 | -6.80 | 2.09e-04 | -5.62 | -5.44 | 0.00 | 0.08 | 0.05 |
| Preonzo (reflector 9) | 2.22e-05 | -8.15 | -7.92 | 9.31e-05 | -6.81 | -6.68 | 0.00 | 0.30 | 0.29 |
| Puigcercós (area 4) | 1.53e-04 | -6.82 | -6.47 | 2.40e-04 | -6.68 | -6.48 | 0.35 | 0.46 | 0.54 |
| Puigcercós (area 6) | 6.97e-05 | -6.95 | -6.60 | 2.17e-04 | -6.14 | -5.94 | 0.03 | 0.43 | 0.16 |
| Puigcercós (area 7) | 1.11e-04 | -6.84 | -6.49 | 3.51e-04 | -5.99 | -5.79 | 0.00 | 0.13 | 0.06 |
| Puigcercós (area 9) | 9.27e-05 | -7.15 | -6.80 | 2.31e-04 | -6.57 | -6.37 | 0.05 | 0.71 | 0.35 |
| Road slope (event 1) | 1.11e-06 | -12.62 | -12.35 | 2.54e-05 | -9.62 | -9.46 | 0.00 | 0.00 | 0.00 |
| Road slope (event 2) | 2.01e-04 | -5.91 | -5.80 | 7.46e-04 | -4.62 | -4.56 | 0.00 | 0.00 | 0.00 |
| Road slope (event 3) | 4.88e-06 | -13.13 | -12.86 | 1.41e-05 | -12.19 | -12.04 | 0.00 | 0.02 | 0.01 |
| Road slope (event 4) | 1.82e-05 | -11.60 | -11.33 | 1.77e-04 | -9.45 | -9.30 | 0.00 | 0.00 | 0.00 |
| Road slope (event 5) | 2.49e-05 | -11.51 | -11.30 | 5.67e-05 | -10.77 | -10.65 | 0.00 | 0.00 | 0.00 |

**Table S9 (continued). Comparison between the LPPLS and PLS models for landslide data (49 events and 94 time series monitoring data in total).**

| Data | LPPLS | | | PLS | | | p-value | | |
|---|---|---|---|---|---|---|---|---|---|
| | NRMSE | NAIC | NBIC | NRMSE | NAIC | NBIC | Wilks | KS | AD |
| Road slope (event 6) | 2.07e-05 | -11.14 | -10.93 | 4.87e-05 | -10.37 | -10.25 | 0.00 | 0.01 | 0.00 |
| Road slope (event 7) | 5.26e-05 | -9.05 | -8.75 | 7.54e-05 | -8.84 | -8.67 | 0.02 | 0.89 | 0.55 |
| Road slope (event 8) | 1.01e-05 | -11.11 | -10.86 | 2.17e-05 | -10.44 | -10.30 | 0.00 | 0.48 | 0.18 |
| Road slope (event 9) | 1.15e-04 | -6.29 | -6.19 | 2.50e-04 | -5.54 | -5.48 | 0.00 | 0.00 | 0.00 |
| Road slope (event 10) | 4.65e-05 | -8.93 | -8.75 | 1.16e-04 | -8.08 | -7.97 | 0.00 | 0.00 | 0.00 |
| Roesgrenda | 3.24e-05 | -10.34 | -10.13 | 4.42e-05 | -10.11 | -9.99 | 0.00 | 0.10 | 0.12 |
| Takabayama | 9.01e-06 | -8.11 | -7.84 | 5.63e-05 | -6.41 | -6.25 | 0.00 | 0.00 | 0.00 |
| Vajont (bench mark 2) | 6.85e-04 | -3.93 | -3.70 | 1.70e-03 | -3.11 | -2.98 | 0.00 | 0.00 | 0.01 |
| Vajont (bench mark 4) | 3.78e-05 | -7.61 | -7.37 | 1.42e-04 | -6.38 | -6.24 | 0.00 | 0.04 | 0.00 |
| Vajont (bench mark 6) | 4.61e-04 | -4.68 | -4.46 | 1.18e-03 | -3.82 | -3.69 | 0.00 | 0.00 | 0.00 |
| Vajont (bench mark 58) | 9.32e-05 | -6.42 | -6.23 | 2.95e-04 | -5.32 | -5.22 | 0.00 | 0.00 | 0.00 |
| Veslemannen (radar point 1) | 1.14e-03 | -2.02 | -1.85 | 1.29e-02 | 0.34 | 0.44 | 0.00 | 0.00 | 0.00 |
| Veslemannen (radar point 2) | 5.55e-04 | -2.84 | -2.64 | 3.73e-03 | -1.00 | -0.89 | 0.00 | 0.00 | 0.00 |
| Veslemannen (radar point 3) | 4.34e-04 | -3.28 | -3.11 | 3.57e-03 | -1.23 | -1.13 | 0.00 | 0.00 | 0.00 |
| Veslemannen (radar point 4) | 3.42e-04 | -4.08 | -3.91 | 2.53e-03 | -2.14 | -2.04 | 0.00 | 0.00 | 0.00 |
| Veslemannen (radar point 5) | 2.69e-04 | -4.69 | -4.51 | 2.10e-03 | -2.68 | -2.58 | 0.00 | 0.00 | 0.00 |
| Veslemannen (radar point 6) | 1.93e-04 | -5.00 | -4.82 | 9.47e-04 | -3.47 | -3.37 | 0.00 | 0.00 | 0.00 |
| Veslemannen (radar point 7) | 2.51e-04 | -4.93 | -4.75 | 1.71e-03 | -3.07 | -2.96 | 0.00 | 0.00 | 0.00 |
| Welland (point 1) | 2.46e-06 | -14.03 | -13.80 | 1.91e-05 | -12.06 | -11.93 | 0.00 | 0.00 | 0.00 |
| Welland (point 2) | 6.32e-06 | -13.08 | -12.91 | 1.95e-05 | -12.00 | -11.91 | 0.00 | 0.04 | 0.00 |
| Xinmo | 9.00e-05 | -8.89 | -8.57 | 1.14e-04 | -8.83 | -8.65 | 0.40 | 0.16 | 0.28 |
| Xintan | 3.44e-03 | -1.11 | -0.78 | 6.74e-03 | -0.63 | -0.45 | 0.00 | 0.36 | 0.21 |

Note: NRMSE is displayed using scientific exponential notation, where 1.00e-02 represents $1.00\times10^{-2}$.

**Table S10. Comparison between the LPPLS and PLS models for rockburst data (11 events and 11 time series monitoring data in total).**

| Data | LPPLS | | | PLS | | | p-value | | |
|---|---|---|---|---|---|---|---|---|---|
| | NRMSE | NAIC | NBIC | NRMSE | NAIC | NBIC | Wilks | KS | AD |
| Coal mine (cut-through #4) | 1.09e-01 | 5.86 | 6.10 | 1.89e-01 | 6.32 | 6.45 | 0.00 | 0.67 | 0.42 |
| Coal mine (cut-through #5) | 4.34e-05 | -8.85 | -8.61 | 1.42e-04 | -7.76 | -7.62 | 0.00 | 0.03 | 0.00 |
| Gold mine (event 1) | 4.60e-01 | 9.82 | 10.16 | 1.42e+00 | 10.66 | 10.86 | 0.00 | 0.53 | 0.33 |
| Gold mine (event 2) | 1.44e+00 | 10.89 | 11.21 | 2.02e+00 | 10.80 | 10.99 | 0.86 | 0.86 | 0.70 |
| Gold mine (event 3) | 4.71e+00 | 13.78 | 13.95 | 7.84e+00 | 14.23 | 14.33 | 0.00 | 0.09 | 0.03 |
| Gold mine (event 4) | 2.46e+00 | 13.00 | 13.16 | 3.32e+00 | 13.25 | 13.34 | 0.00 | 0.39 | 0.31 |
| Gold mine (event 5) | 3.52e+00 | 13.10 | 13.32 | 7.25e+00 | 13.75 | 13.87 | 0.00 | 0.20 | 0.07 |
| Gold mine (event 6) | 2.11e+00 | 11.91 | 12.21 | 3.19e+00 | 12.17 | 12.34 | 0.01 | 0.71 | 0.46 |
| Gold mine (event 7) | 3.68e+00 | 12.67 | 12.95 | 5.29e+00 | 12.90 | 13.06 | 0.01 | 0.94 | 0.73 |
| Gold mine (event 8) | 1.08e+01 | 13.86 | 14.20 | 2.63e+01 | 14.49 | 14.69 | 0.01 | 0.20 | 0.10 |
| Platinum mine | 9.62e-06 | -11.37 | -11.26 | 1.58e-05 | -10.91 | -10.84 | 0.00 | 0.00 | 0.00 |

Note: NRMSE is displayed using scientific exponential notation, where 1.00e-02 represents $1.00\times10^{-2}$.

**Table S11. Comparison between the LPPLS and PLS models for glacier data (17 events and 21 time series monitoring data in total).**

| Data | LPPLS | | | PLS | | | $p$-value | | |
|---|---|---|---|---|---|---|---|---|---|
| | NRMSE | NAIC | NBIC | NRMSE | NAIC | NBIC | Wilks | KS | AD |
| Amery | 1.90e-02 | 2.44 | 2.52 | 2.05e-02 | 2.50 | 2.54 | 0.00 | 0.05 | 0.10 |
| Eiger glacier (2001 event) | 1.71e-05 | -7.23 | -7.00 | 2.19e-05 | -7.08 | -6.95 | 0.00 | 0.82 | 0.67 |
| Eiger glacier (2016 event) | 9.85e-05 | -5.09 | -4.80 | 1.29e-04 | -4.97 | -4.80 | 0.05 | 0.36 | 0.32 |
| Grandes Jorasses (2014 event, prism 13) | 2.56e-04 | -2.80 | -2.68 | 1.01e-03 | -1.46 | -1.39 | 0.00 | 0.00 | 0.00 |
| Grandes Jorasses (2014 event, prism 14) | 7.37e-05 | -4.11 | -3.98 | 1.10e-04 | -3.75 | -3.67 | 0.00 | 0.00 | 0.00 |
| Grandes Jorasses (2020 event) | 5.02e-06 | -9.00 | -8.88 | 1.62e-05 | -7.86 | -7.79 | 0.00 | 0.00 | 0.00 |
| Gruben | 2.00e-05 | -6.17 | -5.89 | 3.34e-05 | -6.16 | -5.99 | 0.71 | 0.43 | 0.46 |
| Mönch | 2.57e-04 | -3.73 | -3.50 | 6.46e-04 | -2.90 | -2.77 | 0.00 | 0.00 | 0.00 |
| Planpincieux (event 1) | 4.96e-05 | -4.17 | -3.94 | 1.17e-04 | -3.41 | -3.28 | 0.00 | 0.31 | 0.14 |
| Planpincieux (event 2) | 8.06e-04 | -1.14 | -0.89 | 2.28e-03 | -0.20 | -0.06 | 0.00 | 0.02 | 0.00 |
| Planpincieux (event 3) | 1.43e-04 | -2.90 | -2.64 | 4.75e-04 | -1.81 | -1.66 | 0.00 | 0.11 | 0.01 |
| Planpincieux (event 4) | 2.43e-04 | -2.36 | -2.01 | 4.23e-03 | 0.22 | 0.41 | 0.00 | 0.07 | 0.01 |
| Planpincieux (event 5) | 1.97e-04 | -2.92 | -2.60 | 8.03e-04 | -1.71 | -1.53 | 0.00 | 0.06 | 0.04 |
| Planpincieux (event 6) | 1.00e-04 | -3.76 | -3.49 | 6.15e-04 | -2.08 | -1.92 | 0.00 | 0.04 | 0.00 |
| UK211 | 1.95e-01 | 6.62 | 6.94 | 3.68e-01 | 7.05 | 7.24 | 0.01 | 0.76 | 0.40 |
| Weisshorn (1973 event) | 5.31e-03 | 1.38 | 1.71 | 1.04e-02 | 1.86 | 2.04 | 0.00 | 0.12 | 0.08 |
| Weisshorn (2005 event, prism 103) | 1.46e-05 | -6.17 | -5.92 | 5.92e-05 | -4.88 | -4.74 | 0.00 | 0.00 | 0.00 |
| Weisshorn (2005 event, prism 104) | 1.10e-05 | -6.44 | -6.19 | 3.54e-05 | -5.37 | -5.23 | 0.00 | 0.06 | 0.01 |
| Weisshorn (2005 event, prism 105) | 2.45e-05 | -5.58 | -5.34 | 6.35e-05 | -4.73 | -4.59 | 0.00 | 0.04 | 0.01 |
| Weisshorn (2005 event, prism 106) | 1.37e-05 | -6.36 | -6.11 | 3.29e-05 | -5.59 | -5.44 | 0.00 | 0.08 | 0.04 |
| Weissmies | 1.93e-03 | 0.40 | 0.62 | 2.60e-03 | 0.62 | 0.75 | 0.00 | 0.28 | 0.27 |

Note: NRMSE is displayed using scientific exponential notation, where 1.00e-02 represents $1.00\times10^{-2}$.

**Table S12. Comparison between the LPPLS and PLS models for volcano data (32 events and 34 time series monitoring data in total).**

| Data | LPPLS | | | PLS | | | *p*-value | | |
|---|---|---|---|---|---|---|---|---|---|
| | NRMSE | NAIC | NBIC | NRMSE | NAIC | NBIC | Wilks | KS | AD |
| Adatara | 1.37e-02 | 3.11 | 3.24 | 5.67e-02 | 4.49 | 4.57 | 0.00 | 0.00 | 0.00 |
| Asama | 1.28e+00 | 12.53 | 12.63 | 5.09e+00 | 13.88 | 13.94 | 0.00 | 0.00 | 0.00 |
| Augustine | 4.00e-01 | 9.49 | 9.62 | 5.95e-01 | 9.85 | 9.93 | 0.00 | 0.01 | 0.07 |
| Axial Seamount | 6.02e-04 | -4.31 | -4.24 | 1.55e-03 | -3.37 | -3.34 | 0.00 | 0.00 | 0.00 |
| Bezymianny | 5.88e-03 | 2.64 | 2.99 | 1.47e-02 | 3.25 | 3.45 | 0.03 | 0.28 | 0.25 |
| Etna (1989 event) | 2.49e-02 | 4.67 | 4.95 | 4.89e-02 | 5.22 | 5.38 | 0.00 | 0.81 | 0.44 |
| Etna (2013 event) | 8.84e-03 | 5.55 | 5.69 | 7.78e-02 | 7.68 | 7.76 | 0.00 | 0.00 | 0.00 |
| Hierro | 4.33e+00 | 13.69 | 13.86 | 3.50e+01 | 15.73 | 15.82 | 0.00 | 0.00 | 0.00 |
| Kilauea (1971 event) | 3.14e-01 | 8.84 | 9.07 | 1.11e+00 | 10.02 | 10.15 | 0.00 | 0.02 | 0.01 |
| Kilauea (1972 event) | 9.35e-01 | 11.77 | 12.02 | 1.44e+01 | 14.41 | 14.55 | 0.00 | 0.00 | 0.00 |
| Kilauea (1983 event) | 2.03e+00 | 12.19 | 12.47 | 8.29e+00 | 13.47 | 13.63 | 0.00 | 0.01 | 0.01 |
| Kujusan | 2.75e-02 | 2.27 | 2.45 | 8.43e-02 | 3.33 | 3.44 | 0.00 | 0.05 | 0.00 |
| Mauna Loa | 2.90e-04 | -4.42 | -4.27 | 5.01e-04 | -3.92 | -3.83 | 0.00 | 0.04 | 0.02 |
| Merapi (2006 event) | 2.75e+00 | 13.73 | 13.97 | 5.48e+00 | 14.32 | 14.46 | 0.00 | 0.03 | 0.03 |
| Merapi (2010 event) | 8.49e-02 | 7.29 | 7.60 | 1.44e-01 | 7.65 | 7.82 | 0.01 | 0.64 | 0.47 |
| Pinatubo | 2.08e-01 | 8.99 | 9.31 | 9.89e-01 | 10.37 | 10.55 | 0.00 | 0.40 | 0.11 |
| Redoubt (1989 event) | 3.82e-02 | 3.96 | 4.26 | 1.07e-01 | 4.84 | 5.01 | 0.00 | 0.22 | 0.07 |
| Redoubt (2009 event) | 6.27e-01 | 9.70 | 9.95 | 1.31e+00 | 10.34 | 10.48 | 0.00 | 0.03 | 0.01 |
| Ruapehu (1995 event) | 8.01e-02 | 5.72 | 5.86 | 3.76e-01 | 7.23 | 7.30 | 0.00 | 0.00 | 0.00 |
| Ruapehu (1996 event) | 4.78e-01 | 9.12 | 9.30 | 1.67e+00 | 10.31 | 10.42 | 0.00 | 0.00 | 0.00 |
| Ruapehu (2006 event) | 2.47e-02 | 3.27 | 3.46 | 3.35e-02 | 3.52 | 3.62 | 0.00 | 0.87 | 0.58 |
| Sakurajima | 5.03e-01 | 9.11 | 9.30 | 1.17e+00 | 9.88 | 9.99 | 0.00 | 0.30 | 0.15 |
| Sierra Negra (2005 event) | 7.07e-04 | -3.59 | -3.35 | 1.03e-03 | -3.31 | -3.17 | 0.00 | 0.20 | 0.23 |
| Sierra Negra (2018 event) | 3.04e-04 | -4.21 | -4.05 | 3.92e-04 | -4.01 | -3.92 | 0.00 | 0.04 | 0.13 |
| St. Helens (1980 event) | 1.75e-03 | 1.22 | 1.46 | 6.85e-03 | 2.49 | 2.63 | 0.00 | 0.48 | 0.32 |

**Table S12 (continued). Comparison between the LPPLS and PLS models for volcano data (32 events and 34 time series monitoring data in total).**

| Data | LPPLS | | | PLS | | | $p$-value | | |
|---|---|---|---|---|---|---|---|---|---|
| | NRMSE | NAIC | NBIC | NRMSE | NAIC | NBIC | Wilks | KS | AD |
| St. Helens (1981 event) | 8.66e-02 | 6.81 | 6.96 | 2.97e-01 | 8.00 | 8.08 | 0.00 | 0.01 | 0.00 |
| St. Helens (1982 event, seismic data) | 4.27e-01 | 8.40 | 8.66 | 1.24e+00 | 9.35 | 9.50 | 0.00 | 0.01 | 0.01 |
| St. Helens (1982 event, radial tilt) | 1.32e-01 | 8.29 | 8.54 | 3.77e-01 | 9.24 | 9.38 | 0.00 | 0.15 | 0.06 |
| St. Helens (1982 event, tangential tilt) | 6.55e-02 | 7.82 | 8.06 | 1.66e-01 | 8.65 | 8.79 | 0.00 | 0.16 | 0.10 |
| St. Helens (1985 event) | 2.77e-02 | 5.28 | 5.58 | 1.53e-01 | 6.83 | 7.00 | 0.00 | 0.69 | 0.53 |
| Soufriere Hills | 4.00e-02 | 3.45 | 3.78 | 9.03e-02 | 4.07 | 4.25 | 0.00 | 0.54 | 0.43 |
| Tokachidake | 2.49e+00 | 11.84 | 12.16 | 9.57e+00 | 13.01 | 13.19 | 0.00 | 0.14 | 0.04 |
| Unzen | 5.30e-02 | 4.69 | 5.01 | 3.41e-01 | 6.39 | 6.56 | 0.00 | 0.05 | 0.01 |
| Yakedake | 3.52e-02 | 3.69 | 3.90 | 1.01e-01 | 4.67 | 4.79 | 0.00 | 0.04 | 0.01 |

Note: NRMSE is displayed using scientific exponential notation, where 1.00e-02 represents $1.00 \times 10^{-2}$.

**Table S13. Parameters of the LPPLS and PLS calibration to landslide data (49 events and 94 time series monitoring data in total).**

|  | Calibration window size (day) | Number of data points | Aggregation interval (day) |
| --- | --- | --- | --- |
| Abbotsford | 19.04 | 97 | N/A |
| Achoma | 114.76 | 34 | N/A |
| Agoyama | 42.00 | 43 | 1 |
| Arvigo | 63.00 | 63 | 1 |
| Baishi | 31.00 | 32 | 1 |
| Baiyan | 394.75 | 30 | N/A |
| Brienz/Brinzauls (reflector 715) | 139.00 | 140 | 1 |
| Brienz/Brinzauls (reflector 719) | 152.00 | 153 | 1 |
| Brienz/Brinzauls (reflector 725) | 144.00 | 145 | 1 |
| Cadia | 298.86 | 23 | N/A |
| Copper open pit | 241.73 | 24 | N/A |
| Dosan (point 1) | 2.17 | 53 | 0.042 |
| Dosan (point 2) | 1.75 | 43 | 0.042 |
| Dosan (point 3) | 1.92 | 47 | 0.042 |
| Gallivaggio | 26.18 | 45 | 1 |
| Galterengraben (TJM1) | 360.00 | 361 | 1 |
| Galterengraben (TJM2) | 346.00 | 347 | 1 |
| Galterengraben (TJM6) | 290.00 | 291 | 1 |
| Grabengufer (GNSS) | 2.83 | 66 | 0.042 |
| Grabengufer (inclinometer W) | 2.21 | 53 | 0.042 |
| Grabengufer (inclinometer N) | 1.75 | 43 | 0.042 |
| Hogarth (extensometer 1) | 35.00 | 27 | 1 |
| Hogarth (extensometer 2) | 45.00 | 38 | 1 |
| Hogarth (extensometer 3) | 76.00 | 68 | 1 |
| Hogarth (extensometer 4) | 72.00 | 65 | 1 |
| Hogarth (extensometer 6) | 80.00 | 73 | 1 |
| Iron mine | 39.00 | 40 | 1 |
| Jinlonggou | 65.00 | 66 | 1 |
| Kagemori (point 1) | 35.00 | 36 | 1 |
| Kagemori (point 3) | 140.00 | 141 | 1 |
| Kagemori (point 13) | 32.00 | 28 | 1 |
| Kagemori (point 15) | 51.00 | 49 | 1 |
| Kagemori (point 17) | 42.00 | 43 | 1 |
| Kagemori (point 18) | 39.00 | 39 | 1 |
| Kagemori (point 21) | 50.00 | 50 | 1 |
| Kagemori (point 23) | 130.00 | 129 | 1 |
| La Saxe | 63.00 | 63 | 1 |
| Letlhakane diamond mine | 97.22 | 105 | N/A |

**Table S13 (continued). Parameters of the LPPLS and PLS calibration to landslide data (49 events and 94 time series monitoring data in total).**

|  | Calibration window size (day) | Number of data points | Aggregation interval (day) |
|---|---|---|---|
| Longjing | 14.29 | 75 | N/A |
| Maoxian (point 1) | 656.50 | 31 | N/A |
| Maoxian (point 2) | 656.50 | 31 | N/A |
| Maoxian (point 3) | 656.50 | 31 | N/A |
| Mt. Beni | 233.23 | 32 | N/A |
| Mud Greek | 472.80 | 38 | N/A |
| Nevis Bluff (point 1) | 100.00 | 101 | 1 |
| Nevis Bluff (point 2) | 81.00 | 82 | 1 |
| Nevis Bluff (point A) | 97.00 | 98 | 1 |
| New Tredegar | 37.94 | 39 | N/A |
| Northern Bohemia | 36.58 | 36 | N/A |
| Open pit mine (event 3) | 0.46 | 44 | 0.01 |
| Open pit mine (event 4) | 1.40 | 68 | 0.021 |
| Open pit mine (event 5) | 2.79 | 67 | 0.042 |
| Otomura | 48.00 | 49 | 1 |
| Preonzo (extensometer 1) | 57.00 | 57 | 1 |
| Preonzo (extensometer 2) | 41.00 | 41 | 1 |
| Preonzo (extensometer 3) | 52.00 | 52 | 1 |
| Preonzo (extensometer 4) | 51.00 | 51 | 1 |
| Preonzo (extensometer 5) | 64.00 | 65 | 1 |
| Preonzo (reflector 2) | 36.00 | 36 | 1 |
| Preonzo (reflector 4) | 34.00 | 34 | 1 |
| Preonzo (reflector 5) | 33.00 | 33 | 1 |
| Preonzo (reflector 8) | 33.00 | 33 | 1 |
| Preonzo (reflector 9) | 66.00 | 67 | 1 |
| Puigcercós (area 4) | 2116.87 | 19 | N/A |
| Puigcercós (area 6) | 1753.69 | 18 | N/A |
| Puigcercós (area 7) | 2120.11 | 20 | N/A |
| Puigcercós (area 9) | 1744.84 | 18 | N/A |
| Road slope (event 1) | 1.96 | 48 | 0.042 |
| Road slope (event 2) | 9.33 | 225 | 0.042 |
| Road slope (event 3) | 0.17 | 49 | 0.0035 |
| Road slope (event 4) | 0.08 | 48 | 0.0017 |
| Road slope (event 5) | 0.26 | 76 | 0.0035 |
| Road slope (event 6) | 1.58 | 76 | 0.021 |
| Road slope (event 7) | 0.83 | 40 | 0.021 |
| Road slope (event 8) | 0.21 | 60 | 0.0035 |
| Road slope (event 9) | 9.92 | 238 | 0.042 |
| Road slope (event 10) | 0.68 | 99 | 0.0069 |

**Table S13 (continued). Parameters of the LPPLS and PLS calibration to landslide data (49 events and 94 time series monitoring data in total).**

|  | Calibration window size (day) | Number of data points | Aggregation interval (day) |
|---|---|---|---|
| Roesgrenda | 35.55 | 79 | N/A |
| Takabayama | 46.00 | 47 | 1 |
| Vajont (bench mark 2) | 69.00 | 70 | 1 |
| Vajont (bench mark 4) | 61.00 | 62 | 1 |
| Vajont (bench mark 6) | 72.00 | 73 | 1 |
| Vajont (bench mark 58) | 98.00 | 99 | 1 |
| Veslemannen (radar point 1) | 108.00 | 108 | 1 |
| Veslemannen (radar point 2) | 91.00 | 91 | 1 |
| Veslemannen (radar point 3) | 108.00 | 108 | 1 |
| Veslemannen (radar point 4) | 107.00 | 107 | 1 |
| Veslemannen (radar point 5) | 104.00 | 104 | 1 |
| Veslemannen (radar point 6) | 102.00 | 102 | 1 |
| Veslemannen (radar point 7) | 101.00 | 101 | 1 |
| Welland (point 1) | 2.96 | 71 | 0.042 |
| Welland (point 2) | 4.83 | 116 | 0.042 |
| Xinmo | 728.27 | 34 | N/A |
| Xintan | 221.87 | 31 | 7 |

Note: if no aggregation treatment is applied to the data, the aggregation interval is indicated as N/A.

**Table S14. Parameters of the LPPLS and PLS calibration to rockburst data (11 events and 11 time series monitoring data in total).**

|  | Calibration window size (day) | Number of data points | Aggregation interval (day) |
|---|---|---|---|
| Coal mine (cut-through #4) | 1.33 | 64 | 0.021 |
| Coal mine (cut-through #5) | 2.58 | 63 | 0.042 |
| Gold mine (event 1) | 4.41 | 21 | N/A |
| Gold mine (event 2) | 1.34 | 14 | N/A |
| Gold mine (event 3) | 11.04 | 108 | N/A |
| Gold mine (event 4) | 43.01 | 125 | N/A |
| Gold mine (event 5) | 12.41 | 77 | N/A |
| Gold mine (event 6) | 5.77 | 39 | N/A |
| Gold mine (event 7) | 63.18 | 46 | N/A |
| Gold mine (event 8) | 114.20 | 23 | N/A |
| Platinum mine | 8.08 | 195 | 0.042 |

Note: if no aggregation treatment is applied to the data, the aggregation interval is indicated as N/A.

**Table S15. Parameters of the LPPLS and PLS calibration to glacier data (17 events and 21 time series monitoring data in total).**

|  | Calibration window size (day) | Number of data points | Aggregation interval (day) |
|---|---|---|---|
| Amery | 2240.00 | 321 | 7 |
| Eiger glacier (2001 event) | 5.15 | 64 | N/A |
| Eiger glacier (2016 event) | 39.00 | 40 | N/A |
| Grandes Jorasses (2014 event, prism 13) | 178.00 | 179 | 1 |
| Grandes Jorasses (2014 event, prism 14) | 161.00 | 162 | 1 |
| Grandes Jorasses (2020 event) | 7.75 | 187 | N/A |
| Gruben | 3.21 | 12 | N/A |
| Mönch | 66.00 | 67 | 1 |
| Planpincieux (event 1) | 67.00 | 68 | 1 |
| Planpincieux (event 2) | 57.00 | 58 | 1 |
| Planpincieux (event 3) | 52.00 | 53 | 1 |
| Planpincieux (event 4) | 20.00 | 21 | 1 |
| Planpincieux (event 5) | 30.00 | 31 | 1 |
| Planpincieux (event 6) | 46.00 | 47 | 1 |
| UK211 | 428.43 | 30 | N/A |
| Weisshorn (1973 event) | 252.70 | 31 | 7 |
| Weisshorn (2005 event, #103) | 23.26 | 54 | N/A |
| Weisshorn (2005 event, #104) | 23.26 | 59 | N/A |
| Weisshorn (2005 event, #105) | 23.26 | 62 | N/A |
| Weisshorn (2005 event, #106) | 23.26 | 56 | N/A |
| Weissmies | 75.00 | 76 | 1 |

Note: if no aggregation treatment is applied to the data, the aggregation interval is indicated as N/A.

**Table S16. Parameters of the LPPLS and PLS calibration to volcano data (32 events and 34 time series monitoring data in total).**

|  | Calibration window size (day) | Number of data points | Aggregation interval (day) |
|---|---|---|---|
| Adatara | 1127.00 | 162 | 7 |
| Asama | 238.00 | 239 | 1 |
| Augustine | 1141.00 | 164 | 7 |
| Axial Seamount | 3171.00 | 454 | 7 |
| Bezymianny | 29.72 | 20 | N/A |
| Etna (1989 event) | 322.00 | 47 | 7 |
| Etna (2013 event) | 140.00 | 141 | 1 |
| Hierro | 112.00 | 113 | 1 |
| Kilauea (1971 event) | 34.00 | 69 | 0.5 |
| Kilauea (1972 event) | 392.00 | 57 | 7 |
| Kilauea (1983 event) | 322.00 | 47 | 7 |
| Kujusan | 101.00 | 102 | 1 |
| Mauna Loa | 129.00 | 130 | 1 |
| Merapi (2006 event) | 62.00 | 63 | 1 |
| Merapi (2010 event) | 34.00 | 35 | 1 |
| Pinatubo | 32.00 | 33 | 1 |
| Redoubt (1989 event) | 266.00 | 39 | 7 |
| Redoubt (2009 event) | 57.00 | 58 | 1 |
| Ruapehu (1995 event) | 153.00 | 154 | 1 |
| Ruapehu (1996 event) | 102.00 | 103 | 1 |
| Ruapehu (2006 event) | 282.00 | 95 | 3 |
| Sakurajima | 92.00 | 93 | 1 |
| Sierra Negra (2005 event) | 910.00 | 66 | 14 |
| Sierra Negra (2018 event) | 1638.00 | 118 | 14 |
| St. Helens (1980 event) | 59.00 | 60 | 1 |
| St. Helens (1981 event) | 142.00 | 143 | 1 |
| St. Helens (1982 event, seismic data) | 52.00 | 53 | 1 |
| St. Helens (1982 event, radial tilt) | 58.00 | 59 | 1 |
| St. Helens (1982 event, tangential tilt) | 59.00 | 60 | 1 |
| St. Helens (1985 event) | 37.00 | 38 | 1 |
| Soufriere Hills | 265.86 | 30 | 1 |
| Tokachidake | 224.00 | 33 | 7 |
| Unzen | 238.00 | 35 | 7 |
| Yakedake | 234.00 | 79 | 3 |

Note: if no aggregation treatment is applied to the data, the aggregation interval is indicated as N/A.